\documentclass[article,aps, prd,  superscriptaddress, showpacs, longbibliography,floatfix,nofootinbib]{revtex4-2}

\usepackage{bm, amsmath, amssymb, relsize, amsfonts, mathrsfs, multirow, braket, siunitx, color, booktabs, arydshln, tabularx}
\usepackage{graphicx}
\graphicspath{{./figures/}}

\DeclareSIUnit \parsec {pc}

\usepackage[dvipsnames]{xcolor}
\usepackage[unicode]{hyperref}
\hypersetup{
  colorlinks=true,
  citecolor=RoyalBlue,
  linkcolor=blue,
  urlcolor=blue
}

\usepackage{array}
\usepackage{cancel}
\usepackage{orcidlink}
\usepackage{dcolumn}
\usepackage{bm}
\usepackage{paralist}
\usepackage{epsfig}
\usepackage{mathrsfs}
\usepackage{comment}
\usepackage{color, colortbl}
\usepackage[inline]{enumitem}
\usepackage{soul}
\usepackage{adjustbox}
\usepackage{float}
\usepackage{amsmath}
\usepackage[toc,page]{appendix}

\usepackage{tikz}
\usetikzlibrary{shapes.geometric, arrows, positioning}

\tikzstyle{startstop} = [ellipse, draw, text centered, minimum width=3.5cm, minimum height=1cm]
\tikzstyle{process} = [rectangle, draw, text centered, minimum width=3.5cm, minimum height=1cm]
\tikzstyle{arrow} = [thick, ->, >=stealth]

\DeclareMathAlphabet{\mathpzc}{OT1}{pzc}{m}{it}
	
\definecolor{LightCyan}{rgb}{0.88,1,1}
\definecolor{lightgray}{gray}{0.9}

\def \IITGn     {Department of Physics, Indian Institute of Technology Gandhinagar, Gandhinagar, Gujarat 382355, India.\vspace*{4pt}}
\def \IITGnCS     {Department of Computer Science \& Engineering, Indian Institute of Technology Gandhinagar, Gandhinagar, Gujarat 382355, India.\vspace*{4pt}}
\def \HRI     {Regional Centre for Accelerator-based Particle Physics, Harish-Chandra Research
Institute, A CI of Homi Bhabha National Institute, Chhatnag Road, Jhunsi, Prayagraj 211019, India}

\begin{document}

\title{Constraining the 3HDM Parameter Space using Active Learning}

\author{\textsc{Nipun Batra}\orcidlink{0000-0002-0736-7169}\vspace*{7pt}}
\email{nipun.batra@iitgn.ac.in}
\affiliation{\IITGnCS}

\author{\textsc{Baradhwaj Coleppa}\orcidlink{0000-0002-8761-3138}}
\email{baradhwaj@iitgn.ac.in }
\affiliation{\IITGn}

\author{\textsc{Akshat Khanna}\orcidlink{0000-0002-2322-5929}}
\email{khanna\textunderscore akshat@iitgn.ac.in}
\affiliation{\IITGn}

\author{\textsc{Santosh Kumar Rai}\orcidlink{0000-0002-4130-6992}\vspace*{7pt}}
\email{skrai@hri.res.in}
\affiliation{\HRI}

\author{\textsc{Agnivo Sarkar}\orcidlink{0000-0001-9596-1936}\vspace*{7pt}}
\email{agnivosarkar@hri.res.in}
\affiliation{\HRI}

\begin{flushright}
    HRI-RECAPP-2025-03
\end{flushright}

\begin{abstract}
One of the standard ways to study scenarios beyond the Standard Model involves extending the Higgs Sector. This work examines the Three Higgs Doublet Model (3HDM) in a Type-Z or democratic setup, where each Higgs doublet couples exclusively to a specific type of fermion. The particle spectrum of the 3HDM includes four charged Higgs bosons, two CP-odd scalars, and three CP-even scalars. This work investigates the allowed mass and coupling parameter space in the Type-Z 3HDM after imposing all theoretical and experimental constraints. We extract the allowed parameter space under three distinct alignment-limit conditions or mass hierarchies leveraging machine learning techniques. Specifically, we analyze scenarios where the 125 GeV Higgs is the lightest, an intermediary, or the heaviest CP-even Higgs boson. Our findings indicate that while a single lighter CP-even Higgs boson below 125 GeV still remains a possibility, the presence of two lighter Higgses is ruled out.
\end{abstract}
\maketitle
\section{Introduction}
\label{sec:intro}
The Standard Model (SM) of particle physics \cite{Salam:1968rm,Weinberg:1967tq,Glashow:1961tr} based on the $SU(3)_c\times SU(2)_L\times U(1)_Y$ gauge group, has enjoyed tremendous theoretical and experimental success thus far. With the discovery of a Higgs boson of mass 125 GeV with properties consistent with those of the SM-Higgs \cite{ATLAS:2012yve,CMS:2012qbp}, the particle spectrum of the SM has been firmly established. There is, still, some room for error when it comes to the couplings of the discovered Higgs boson with the SM particles which keeps alive the hope of constructing models that supplant the SM dynamics - these collectively fall under the umbrella of ``Beyond the Standard Model" scenarios or BSM for short. We also know by now that the SM cannot be a complete theory - even leaving aside the fact that there is no room for gravity in the SM, there are still tantalizing experimental evidences like the origins of neutrino mass and dark matter that undoubtedly tell us that there needs to be physics beyond the SM. In addition, vexing theoretical questions of the naturalness problem (related to the stability of the weak scale) that have long prompted theorists to look for possible solutions extending the SM dynamics.

One of the earliest proposed BSM scenarios was Supersymmetry (SUSY) - in addition to stabilizing the running of the Higgs mass, SUSY theories also offered the prospect of gauge coupling unification. Considerations of anomaly cancellation necessitated that these theories necessarily invoke two Higgs doublets to engineer electroweak symmetry breaking (EWSB) and provide masses to the various gauge bosons and fermions. Along these lines, models with extended scalar sectors came to be also constructed as these were interesting in their own right. The earliest such attempts were along the lines of the ``Two Higgs Doublet Models" (2HDM) \cite{Branco:2011iw} - as the name suggests, these have two Higgs doublets and have been studied extensively in the literature \cite{Coleppa:2013dya,Chang:2012ve,Grinstein:2013npa}. The study of extended Higgs sectors is by no means confined to only the 2HDMs - the Georgi-Machacek model (which includes a Higgs triplet as well) \cite{Georgi:1985nv,Chanowitz:1985ug,deLima:2022yvn}, various other avatars of the 2HDM with extra inert scalars to serve as dark matter candidates \cite{Drozd:2014yla,Bhattacharya:2023qfs,Muhlleitner:2016mzt,Keus:2017ioh} etc. have also been studied. Interestingly, the Higgs data still leaves room for many of these extended Higgs sector scenarios and these are not (yet) completely ruled out.

Recently, the ``Three Higgs Doublet" models (3HDM) have also been introduced. These have the attractive possiblity of allowing for CP violation \cite{Logan:2020mdz,Akeroyd:2021fpf} in addition to offering a rich scope for phenomenology \cite{Dey:2023exa,CarcamoHernandez:2022vjk} given that they have many as yet undiscovered scalar particles in their spectrum, and dark matter prospects \cite{Boto:2024tzp,Kuncinas:2024zjq,Deng:2025dcq,Cordero:2017owj,Cordero-Cid:2018man,Aranda:2019vda,Hernandez-Sanchez:2020aop,Keus:2014jha,Keus:2014isa,Cordero-Cid:2016krd,Dey:2024epo}. There are many versions of 3HDM one can construct based on whether one wants EWSB to proceed via all three doublets or just one or two of them. In addition, various possibilities exist for patterns of Yukawa couplings in these models each with their own unique phenomenology. There are many papers in the literature that look at one or more aspects of the 3HDM \cite{Das:2025mqs,Altmannshofer:2025pjj,Romao:2024gjx,Bento_2022,Boto_2021,Das:2022gbm,Chakraborti:2021bpy,Akeroyd:2016ssd,Akeroyd:2018axd,Akeroyd:2019mvt,Akeroyd:2022ouy,Das:2019yad,Cree:2011uy,Akeroyd:2021fpf,Keus:2013hya}, and also address the possibility of whether the parameter space of the 3HDM is allowed with respect to specific theoretical (like unitarity and perturbativity considerations) and experimental constraints. Our goal in this paper is to look at a particular version of the 3HDM and analyze its parameter space, subjecting it to all known theoretical and experimental constraints. As we will lay out in the paper, this is a rather daunting task given the large number of parameters in the model made only tractable by the use of active learning algorithms. 

The paper is organized as follows: In Sec.~\ref{sec:model}, we lay out the specifics of the variant of 3HDM we work with, detailing the scalar spectrum of the model followed by the Yukawa sector and the relevant couplings. Then, in Sec.~\ref{sec:alignment}, we discuss the alignment limit in this model - given that there are three CP-even Higgs bosons in the spectrum, any one of them can be SM-like. Then in Sec.~\ref{sec:constraints}, we lay out all the theoretical and experimental constraints of relevance that we will use to isolate the surviving regions of the 3HDM parameter space, followed in Sec.~\ref{sec:methodology} by a discussion of the active learning strategy we adopt. We present our results in Sec.~\ref{sec:results} and conclude in Sec.~\ref{sec:conclusions}.
\section{3HDM: Masses And Couplings}
\label{sec:model}
In this section, we discuss in detail the basic construction of the 3HDM - we begin with the scalar sector and carefully diagonalize the various mass matrices isolating the Higgs spectra. We then discuss Electroweak Symmetry Breaking (EWSB) in the model and briefly discuss the fermion sector. We finally end this section with a calculation of the relevant couplings of interest.
\subsection{Scalar Spectrum}
As the name implies, the Three Higgs Doublet Model (3HDM) involves extending the SM scalar sector by the addition of two more $SU(2)_L$ scalar doublets - these can be represented in the usual form 

\begin{equation}
		\Phi_k = \begin{pmatrix}
			\phi_k^+ \\ \frac{v_k+p_k+in_k}{\sqrt{2}}
		\end{pmatrix},
  \label{eq:phi_form}
\end{equation}
where $k=1,2,3$. While one could construct the most general $SU(2)_L \times U(1)_Y$ invariant Lagrangian with these three fields, we first restrict the possible terms by imposing an additional $Z_3$ symmetry under which the three Higgs fields transform as
\begin{equation}
     	\label{eq:phitrans}
     	\Phi_1 \rightarrow \omega \Phi_1, \; \; \Phi_2 \rightarrow \omega^2 \Phi_2, \;\textrm{and} \;\Phi_3 \rightarrow \Phi_3,
\end{equation}
where $\omega = e^{2\pi i/3}$, are the cube roots of unity. The most general $SU(2)_L \times U(1)_Y$ invariant potential that admits this $Z_3$ symmetry is given by

\begin{equation}
     	\label{eq:scalarpot}
     	\begin{split}
     		V & = m_{11}^2(\Phi_1^\dagger\Phi_1) + m_{22}^2(\Phi_2^\dagger\Phi_2) + m_{33}^2(\Phi_3^\dagger\Phi_3)
     		\\ & + \lambda_1(\Phi_1^\dagger\Phi_1)^2 + \lambda_2(\Phi_2^\dagger\Phi_2)^2 + \lambda_3(\Phi_3^\dagger\Phi_3)^2 \\ & + \lambda_{4}(\Phi_1^\dagger\Phi_1)(\Phi_2^\dagger\Phi_2) + \lambda_{5}(\Phi_1^\dagger\Phi_1)(\Phi_3^\dagger\Phi_3) + \lambda_{6}(\Phi_2^\dagger\Phi_2)(\Phi_3^\dagger\Phi_3) \\ & + \lambda_{7}(\Phi_1^\dagger\Phi_2)(\Phi_2^\dagger\Phi_1) + \lambda_{8}(\Phi_1^\dagger\Phi_3)(\Phi_3^\dagger\Phi_1) +  \lambda_{9}(\Phi_2^\dagger\Phi_3)(\Phi_3^\dagger\Phi_2) \\ & + [\lambda_{10}(\Phi_1^\dagger\Phi_2)(\Phi_1^\dagger\Phi_3) + \lambda_{11}(\Phi_1^\dagger\Phi_2)(\Phi_3^\dagger\Phi_2) + \lambda_{12}(\Phi_1^\dagger\Phi_3)(\Phi_2^\dagger\Phi_3) + h.c.]. \\ &
     	\end{split} 
\end{equation}
Here, in general $\lambda_{1,...,9}$ are real parameters (to guarantee Hermiticity of the Lagrangian), while $ \lambda_{10}, \lambda_{11}, \lambda_{12}$ can be complex. The complex phases in the potential induce mixing between the CP-odd and CP-even terms, and there is no intrinsic phase relationship to eliminate these mixings. The only way to suppress them is to set the imaginary component of $\lambda_{12}$ to zero, which consequently nullifies\footnote{The details of these calculations are given in Appendix \ref{sec:appmixing}.} the complex parts of $\lambda_{10}$ and $\lambda_{11}$ as well.

 Before proceeding, we note here for completeness' sake that by applying the minimization conditions to the potential, the three real parameters $m_{11}^2$, $m_{22}^2$ and $m_{33}^2$ can be traded for the couplings $\lambda_{1,...,12}$. 

\begin{equation}
     	\begin{split}
     		m_{11}^2 & = -\lambda_{1}v_1^2 - \frac{v_2^2}{2}(\lambda_{4}+\lambda_{7}) - \frac{v_3^2}{2}(\lambda_{5}+\lambda_{8}) -Re(\lambda_{10})v_2v_3 - \frac{v_2v_3}{2v_1}[Re(\lambda_{11})v_2+Re(\lambda_{12})v_3],\\
     		m_{22}^2 & = -\lambda_{2}v_2^2 - \frac{v_1^2}{2}(\lambda_{4}+\lambda_{7}) - \frac{v_3^2}{2}(\lambda_{6}+\lambda_{9}) -Re(\lambda_{11})v_1v_3 - \frac{v_1v_3}{2v_2}[Re(\lambda_{10})v_1+Re(\lambda_{12})v_3],\,\textrm{and} \\
     		m_{33}^2 & = -\lambda_{3}v_3^2 - \frac{v_1^2}{2}(\lambda_{5}+\lambda_{8}) - \frac{v_2^2}{2}(\lambda_{6}+\lambda_{9}) -Re(\lambda_{12})v_1v_2 - \frac{v_1v_2}{2v_3}[Re(\lambda_{10})v_1+Re(\lambda_{11})v_2]. 
     	\end{split}
\end{equation}

We begin with the  relevant mass terms for the CP-even Higgs bosons that are readily deduced from the scalar potential in Eqn.~\ref{eq:scalarpot} and can be written as
	
\begin{equation*}
		V_p^{mass} \supset \begin{pmatrix}
			p_1 & p_2 & p_3
		\end{pmatrix} \frac{\mathcal{M}^2_S}{2} \begin{pmatrix}
			p_1 \\ p_2 \\ p_3 
		\end{pmatrix},
\end{equation*}
with the elements of the $\mathcal{M}^2_S$ given by

\begin{equation}
    \begin{split}
        (\mathcal{M}^2_S)_{11} & = 2\lambda_{1}v_1^2 - \frac{v_2v_3}{2v_1}[Re(\lambda_{11})v_2+Re(\lambda_{12})v_3]\\     (\mathcal{M}^2_S)_{12} & = (\lambda_{4}+\lambda_{7})v_1v_2 + Re(\lambda_{10})v_1v_3 + Re(\lambda_{11})v_2v_3 + \frac{v_3^2}{2}Re(\lambda_{12})\\
        (\mathcal{M}^2_S)_{13} & = (\lambda_{5}+\lambda_{8})v_1v_3 + Re(\lambda_{10})v_1v_2 + Re(\lambda_{12})v_2v_3 + \frac{v_2^2}{2}Re(\lambda_{11})\\
        (\mathcal{M}^2_S)_{22} & =  2\lambda_{2}v_2^2 - \frac{v_1v_3}{2v_2}[Re(\lambda_{10})v_1+Re(\lambda_{12})v_3] \\
        (\mathcal{M}^2_S)_{23} & = (\lambda_{6}+\lambda_{9})v_2v_3 + Re(\lambda_{11})v_1v_2 + Re(\lambda_{12})v_1v_3 + \frac{v_1^2}{2}Re(\lambda_{10}) \\
        (\mathcal{M}^2_S)_{33} & = 2\lambda_{3}v_3^2 - \frac{v_1v_2}{2v_3}[Re(\lambda_{10})v_1+Re(\lambda_{11})v_2]
    \end{split}
    \label{eq:scalarmassterms}
\end{equation}	
This real symmetric mass matrix can be diagonalized by an orthogonal transformation by a matrix $O_\alpha$, defined as
\begin{equation}
    \label{eq:matalphtrans}
    O_\alpha = R_3.R_2.R_1,
\end{equation}
where 
\begin{equation}
    R_1 =  \begin{pmatrix} c_{\alpha 1} & s_{\alpha 1} & 0 \\ -s_{\alpha 1} & c_{\alpha 1} & 0 \\ 0 & 0 & 1
    \end{pmatrix}, \; \;R_2 = \begin{pmatrix}
        c_{\alpha_2} & 0 & s_{\alpha_2} \\ 0 & 1 & 0 \\ -s_{\alpha_2} & 0 & c_{\alpha_2}
    \end{pmatrix}, \; \textrm{and} \; R_3 = \begin{pmatrix}
        1 & 0 & 0 \\ 0 & c_{\alpha_3} & s_{\alpha_3} \\ 0 & -s_{\alpha_3} & c_{\alpha_3}
    \end{pmatrix},
\end{equation}
and thus
\begin{equation}
    \label{eq:matalphtrans}
    O_\alpha = \begin{pmatrix}
        c_{\alpha 1} c_{\alpha 2} & c_{\alpha 2} s_{\alpha 1} & s_{\alpha 2} \\ -c_{\alpha 3} s_{\alpha 1} - s_{\alpha 3} s_{\alpha 2} c_{\alpha 1} & c_{\alpha 3} c_{\alpha 1} - s_{\alpha 3} s_{\alpha 2} s_{\alpha 1} & s_{\alpha 3} c_{\alpha 2} \\ s_{\alpha 3} s_{\alpha 1} - c_{\alpha 3} s_{\alpha 2} c_{\alpha 1} & -s_{\alpha 3} c_{\alpha 1} - c_{\alpha 3} s_{\alpha 2} s_{\alpha 1} & c_{\alpha 3} c_{\alpha 2} 
    \end{pmatrix}.
\end{equation}
The diagonalization condition implies
\begin{equation}
    O_\alpha.\mathcal{M}^2_S.O_\alpha^T = \begin{pmatrix}
        m_{H 1}^2 & 0 & 0 \\ 0 & m_{H 2}^2 & 0 \\ 0 & 0 & m_{H 3}^2
    \end{pmatrix}
\end{equation}
from which one can write down the eigenvalues $m_{Hk}^2$. However these expressions are a little unwieldy, and for the purposes of this paper, it is more convenient to set the masses as external tunable parameters and rewrite the couplings $\lambda_i$'s in terms of them. This can be accomplished in a more straightforward manner by noting that 
\begin{equation}
    \mathcal{M}^2_S = O_\alpha^T.\begin{pmatrix}
        m_{H 1}^2 & 0 & 0 \\ 0 & m_{H 2}^2 & 0 \\ 0 & 0 & m_{H 3}^2
    \end{pmatrix}.O_\alpha
    \label{eq:ms2}
\end{equation}
and equating Eqns.~\ref{eq:scalarmassterms} and \ref{eq:ms2}. The resulting equations expressing the couplings $\lambda_{1....6}$ in terms of the other couplings and the eigenvalues $m_{Hk}^2$ are given below:
\begin{equation}
    \begin{split}
        \lambda_{1} & = m_{H1}^2\left(\frac{c_{\alpha 1}^2c_{\alpha 2}^2}{2c_{\beta 1}^2c_{\beta 2}^2v^2}\right) + m_{H2}^2\left(\frac{(c_{\alpha 3}s_{\alpha 1}+c_{\alpha 1}s_{\alpha 2}s_{\alpha 3})^2}{2c_{\beta 1}^2c_{\beta 2}^2v^2}\right) + m_{H3}^2\left(\frac{(c_{\alpha 1}c_{\alpha 3}s_{\alpha 2}-s_{\alpha 1}s_{\alpha 3})^2}{2c_{\beta 1}^2c_{\beta 2}^2v^2}\right)\\ & + Re(\lambda_{11})\frac{s_{\beta 1}^2s_{\beta 2}}{4c_{\beta 1}^3c_{\beta 2}} + Re(\lambda_{12})\frac{s_{\beta 1}s_{\beta 2}^2}{4c_{\beta 1}^3c_{\beta 2}^2} \\ \lambda_{2} & = m_{H1}^2\left(\frac{s_{\alpha 1}^2c_{\alpha 2}^2}{2s_{\beta 1}^2c_{\beta 2}^2v^2}\right) + m_{H2}^2\left(\frac{(c_{\alpha 1}c_{\alpha 3}-s_{\alpha 1}s_{\alpha 2}s_{\alpha 3})^2}{2s_{\beta 1}^2c_{\beta 2}^2v^2}\right) + m_{H3}^2\left(\frac{(c_{\alpha 1}s_{\alpha 3}+s_{\alpha 1}s_{\alpha 2}c_{\alpha 3})^2}{2s_{\beta 1}^2c_{\beta 2}^2v^2}\right) \\ & + Re(\lambda_{10})\frac{c_{\beta 1}^2s_{\beta 2}}{4s_{\beta 1}^3c_{\beta 2}} + Re(\lambda_{12})\frac{c_{\beta 1}s_{\beta 2}^2}{4s_{\beta 1}^3c_{\beta 2}^2}  \\ \lambda_{3} & = m_{H1}^2\left(\frac{s_{\alpha 2}^2}{2s_{\beta 2}^2v^2}\right) + m_{H2}^2\left(\frac{c_{\alpha 2}^2s_{\alpha 3}^2}{2s_{\beta 2}^2v^2}\right) +m_{H3}^2\left(\frac{c_{\alpha 2}^2c_{\alpha 3}^2}{2s_{\beta 2}^2v^2}\right) + Re(\lambda_{10})\left(\frac{c_{\beta 1}^2c_{\beta 2}^3s_{\beta 1}}{4s_{\beta 2}^3}\right) + Re(\lambda_{11})\left(\frac{s_{\beta 1}^2c_{\beta 2}^3c_{\beta 1}}{4s_{\beta 2}^3}\right)  \\ \lambda_{4} & = m_{H1}^2\left(\frac{c_{\alpha 1}s_{\alpha 1}c_{\alpha 2}^2}{c_{\beta 1}s_{\beta 1}c_{\beta 2}^2v^2}\right) - m_{H2}^2\left(\frac{(s_{\alpha 1}c_{\alpha 3}+c_{\alpha 1}s_{\alpha 2}s_{\alpha 3})(c_{\alpha 1}c_{\alpha 3}-s_{\alpha 1}s_{\alpha 2}s_{\alpha 3})}{c_{\beta 1}s_{\beta 1}c_{\beta 2}^2v^2}\right)  \\ & - m_{H3}^2\left(\frac{(c_{\alpha 1}s_{\alpha 3}+s_{\alpha 1}s_{\alpha 2}c_{\alpha 3})(s_{\alpha 1}s_{\alpha 3}-c_{\alpha 1}s_{\alpha 2}c_{\alpha 3})}{c_{\beta 1}s_{\beta 1}c_{\beta 2}^2v^2}\right) - Re(\lambda_{10})\frac{s_{\beta 2}}{s_{\beta 1}c_{\beta 2}} \\ &  - Re(\lambda_{11})\frac{s_{\beta 2}}{c_{\beta 1}c_{\beta 2}} - Re(\lambda_{12})\frac{s_{\beta 2}^2}{2c_{\beta 1}s_{\beta 1}c_{\beta 2}^2} - \lambda_{7}\\
        \lambda_{5} & = m_{H1}^2\left(\frac{c_{\alpha 1}s_{\alpha 2}c_{\alpha 2}}{c_{\beta 1}s_{\beta 2}c_{\beta 2}v^2}\right) - m_{H2}^2\left(\frac{c_{\alpha 2}s_{\alpha 3}(s_{\alpha 1}c_{\alpha 3}+c_{\alpha 1}s_{\alpha 2}s_{\alpha 3})}{c_{\beta 1}s_{\beta 2}c_{\beta 2}v^2}\right) \\ & + m_{H3}^2\left(\frac{c_{\alpha 2}c_{\alpha 3}(s_{\alpha 1}s_{\alpha 3}-c_{\alpha 1}s_{\alpha 2}c_{\alpha 3})}{c_{\beta 1}s_{\beta 2}c_{\beta 2}v^2}\right) - Re(\lambda_{10})\frac{s_{\beta 1}c_{\beta 2}}{s_{\beta 2}} - Re(\lambda_{11})\frac{s_{\beta 1}^2c_{\beta 2}}{2c_{\beta 1}s_{\beta 2}} - Re(\lambda_{12})\frac{s_{\beta 1}}{c_{\beta 1}} - \lambda_{8} \\ \lambda_{6} & = m_{H1}^2\left(\frac{s_{\alpha 1}s_{\alpha 2}c_{\alpha 2}}{s_{\beta 1}s_{\beta 2}c_{\beta 2}v^2}\right) + m_{H2}^2\left(\frac{c_{\alpha 2}s_{\alpha 3}(c_{\alpha 1}c_{\alpha 3}-s_{\alpha 1}s_{\alpha 2}s_{\alpha 3})}{s_{\beta 1}s_{\beta 2}c_{\beta 2}v^2}\right) \\ & - m_{H3}^2\left(\frac{c_{\alpha 2}c_{\alpha 3}(c_{\alpha 1}s_{\alpha 3}+s_{\alpha 1}s_{\alpha 2}c_{\alpha 3})}{s_{\beta 1}s_{\beta 2}c_{\beta 2}v^2}\right) - Re(\lambda_{10})\frac{c_{\beta 1}^2c_{\beta 2}}{2s_{\beta 1}s_{\beta 2}} - Re(\lambda_{11})\frac{c_{\beta 1}c_{\beta 2}}{s_{\beta 2}} - Re(\lambda_{12})\frac{c_{\beta 1}}{s_{\beta 1}} - \lambda_{9}. 
    \end{split}
    \label{eqn:lambda-masses}
\end{equation}
The gauge eigenstates can be written down in terms of the mass eigenstates using the transformation matrix in a straightforward manner: 

\begin{equation}
    \begin{split}
        H_1 & = c_{\alpha_2} c_{\alpha_1} p_1 + c_{\alpha_2} s_{\alpha_1} p_2 + s_{\alpha_2}p_3, \\
        H_2 & = -(c_{\alpha_3} s_{\alpha_1} + s_{\alpha_3} s_{\alpha_2} c_{\alpha_1})p_1 + (c_{\alpha_3} c_{\alpha_1} - s_{\alpha_3} s_{\alpha_2} s_{\alpha_1})p_2+(s_{\alpha_3} c_{\alpha_2})p_3,\,\textrm{and} \\
        H_3 & =  (s_{\alpha_3} s_{\alpha_1} - c_{\alpha_3} s_{\alpha_2} c_{\alpha_1})p_1 -  (s_{\alpha_3} c_{\alpha_1} + c_{\alpha_3} s_{\alpha_2} s_{\alpha_1})p_2 + (c_{\alpha_3} c_{\alpha_2}) p_3.
    \end{split}
\end{equation}
Having thus diagnolized the CP-even states, we now move on to the charged Higgs sector.

The mass terms for the charged Higgses can similarly be extracted from the scalar potential given in equation ~\ref{eq:scalarpot} - we write the terms symbolically as
\begin{equation*}
     	V_C^{mass} \supset \begin{pmatrix}
     		\phi_1^- & \phi_2^- & \phi_3^-
     	\end{pmatrix} \frac{\mathcal{M}^2_{\phi^{\pm}}}{2} \begin{pmatrix}
     		\phi_1^+ \\ \phi_2^+ \\ \phi_3^+ 
     	\end{pmatrix},
\end{equation*}
where, $\mathcal{M}^2_{\phi^{\pm}}$ is the $3 \times 3$ charged Higgs mass matrix. To diagonalize this, we first employ a similarity transformation using the matrix $O_\beta$: 
\begin{equation*}
    (B_C)^2 = O_\beta.\mathcal{M}^2_{\phi^{\pm}}.O_\beta^T,
\end{equation*}
where
\begin{eqnarray}
    \label{eq:betatrans}
    O_\beta &=& \begin{pmatrix}
        c_{\beta 2} & 0 & s_{\beta 2} \\ 0 & 1 & 0 \\ -s_{\beta 2} & 0 & c_{\beta 2}
    \end{pmatrix} \begin{pmatrix} c_{\beta 1} & s_{\beta 1} & 0 \\ -s_{\beta 1} & c_{\beta 1} & 0 \\ 0 & 0 & 1
    \end{pmatrix} \nonumber\\
     &=& \begin{pmatrix}
        c_{\beta 2}c_{\beta 1} & c_{\beta 2}s_{\beta 1} & s_{\beta 2} \\ -s_{\beta 1} & c_{\beta 1} & 0 \\ -c_{\beta 1}s_{\beta 2} & -s_{\beta 1}s_{\beta 2} & c_{\beta 2}
    \end{pmatrix}
\end{eqnarray}
Here, $\tan\beta_1=v_2/v_1$ and $\tan\beta_2=v_3/\sqrt{v_1^2+v_2^2}$. This transformation simplifies $\mathcal{M}^2_{\phi^{\pm}}$ into a block diagonal form, isolating the mass subspaces. The presence of the zero eigenvalue corresponds to the Goldstone boson to be eaten by the SM $W^\pm$ via the Higgs mechanism.     
\begin{equation*}
     	(B_C)^2 =  \begin{pmatrix}
     		0 & 0 & 0 \\ 0 & \mathcal{M}^2_{22} & \mathcal{M}^2_{23} \\ 0 & \mathcal{M}^{2*}_{23} & \mathcal{M}^2_{33}\\
     	\end{pmatrix}
\end{equation*} 
The above mass matrix can now be fully diagonalized emplying another orthogonal rotation:
     
\begin{equation*}
     	O_{\gamma 2}.(B_C)^2.O_{\gamma 2}^\dagger = \begin{pmatrix}
     		0 & 0 & 0 \\ 0 & m_{H^{\pm}_2}^2 & 0 \\ 0 & 0 & m_{H^{\pm}_3}^2
     	\end{pmatrix}
\end{equation*}
As in the case of the CP-even scalars, we can derive explicit relationships between the scalar potential parameters $\lambda_{7}$, $\lambda_{8}$, and $\lambda_{9}$ and the masses of the charged Higgs bosons, mixing angles and the vacuum expectation values (vevs). These relationships are crucial for connecting theoretical parameters with experimental measurements, providing a pathway to validate the model against observed data. The detailed steps and calculations involved in this diagonalization process, including the specific forms of the matrices $O_\beta$ and $O_{\gamma 2}$, and the resulting expressions for the masses in terms of $\lambda_{7}$, $\lambda_{8}$, and $\lambda_{9}$, are outlined in Appendix~\ref{sec:appcharged}.

Finally, writing the mass terms for the CP-Odd Higgs in a similar fashion
\begin{equation*}
        V_n^{mass} \supset \begin{pmatrix}
            n_1 & n_2 & n_3
        \end{pmatrix} \frac{\mathcal{M}^2_n}{2} \begin{pmatrix}
            n_1 \\ n_2 \\ n_3 
        \end{pmatrix}, 
\end{equation*}
we can diagonalize the pseudoscalar mass matrix $\mathcal{M}^2_n$ exactly like in the previous case, i.e., do an orthogonal rotation with $O_\beta$ to get it into a block diagonal form followed by a rotation by a matrix $O_{\gamma 1}$ to fully diagonalize the matrix\footnote{Of course, it is entirely possible to do a single rotation like in Eqn.~\ref{eq:matalphtrans}.} to yield the physical masses and states. The detailed steps and calculations involved in this diagonalization process are documented in Appendix~\ref{sec:appendpseudo}. Once again, we trade the Lagrangian parameters for the masses ($m_{A_1}, m_{A_2}$), vevs, and the mixing angles.
\subsection{Yukawa Sector}

Tree-level Flavor-Changing Neutral Currents (FCNCs) are tightly constrained by experimental observations, necessitating mechanisms to suppress them effectively in theoretical models. To address this, we impose the Natural Flavor Conservation (NFC) criterion - this dictates that each type of fermion couples exclusively to a single Higgs doublet, thereby preventing FCNCs at tree level \cite{Glashow:1976nt}. In constructing the Yukawa Lagrangian, we adopt the Type-Z or the democratic setup, wherein each Higgs doublet contributes to the mass generation of a different type of fermion (i.e., one each for up-type quarks, down-type quarks, and leptons). This approach not only simplifies the theoretical framework but also aligns with the observed mass hierarchies and mixing angles in the fermion sector. 

\begin{equation}
		\label{eq:yukeq}
		\mathcal{L}_{Yukawa} = -[\bar{L}_L \Phi_1 \mathcal{G}_l l_R+\bar{Q}_L \Phi_2 \mathcal{G}_d d_R + \bar{Q}_L \tilde{\Phi}_3 \mathcal{G}_u u_R  + h.c].
\end{equation}
The $\mathcal{G}_f$ are the Yukawa matrices. In terms of the fermion mass matrices they can be written as
\begin{equation*}
		\mathcal{G}_f = \frac{\sqrt{2} \mathcal{M}_f}{v_i}.
\end{equation*}
We work with a $Z_3$ symmetric potential as given in Eqn.~\ref{eq:phitrans} - for the Yukawa Lagrangian to remain invariant under the same, the right handed fermion fields transform as 
\begin{equation}
		\label{eq:fermtrans}
		d_R  \rightarrow \omega d_R , \; \; \; \; l_R \rightarrow \omega^2 l_R, \; \; \; \; u_R \rightarrow  u_R.
\end{equation}
\subsection{Gauge Boson Masses}
EWSB proceeds in the 3HDM in a fashion analagous to the SM and the 2HDM. Each of the three fields $\Phi_i$ (charged under the SM gauge symmetry) develop a vev $v_i$ (Eqn.~\ref{eq:phi_form}) to break the $SU(2)\times U(1)_Y$ down to $U(1)_{\textrm{em}}$. The covariant derivatives can be written as
\begin{equation}
    D_\mu\Phi_i=\partial_\mu\Phi_i+i\frac{g}{2}\sigma^a W_\mu^a\Phi_i+ i\frac{g'}{2}Y B_\mu\Phi_i,
    \label{eq:phi_covariant}
\end{equation}
where $g$ and $g'$ are the $SU(2)$ and $U(1)_Y$ couplings respectively. After symmetry breaking, the $W$ boson mass can be read off and is given by
\begin{equation*}
    m_W^2=\frac{g^2}{4}(v_1^2+v_2^2+v_3^2),
\end{equation*}
thus constraining the three vevs to obey $\sqrt{v_1^2+v_2^2+v_3^2}=v=$ 246 GeV.
\subsection{Couplings}
With the masses and eigenstates of the various Higgses in the model firmly in place, we are now in a position to calculate all the relevant couplings. For our purposes here, we stick to 3-point couplings. Defining
 \begin{equation}
    \begin{split}
     		k_1 & = c_{\beta 1}c_{\alpha 1}+s_{\beta 1}s_{\alpha 1} = cos(\alpha_1-\beta_1),\,\textrm{and}\\
     		k_2 & = s_{\beta 1}c_{\alpha 1}-c_{\beta 1}s_{\alpha 1} = sin(\beta_1-\alpha_1),
    \end{split}
\end{equation}
we tabulate, in Table~\ref{tab:couplings}, all couplings of the form $HVV$, $Hf\bar{f}$, and $H^+t\bar{b}$, where $H$ denotes any one of the CP-even or CP-odd scalars, and $V$ and $f$ are generic SM gauge boson and fermion respectively. The couplings in Table~\ref{tab:couplings} are presented as the scale factor with respect to the corresponding SM coupling (except for the charged Higgs, which, of course, has no analogue in the SM). The list of couplings of the form $HHV$ (with $H$ being a generic scalar/pseudoscalar) is given in Appendix \ref{sec:couplings-app}.
\begin{table}[h]
    \centering
    \begin{tabular}{l l}
    \toprule[1pt]
        Coupling & \ \ \ \ \ \ \ Coefficient \\
        \midrule[1pt]
        $\xi^{Z Z}_{H_1}$ & \ \ \ \ \ \ \ $c_{\beta 2}c_{\alpha 2}k_1+s_{\beta 2}s_{\alpha 2}$ \\
        $\xi^{Z Z}_{H_2}$ & \ \ \ \ \ \ \ $c_{\beta 2}c_{\alpha 3}k_2+(s_{\beta 2}c_{\alpha 2}-c_{\beta 2}k_1s_{\alpha 2})s_{\alpha 3}$ \\
        $\xi^{Z Z}_{H_3}$ & \ \ \ \ \ \ \ $-c_{\beta 2}s_{\alpha 3}k_2+(s_{\beta 2}c_{\alpha 2}-c_{\beta 2}k_1s_{\alpha 2})c_{\alpha 3}$ \\
        $\xi^{\bar{b} b}_{H_1}$ & \ \ \ \ \ \ \ $\frac{s_{\alpha_1}c_{\alpha_2}}{s_{\beta_1}c_{\beta_2}}$ \\
        $\xi^{\bar{b} b}_{H_2}$ & \ \ \ \ \ \ \ $\frac{(c_{\alpha_1}c_{\alpha_3}-s_{\alpha_1}s_{\alpha_2}s_{\alpha_3})}{s_{\beta_1}c_{\beta_2}}$ \\
        $\xi^{\bar{b} b}_{H_3}$ & \ \ \ \ \ \ \ $\frac{(c_{\alpha_1}s_{\alpha_3}+s_{\alpha_1}s_{\alpha_2}c_{\alpha_3})}{s_{\beta_1}c_{\beta_2}}$ \\
        $\xi^{\bar{t} t}_{H_1}$ & \ \ \ \ \ \ \ $\frac{s_{\alpha_2}}{s_{\beta_2}}$ \\
        $\xi^{\bar{t} t}_{H_2}$ & \ \ \ \ \ \ \ $\frac{c_{\alpha_2}s_{\alpha_3}}{s_{\beta_2}}$ \\
        $\xi^{\bar{t} t}_{H_3}$ & \ \ \ \ \ \ \ $\frac{c_{\alpha_2}c_{\alpha_3}}{s_{\beta_2}}$ \\
        $\xi^{\bar{e} e}_{H_1}$ & \ \ \ \ \ \ \ $\frac{c_{\alpha_1}c_{\alpha_2}}{c_{\beta_1}c_{\beta_2}}$ \\
        $\xi^{\bar{e} e}_{H_2}$ & \ \ \ \ \ \ \ $\frac{(s_{\alpha_1}c_{\alpha_3}+c_{\alpha_1}s_{\alpha_2}s_{\alpha_3})}{c_{\beta_1}c_{\beta_2}}$ \\
        $\xi^{\bar{e} e}_{H_3}$ & \ \ \ \ \ \ \ $\frac{(-s_{\alpha_1}s_{\alpha_3}+c_{\alpha_1}s_{\alpha_2}c_{\alpha_3})}{c_{\beta_1}c_{\beta_2}}$ \\
        $\xi^{\bar{b} b}_{A_2}$ & \ \ \ \ \ \ \ $\frac{c_{\beta_1}c_{\gamma_1}+s_{\beta_1}s_{\beta_2}s_{\gamma_1}}{s_{\beta_1}c_{\beta_2}}$ \\
        $\xi^{\bar{b} b}_{A_3}$ & \ \ \ \ \ \ \ $\frac{-c_{\beta_1}s_{\gamma_1}+s_{\beta_1}s_{\beta_2}c_{\gamma_1}}{s_{\beta_1}c_{\beta_2}}$ \\
        $\xi^{\bar{t} t}_{A_2}$ & \ \ \ \ \ \ \ $\frac{c_{\beta_2}s_{\gamma_1}}{s_{\beta_2}}$ \\
        $\xi^{\bar{t} t}_{A_3}$ & \ \ \ \ \ \ \ $\frac{c_{\beta_2}c_{\gamma_1}}{s_{\beta_2}}$ \\
        $\xi^{\bar{e} e}_{A_2}$ & \ \ \ \ \ \ \ $\frac{s_{\beta_1}c_{\gamma_1}-c_{\beta_1}s_{\beta_2}s_{\gamma_1}}{c_{\beta_1}c_{\beta_2}}$ \\
        $\xi^{\bar{e} e}_{A_3}$ & \ \ \ \ \ \ \ $\frac{s_{\beta_1}s_{\gamma_1}+c_{\beta_1}s_{\beta_2}c_{\gamma_1}}{c_{\beta_1}c_{\beta_2}}$ \\
        $\bar{t} b H_2^+$ & \ \ \ \ \ \ \ $-c_{\beta_2}s_{\gamma_2}y^{u}_{3,3}V_{CKM(3,3)}P_{L} - (c_{\beta_1}c_{\gamma_2}+s_{\beta_1}s_{\beta_2}s_{\gamma_2})y^{d*}_{3,3}V_{CKM(3,3)}P_{R}$ \\
        $\bar{t} b H_3^+$ & \ \ \ \ \ \ \ $c_{\beta_2}c_{\gamma_2}y^{u}_{3,3}V_{CKM(3,3)}P_{L} - (c_{\beta_1}s_{\gamma_2}-s_{\beta_1}s_{\beta_2}c_{\gamma_2})y^{d*}_{3,3}V_{CKM(3,3)}P_{R}$ \\
        $\bar{\nu} e H_2^+$ & \ \ \ \ \ \ \ $ (c_{\gamma_2}s_{\beta_1}-c_{\beta_1}s_{\beta_2}s_{\gamma_2})y^{e*}_1P_{R}$ \\
        $\bar{\nu} e H_3^+$ & \ \ \ \ \ \ \ $ (s_{\gamma_2}s_{\beta_1}+c_{\beta_1}s_{\beta_2}c_{\gamma_2})y^{e}_1P_{R}$ \\
    \bottomrule[1pt]
    \bottomrule[1pt]
    \end{tabular}
    \caption{The table presents all the relevant three point couplings scaled to the SM value (except for the charged Higgs couplings) involving the various Higgses and the SM gauge bosons and fermions. All these couplings play direct roles in calculating the various experimental and theoretical constraints on the model.}
    \label{tab:couplings}
\end{table}
\section{Alignment Limit in the 3HDM}
\label{sec:alignment}

The discovery of a SM-like 125 GeV Higgs boson implies that one of the three CP-even Higgs bosons in the 3HDM needs to be identified with it - in this section, we investigate the constraints arising out of the imposition of this alignment limit. Unlike in the 2HDM where after diagonalization, it is obvious which one of the two CP-even states is lighter, the eigenvalues in the 3HDM do not have a self-imposed hierarchy. This means that the mass ordering of the three CP-even Higgs bosons can change depending on the numerical values chosen for the $\lambda$'s and the vevs. Thus, we carry out the concomitant examination of the alignment limit conditions for three cases, classifying them as Regular Hierarchy (the lightest of the three is identified as the SM-like Higgs), Medial Hierarchy (the second lightest being SM-like), and the Inverted Hierarchy (the heaviest of the three being SM-like). However, even this does not quite exhaust the possibilities as even within one case, say the Regular Hierarchy, we have the freedom to choose any one of the three as the lightest. To streamline the situation a little and for ease of investigation, we will fix $H_1$ as the SM-like boson in the regular hierarchy, $H_2$ in the medial hierarchy, and $H_3$ in the inverted hierarchy. We now proceed to investigate these cases separately.\footnote{We reiterate that the most democratic way of doing the analysis would be to fix one of the three CP-even Higgs bosons, say $H_1$, to be the SM-like Higgs and proceed to constrain all other masses and angles from theoretical and experimental results. However, the computational time for such an analysis proves to be a major limiting factor.}


\subsection{Regular Hierarchy}
\label{sec:regularorder}

The coupling of the $H_1$ with the SM gauge bosons in the democratic 3HDM is given by
\begin{equation}
g_{HZZ}=\frac{ve^2(c_{\beta 2}c_{\alpha 2}\cos(\alpha_1-\beta_1)+s_{\beta 2}s_{\alpha 2})}{2c_w^2s_w^2},
\end{equation}
and thus the alignment limit condition for the regular hierarchy reads
     
 \begin{equation}
     	c_{\beta_2}c_{\alpha_2}\cos(\alpha_1-\beta_1) + s_{\beta_2}s_{\alpha_2} = 1.
      \label{eq:al1}
 \end{equation}
Letting $k=\cos(\alpha_1-\beta_1)$, we now have three broad possibilities: 
\begin{itemize}
    \item $k=1\implies \alpha_1 = \beta_1 + 2n\pi$ and  $\alpha_2 = \beta_2 + 2n\pi$.
    \item $k=-1\implies \alpha_1 = \beta_1 + (2n+1)\pi$ and $ \alpha_2 = - \beta_2 + (2n+1)\pi$. 
    \item $k\neq\pm 1$. In this case, Eqn.~\ref{eq:al1} is satisfied only when the following relationship holds:
    \begin{equation*}
    	\alpha_2 = \beta_2 = \frac{(2n+1)\pi}{2}.
\end{equation*}
\end{itemize}
The last constraint is a special case - as is clear from the definitions of $\tan\beta_1$ and $\tan\beta_2$ below Eqn.~\ref{eq:betatrans}, this constraint on the angles effectively places all the vev in the third doublet and hence, mixing for the charged and CP-Odd Higgs will take place only between the first two doublets. In the rest of the paper, we fix $k=1$ to signify the regular hierarchy.

\subsection{Medial Hierarchy}
\label{sec:medialorder}

For the medial order setup, we align $H_2$ with the SM Higgs Boson, and thus impose a mass hierarchy among the three CP-even Higgs bosons. The motivation behind this choice is to see if the 3HDM could admit one heavier and one lighter Higgses as compared to the SM one. Requiring the coupling of the $H_2$ with the weak gauge bosons be SM-like yields 
\begin{equation}
   	c_{\beta_2}c_{\alpha_3}\sin(\beta_1-\alpha_1) + s_{\beta_2}c_{\alpha_2}s_{\alpha_3} - c_{\beta_2}s_{\alpha_2}s_{\alpha_3}\cos(\alpha_1-\beta_1) = 1.
\end{equation}
Again let $k=\cos(\alpha_1-\beta_1)$, this can be re-written as 
\begin{equation}
    \label{eq:medialorder}
   	\pm c_{\beta_2}c_{\alpha_3}\sqrt{1-k^2} + s_{\beta_2}c_{\alpha_2}s_{\alpha_3} - c_{\beta_2}s_{\alpha_2}s_{\alpha_3}k = 1
\end{equation}
Again, the problem is disbanded into three cases depending on the value of $k$. 
\begin{itemize}
    \item $k=1 \implies \alpha_1 = \beta_1 + 2n\pi$. It is clear from Eqn.~\ref{eq:medialorder} that in this case, we have $\sin\alpha_3\,\sin(\beta_2-\alpha_2)=1$. This has two possible solutions: $\alpha_3 = \frac{\pi}{2}$, $\alpha_2 = \beta_2 - \frac{\pi}{2}$ and $\alpha_3 = \frac{3\pi}{2}$, $\alpha_2 = \beta_2 - \frac{3\pi}{2}$. We can combine these conditions to read $\tan \alpha_2 =  -\cot \beta_2 $.
    \item $k=-1\implies \alpha_1 = \beta_1 + (2n+1)\pi$. Again, from Eqn.~\ref{eq:medialorder}, we see that there are two possible solutions: $\alpha_3=\frac{\pi}{2}$, $\alpha_2 = -\beta_2 + \frac{\pi}{2}$, and $\alpha_3=\frac{3\pi}{2}$,  $\alpha_2 = -\beta_2 + \frac{3\pi}{2}$. Thus, we have in this case, $\tan\alpha_2 =  \cot\beta_2$. 
    \item $k\neq\pm 1$. In this case, the following relations need to hold:
    \begin{equation*}
        \alpha_3 = \left\{\frac{\pi}{2}, \frac{3\pi}{2}\right\}, \,\,
        \alpha_2 = \left\{0, \pi\right\}, \,\,\textrm{and}\,\,
        \beta_2 = \left\{\frac{\pi}{2}, \frac{3\pi}{2}\right\}.
    \end{equation*}
    This is again a special case, as it places all the vev in the third doublet and hence, mixing for the charged and CP-Odd Higgs will take place only between the first two doublets. For our analysis, we proceed with the choice $k=1$.
\end{itemize}

\subsection{Inverted Hierarchy}
\label{sec:oppositeorder}

In this scenario, we align $H_3$ with the SM Higgs Boson, also assuming this to be the heaviest, thus exploring the possibility of 2 CP-Even Higgses lighter than the $125$ GeV SM Higgs. The alignment limit condition for this case is given by
\begin{equation}
	 	c_{\alpha_3}s_{\beta_2}c_{\alpha_2} - c_{\alpha_3}c_{\beta_2}s_{\alpha_2}cos(\alpha_1-\beta_1) - c_{\beta_2}s_{\alpha_3}sin(\beta_1-\alpha_1) = 1,
\end{equation}
or, equivalently,
 \begin{equation}
        \label{eq:oppositeorder}
	 	c_{\alpha_3}s_{\beta_2}c_{\alpha_2} - c_{\alpha_3}c_{\beta_2}s_{\alpha_2}k \pm c_{\beta_2}s_{\alpha_3}\sqrt{1-k^2} = 1.
 \end{equation}
Again, we consider the three different possibilities based on the value of $k$.
\begin{itemize}
    \item $k=1 \implies \alpha_1 = \beta_1 + 2n\pi$. It is clear from Eqn.~\ref{eq:oppositeorder}, that in this case we have $\cos \alpha_3 \, \sin(\beta_2 - \alpha_2) = 1$. This has two possible solutions: $\alpha_3=0$, $\alpha_2 = \beta_2 - \frac{\pi}{2}$ and $\alpha_3=\pi$, $\alpha_2 = \beta_2 - \frac{3\pi}{2}$. We can combine these conditions to read $\tan \alpha_2 = - \cot \beta_2$. 
    
    \item $k=-1 \implies \alpha_1 = \beta_1 + (2n+1)\pi$. Again, from Eqn.~\ref{eq:oppositeorder}, we see that there are two possible solutions: $\alpha_3 = 0$, $\alpha_2 = -\beta_2 + \frac{\pi}{2}$ and $\alpha_3 = \pi$, $\alpha_2 = -\beta_2 + \frac{3\pi}{2}$. Hence for this case, we have $\tan(\alpha_2) =  \cot(\beta_2)$. 
    
    \item $k\neq \pm 1$. It is clear from Eqn.~\ref{eq:oppositeorder} that when the following relations hold, $k$ can take any value:
     \begin{equation*}
        \alpha_3 = \left\{0,\pi\right\}, \,\,
        \alpha_2 = \left\{0, \pi\right\}, \,\,\textrm{and}\,\,
        \beta_2 = \left\{\frac{\pi}{2}, \frac{3\pi}{2}\right\}.
    \end{equation*}
    This is again a special case which places all the vev in the third doublet. For our analysis, we proceed fixing $k=1$.
\end{itemize}
We have not studied the $k=-1$ case separately because there exists a clear mapping between the $k=1$ and $k=-1$ cases that does not affect any of the results in the current work. In the Regular hierarchy, changing $k$ from $+1$ to $-1$ is tantamount to the field redefintions $H_1\to H_1$, $H_2\to -H_2$, and $H_3 \to -H_3$. Similarly in the Medial Hierarchy, this is equivalent to $H_1\to -H_1$, $H_2\to H_2$, and $H_3 \to -H_3$, and in the Inverted Hierarchy case it is $H_1\to -H_1$, $H_2\to -H_2$, and $H_3 \to H_3$. Thus, the $k=-1$ results can be gotten from the $k=1$ results by employing field redefinitions where the two Higgses other than the 125 GeV one change sign. Since the constraints have been arrived at by computing cross-sections of specific processes, all our results depend only on the square of the couplings and will not change. However, the $k=1$ and $-1$ cases can be potentially distinguished if there are more than one Feynamn diagrams contributing to a specific final state and thus, these two cases might contribute constructively or destructively thus altering the rates. However, we have checked that these considerations do not apply to any of the results in the present work.
\section{Constraining the parameter space of 3HDM}
\label{sec:constraints}

In this section, we examine the various theoretical and experimental constraints that bound the values of the various masses and angular parameters in this model. 

\subsection{Theoretical Constraints}

We begin by investigating the theoretical constraints arising from imposing stability of the scalar potential, unitarity, perturbativity, and the electroweak precision observables. 

\subsubsection{Stability Constraints}

The scalar potential must be bounded from below to ensure vacuum stability. In doing so, we constrain the value of the parameters $\lambda_n$ such that the potential given by Eqn.~\ref{eq:scalarpot} is bounded from below in all the directions - we examine only the quartic couplings of the potential: 
    
\begin{equation}
    \label{eq:quartic}
    \begin{split}
        V_4 & =  \lambda_1(\phi_1^\dagger\phi_1)^2 + \lambda_2(\phi_2^\dagger\phi_2)^2 + \lambda_3(\phi_3^\dagger\phi_3)^2 \\ & + \lambda_{4}(\phi_1^\dagger\phi_1)(\phi_2^\dagger\phi_2) + \lambda_{5}(\phi_1^\dagger\phi_1)(\phi_3^\dagger\phi_3) + \lambda_{6}(\phi_2^\dagger\phi_2)(\phi_3^\dagger\phi_3) \\ & + \lambda_{7}(\phi_1^\dagger\phi_2)(\phi_2^\dagger\phi_1) + \lambda_{8}(\phi_1^\dagger\phi_3)(\phi_3^\dagger\phi_1) +  \lambda_{9}(\phi_2^\dagger\phi_3)(\phi_3^\dagger\phi_2) \\ & + \left[\lambda_{10}(\phi_1^\dagger\phi_2)(\phi_1^\dagger\phi_3) + \lambda_{11}(\phi_1^\dagger\phi_2)(\phi_3^\dagger\phi_2) + \lambda_{12}(\phi_1^\dagger\phi_3)(\phi_2^\dagger\phi_3) + h.c.\right]. 
    \end{split}
\end{equation}
Let us examine the asymptotic conditions of the potential by performing the following parameterization of the fields:
\begin{equation*}
    	a \equiv \phi_1^\dagger\phi_1,\, b \equiv \phi_2^\dagger\phi_2,\,\textrm{and}\,\, c \equiv \phi_3^\dagger\phi_3.
\end{equation*}
Further, we define
\begin{equation*}
    \begin{split}
    		d & \equiv Re(\phi_1^\dagger\phi_2), \; \; e \equiv Im(\phi_1^\dagger\phi_2), \\ f & \equiv Re(\phi_1^\dagger\phi_3), \; \; g \equiv Im(\phi_1^\dagger\phi_3), \\
    		h & \equiv Re(\phi_2^\dagger\phi_3), \; \; j \equiv Im(\phi_2^\dagger\phi_3).
    \end{split}
\end{equation*}
With this, it is clear that 
\begin{align*}
    & a = 0 \implies d=e=f=g=0, \\
    & b = 0 \implies d=e=h=j=0,\,\,\textrm{and} \\
    & c = 0 \implies f=g=h=j=0.
\end{align*}
But it is possible to choose arbitrary values for $a$ even if we make $d=e=f=g=0$, for $b$ even if we make $d=e=h=j=0$ and similarly, for $c$ even if we make $f=g=h=j=0$. In terms of the new variables, the potential in Eqn.~\ref{eq:quartic} can be recast as
\begin{equation}
    \begin{split}
        V_4 & = \frac{1}{2}(\sqrt{\lambda_{1}}a-\sqrt{\lambda_{2}}b)^2 + \frac{1}{2}(\sqrt{\lambda_{1}}a-\sqrt{\lambda_{3}}c)^2 + \frac{1}{2}(\sqrt{\lambda_{2}}b-\sqrt{\lambda_{3}}c)^2 \\ & + (\lambda_{4}+\sqrt{\lambda_{1}\lambda_{2}})(ab-d^2-e^2) + (\lambda_{5}+\sqrt{\lambda_{1}\lambda_{3}})(ac-d^2-e^2) \\ & + (\lambda_{6}+\sqrt{\lambda_{2}\lambda_{3}})(bc-h^2-j^2) + 2(\lambda_{4} + \lambda_{7} +\sqrt{\lambda_{1}\lambda_{2}})d^2 \\ & + 2(\lambda_{5} + \lambda_{8} +\sqrt{\lambda_{1}\lambda_{3}})f^2 + 2(\lambda_{6} + \lambda_{9} +\sqrt{\lambda_{2}\lambda_{3}})h^2 \\ & + (-\lambda_{4} - \lambda_{7} -\sqrt{\lambda_{1}\lambda_{2}})(d^2-e^2) + (-\lambda_{5} - \lambda_{8} -\sqrt{\lambda_{1}\lambda_{3}})(f^2-g^2) \\ & + (-\lambda_{6} - \lambda_{9} -\sqrt{\lambda_{2}\lambda_{3}})(h^2-j^2) + 2\,\mathfrak{Re}(\lambda_{10})(df-eg) \\ & - 2\,\mathfrak{Im}(\lambda_{10})(dg+ef) + 2\,\mathfrak{Re}(\lambda_{11})(dh+ej) + ,\mathfrak{Im}(\lambda_{11})(dj-eh) \\ & + 2\,\mathfrak{Re}(\lambda_{12})(fh-gj) - 2\,\mathfrak{Im}(\lambda_{12})(fj+gh).
    \end{split}
\end{equation} 

Let us consider various cases to constrain the $\lambda$ parameters in the potential.

(i) Consider the field direction $b=0$ (hence $d=e=h=j=0$), and $c=0$ (hence $f=g=h=j=0$). Then 
\begin{equation*}
    V_4 = \lambda_{1}a^2.
\end{equation*}
Now in the limit $a \rightarrow \infty$, demanding that $V_4$ does not tend to a large negative value requires
\begin{equation}
    \lambda_{1} \geq 0.
    \label{eq:stability_i}
\end{equation}
(ii) Consider the field direction $a=0$ (hence $d=e=f=g=0$), and $c=0$ (hence $f=g=h=j=0$). Here,
\begin{equation*}
    V_4 = \lambda_{2}b^2,
\end{equation*}
and in the limit $b \rightarrow \infty$, $V_4$ not tending to a large negative value requires
\begin{equation}
    \lambda_{2} \geq 0.
\end{equation}
(iii) Consider the field direction $a=0$ (hence $d=e=f=g=0$), and $b=0$ (hence $d=e=h=j=0$), then 
\begin{equation*}
    V_4 = \lambda_{3}c^2.
\end{equation*}
Now, in the limit $c \rightarrow \infty$, demanding that $V_4$ is bounded from below, we find
\begin{equation}
    \lambda_{3} \geq 0.
\end{equation}
(iv) Consider the field direction $a=\sqrt{\frac{\lambda_{2}}{\lambda_{1}}}b$ and $d=e=0$. Also, let $c=0$ (hence $f=g=h=j=0$). Then,
\begin{equation*}
    V_4 = (\lambda_{1}a + (\lambda_{4} + \sqrt{\lambda_{1}\lambda_{2}})b)a. 
\end{equation*}
Now, in the limit $a \rightarrow \infty$ and $b \rightarrow \infty$, stability of the potential requires
\begin{equation*}
    \lambda_{1}a + (\lambda_{4} + \sqrt{\lambda_{1}\lambda_{2}})b \geq 0.
\end{equation*} 
On substituting the value of a, we get
\begin{equation}
    \lambda_{4} + 2\sqrt{\lambda_{1}\lambda_{2}} \geq 0.
\end{equation}
(v) Consider the field direction $a=\sqrt{\frac{\lambda_{3}}{\lambda_{1}}}c$ and $f=g=0$. Also, let $b=0$ (hence $d=e=h=j=0$). Here
\begin{equation*}
    V_4 = (\lambda_{3}c + (\lambda_{5} + \sqrt{\lambda_{1}\lambda_{3}})a)c. 
\end{equation*}
Now, in the limit $a \rightarrow \infty$ and $c \rightarrow \infty$, we need
\begin{equation*}
    \lambda_{3}c + (\lambda_{5} + \sqrt{\lambda_{1}\lambda_{3}})a \geq 0.
\end{equation*} 
On substituting the value of c, we find
\begin{equation}
    \lambda_{5} + 2\sqrt{\lambda_{1}\lambda_{3}} \geq 0.
\end{equation}
(vi) Consider the field direction $b=\sqrt{\frac{\lambda_{3}}{\lambda_{2}}}c$ and $h=j=0$. Also, let $a=0$ (hence $d=e=f=g=0$). Now
\begin{equation*}
    V_4 = (\lambda_{2}b + (\lambda_{6} + \sqrt{\lambda_{2}\lambda_{3}})c)b.
\end{equation*}
Now, in the limit $b \rightarrow \infty$ and $c \rightarrow \infty$, we require
\begin{equation*}
    \lambda_{2}b + (\lambda_{6} + \sqrt{\lambda_{2}\lambda_{3}})c \geq 0.
\end{equation*} 
On substituting the value of b, we get
\begin{equation}
    \lambda_{6} + 2\sqrt{\lambda_{2}\lambda_{3}} \geq 0.
\end{equation}
(vii) Consider the field direction $a=\sqrt{\frac{\lambda_{2}}{\lambda_{1}}}b$ along with $ab=d^2+e^2$. Also, let $c=0$ (hence $f=g=h=j=0$).Hence
\begin{equation*}
    V_4 = \lambda_{1}a^2 + (\lambda_{4} + \lambda_{7} + \sqrt{\lambda_{1}\lambda_{2}})d^2 + 
    (\lambda_{4} + \lambda_{7} + \sqrt{\lambda_{1}\lambda_{2}})e^2. 
\end{equation*}
On substituting the value of a, we get
\begin{equation*}
    V_4 = (\lambda_{4} + \lambda_{7} + 2\sqrt{\lambda_{1}\lambda_{2}})d^2 + 
    (\lambda_{4} + \lambda_{7} + 2\sqrt{\lambda_{1}\lambda_{2}})e^2. 
\end{equation*}
We see that in both the cases $e=0,\,d \rightarrow \infty$ and $d=0,\,e \rightarrow \infty$, the constraint required so that the potential does not hit large negative values is
\begin{equation}
    \lambda_{4} + \lambda_{7} + 2\sqrt{\lambda_{1}\lambda_{2}} \geq 0.
\end{equation}
(viii) Consider the field direction $a=\sqrt{\frac{\lambda_{3}}{\lambda_{1}}}c$ along with $ac=f^2+g^2$. Also, let $b=0$ (hence $d=e=h=j=0$). Thus,
\begin{equation*}
    V_4 = \lambda_{3}c^2 + (\lambda_{5} + \lambda_{8} + \sqrt{\lambda_{1}\lambda_{3}})f^2 + 
    (\lambda_{5} + \lambda_{8} + \sqrt{\lambda_{1}\lambda_{3}})g^2, 
\end{equation*}
 which can be, on substituting the value of c, written as
\begin{equation*}
    V_4 = (\lambda_{5} + \lambda_{8} + 2\sqrt{\lambda_{1}\lambda_{3}})f^2 + 
    (\lambda_{5} + \lambda_{8} + 2\sqrt{\lambda_{1}\lambda_{3}})g^2.
\end{equation*}
We see that in both the cases $f=0,\,g \rightarrow \infty$ and $g=0,\,f \rightarrow \infty$, demanding stability requires
\begin{equation}
    \lambda_{5} + \lambda_{8} + 2\sqrt{\lambda_{1}\lambda_{3}} \geq 0.
\end{equation}
(ix) Finally, consider the field direction $b=\sqrt{\frac{\lambda_{3}}{\lambda_{2}}}c$ along with $bc=h^2+j^2$, $a=0$ (and hence $d=e=f=g=0$). Thus,
\begin{equation*}
    V_4 = \lambda_{2}b^2 + (\lambda_{6} + \lambda_{9} + \sqrt{\lambda_{2}\lambda_{3}})h^2 + 
    (\lambda_{6} + \lambda_{9} + \sqrt{\lambda_{2}\lambda_{3}})j^2. 
\end{equation*}
On substituting the value of b, we find
\begin{equation*}
    V_4 = (\lambda_{6} + \lambda_{9} + 2\sqrt{\lambda_{2}\lambda_{3}})h^2 + 
    (\lambda_{6} + \lambda_{9} + 2\sqrt{\lambda_{2}\lambda_{3}})j^2. 
\end{equation*}
We see that in both the cases $h=0,\,j \rightarrow \infty$ and $j=0,\,h \rightarrow \infty$, requiring the potential be bounded from below necessitates
\begin{equation}
    \lambda_{6} + \lambda_{9} + 2\sqrt{\lambda_{2}\lambda_{3}} \geq 0.
    \label{eq:stability_f}
\end{equation}

Eqns.~\ref{eq:stability_i}-\ref{eq:stability_f} represent the set of all constraints on the various couplings and their combinations thereof from requirements of the stability of the potential.

\subsubsection{Unitarity and Perturbativity Constraints}

Using partial wave analysis one can express any scattering amplitude in the following manner
\begin{equation}
\mathcal{M}\left(\theta\right) = 16\pi\sum^{\infty}_{\ell = 0}a_{\ell}\left(2\ell + 1\right)P_{\ell}\left(\cos\theta\right).
\label{Eq:amplitude}
\end{equation}
Here $P_{\ell}\left(\cos\theta\right)$ represents the Legendre Polynomials of order $\ell$. Using the orthonormality condition of these Legendre Polynomials, the scattering amplitude corresponding to any $2 \to 2$ process can be related to $a_{\ell}$ in Eqn.~\ref{Eq:amplitude}. To calculate the tree level unitarity constraints, one needs to study the energy growth of $2 \to 2$ scattering amplitudes involving the scalars. In the high energy limit, all these  are proportional to the quartic terms of the potential. After extracting the zeroth order partial wave amplitude $a_{0}$ from Eqn.~\ref{Eq:amplitude}, one can use it to form the S-matrix which has different two-body states as rows and columns. The eigenvalues of this matrix can be bounded using the unitarity constraints $|a_{0}| < 0$ - one can find the necessary details of this procedure in \cite{Akeroyd:2000wc}.

In this paper, we have used the results of ~\cite{Bento_2022,Boto_2021} for unitarity Constraints in the $Z_3$ symmetric 3HDM. The conversion between the parameters of the potential in ~\cite{Bento_2022,Boto_2021} and the ones employed in this work are given below:

\begin{equation}
    \begin{split}
    & r_1 \rightarrow \lambda_{1}, \; \; \; \; r_2 \rightarrow \lambda_{2}, \; \; \; \; r_3 \rightarrow \lambda_{3}, \; \; \; \; r_4 \rightarrow \frac{\lambda_{4}}{2},  \\
    & r_5 \rightarrow \frac{\lambda_{5}}{2}, \; \; \; \; r_6 \rightarrow \frac{\lambda_{6}}{2}, \; \; \; \; r_7 \rightarrow \frac{\lambda_{7}}{2}, \; \; \; \; r_8 \rightarrow \frac{\lambda_{8}}{2}, \\ 
    & r_9 \rightarrow \frac{\lambda_{9}}{2}, \; \; \; \; c_4 \rightarrow \frac{\lambda_{10}}{2}, \; \; \; \; c_{12} \rightarrow \frac{\lambda_{11}}{2}, \; \; \; \; c_{11} \rightarrow \frac{\lambda_{12}}{2}
   \end{split}
\end{equation}

The unitarity condition to be imposed on the $21$ eigenvalues $\Lambda_i$ of the relevant scattering matrices is
\begin{equation}
    |\Lambda_i| \leq 8\pi.
    \label{eq:unitarity}
\end{equation} 

On the other hand, the perturbativity condition on the parameters of the scalar potential demands that, for all $i$,
\begin{equation}
    \lambda_i \leq |4\pi|.
    \label{eq:perturbativity}
\end{equation} 
We remark here that Eqn.~\ref{eq:unitarity} also constrains all the couplings $\lambda_i$ (as the eigenvalues $\Lambda_i$ are indeed functions of the couplings $\lambda_i$), as does Eqn.~\ref{eq:perturbativity} - these two conditions constrain the couplings independently.
\subsection{Experimental Constraints}

In this section, we detail the various experimental constraints analyzed to validate a viable Three Higgs Doublet Model (3HDM). Firstly, we examined the exclusion limits from direct searches for the Higgs boson at the Large Hadron Collider (LHC), the Large Electron-Positron Collider (LEP), and the Tevatron. These exclusion limits were assessed at the $95 \%$ C.L. using the \texttt{HiggsBounds-6} module via the \texttt{HiggsTools} package \cite{Bahl_2023}. This analysis ensures that the model is consistent with the non-observation of additional Higgs bosons in these extensive collider experiments. Next, we tested the compatibility of our aligned $125$ GeV Higgs boson with the Standard Model (SM) Higgs boson using a goodness-of-fit test. Specifically, we calculated the chi-square ($\chi^2$ ) value with \texttt{HiggsSignals-3} via \texttt{HiggsTools}, which compares the predicted signal strengths of our Higgs boson to those observed experimentally. We explore the parameter spaces that fulfill the condition, $\chi_{125}^2 < 189.42$, corresponding to a $95 \%$ C.L. with 159 degrees of freedom for a global fit against these observables \cite{Benbrik:2024ptw,Belyaev:2023xnv}. This step verifies that the predicted properties of our Higgs boson closely match those of the Higgs boson observed at the LHC, ensuring that the model remains viable under current experimental constraints. In addition to direct Higgs searches, we incorporated constraints from B-physics observables, which are sensitive to potential new physics contributions in flavor-changing neutral current processes. Specifically we tested the $\mathcal{BR}(B\rightarrow X_s \gamma)$ using next-to-leading order (NLO) calculations. We closely follow the NLO QCD predictions for 2HDM as described in ~\cite{Borzumati_1998}. The analysis has been extended in ~\cite{Akeroyd_2021,Boto_2021} to incorporate the contributions from an additional charged Higgs Boson. The couplings herein are as defined in Table~\ref{tab:couplings}. 

\begin{align}
    X_1 & = -\frac{\cos \beta_1\cos\gamma_2 + \sin\beta_1\sin\beta_2\sin\gamma_2}{\sin\beta_1\cos\beta_2} \\
    Y_1 & = -\frac{\cos\beta_2\sin\gamma_2}{\sin\beta_2} \\
    X_2 & = -\frac{\cos\beta_1\sin\gamma_2 - \sin\beta_1\sin\beta_2\cos\gamma_2}{\sin\beta_1\cos\beta_2} \\
    Y_2 & = \frac{\cos\beta_2\cos\gamma_2}{\sin\beta_2}
\end{align}

We took our input parameters updated to the most recent values from the Particle Data Group ~\cite{Workman:2022ynf}. 

\begin{align*}
    \alpha_s(M_Z) & = 0.1179 \pm 0.0010 , \ \ \ \ \ \ \ \ \ \ \ \ \ \ m_t = 172.76 \pm 0.3 \\
    \frac{1}{z} & = \frac{m_b}{m_c} = 4.58 \pm 0.01, \ \ \ \ \ \ \ \ \ \ \ \ \ \ \alpha  = \frac{1}{137.036} \\
    BR_{SL} & = 0.1049 \pm 0.0046, \ \ \ \  |\frac{V_{ts}^*V_{tb}}{V_{cb}}|^2 = 0.95 \pm 0.02 \\
    m_b(1S) & = 4.65 \pm 0.03, \ \ \ \ \ \ \ \ \ \ \ \ \ \ m_c = 1.27 \pm 0.02 \\
    m_Z & = 91.1876 \pm 0.0021, \ \ \ \ \ \ \ \ \  m_W = 86.377 \pm 0.012 
\end{align*}
This analysis is critical for ruling out parameter space regions that would lead to significant discrepancies in flavor physics. The following restriction has been imposed which represents the $3 \sigma$ experimental limit, 

\begin{equation*}
    2.87 \times 10^{-4} < \mathcal{BR}(B \rightarrow X_s \gamma) < 3.77 \times 10^{-4}
\end{equation*}

Finally, to check if the model passes all the electroweak precision observable constraints, we calculated the oblique parameters S, T, and U using the \texttt{SPheno} package \cite{Porod_2003,Porod_2012} with the model file written in \texttt{SARAH} \cite{Staub_2014}. These parameters are derived from precision measurements of electroweak interactions and provide a comprehensive test of the model’s consistency with the Standard Model predictions. The oblique parameters offer stringent constraints on new physics scenarios, ensuring that any extensions to the Higgs sector remain within the bounds set by precision electroweak data. The \texttt{SPheno} results are also found to be consistent with the calculated values using the formulation given in ~\cite{Grimus_2008}. The numerical values of the observables that we use to constrain the model are \cite{Workman:2022ynf}

\begin{align*}
    S & = -0.02 \pm 0.10, \\
    T & = 0.03 \pm 0.12,\,\,\textrm{and} \\
    U & = 0.01 \pm 0.11.
\end{align*}
\section{Methodology}
\label{sec:methodology}

The Three Higgs Doublet Model (3HDM) presents a formidable challenge due to the curse of dimensionality, with its scalar potential governed by $15$ independent parameters. Our objective is to efficiently explore this $15$ dimensional parameter space to identify regions that satisfy theoretical constraints. Exhaustively searching for allowed points using a grid-based method would be computationally infeasible, potentially taking years. To circumvent this, we propose training a binary machine learning classifier \footnote{For a comprehensive review of ML techniques in particle physics, the reader is encouraged to consult \cite{Feickert:2021ajf,Plehn:2022ftl}.} to predict whether a given point adheres to the constraints. However, the initial dataset generation reveals a substantial class imbalance, where most points fall into the disallowed category, thus skewing the dataset. The success of the machine learning model hinges on both the quality and representativeness of the training data. Crucially, points near the decision boundary—where the classifier exhibits the greatest uncertainty—carry more informational value than others. To optimize the selection of such data points, we employ an Active Learning \cite{Settles2009ActiveLL} strategy. Our Active Learning algorithm significantly enhances the efficiency of the neural network classifier by focusing on a minimal yet highly informative subset of points. This process unfolds as follows:

\begin{enumerate}
    \item A machine learning model is first trained on a small, labeled dataset, with data points categorized as allowed or disallowed using traditional programming approaches. 
    \item The trained model then predicts outcomes for a large, unlabeled data pool. 
    \item Entropy scores are calculated for each point in the pool, measuring the model’s uncertainty.
    \item Points with the highest entropy, where the model is most uncertain, are added to the labeled dataset.
    \item The model is retrained, and the process repeats, iteratively refining the model's predictions.
    \item This cycle continues until the model achieves the desired level of accuracy.
\end{enumerate}

The Active Learning algorithm adopts a greedy approach to effectively delineate the decision boundary between allowed and disallowed regions, thus mitigating the curse of dimensionality. By selectively focusing on the most informative points, the algorithm minimizes computational costs while maximizing the model’s accuracy. Iteratively refining the decision boundary with fewer samples allows the classifier to converge more efficiently, producing a comprehensive and precise mapping of the parameter space. This approach deepens our understanding of the theoretical constraints governing the 3HDM, while significantly improving both computational efficiency and predictive power.

The training dataset for our model was generated and analyzed using traditional programming techniques combined with Monte Carlo sampling. Although, MC sampling provides a broad exploration of the parameter space but results in a dataset heavily skewed towards disallowed points. This imbalance necessitates careful construction of the training dataset, wherein the ratio of allowed to disallowed points is maintained at a minimum of $0.4$. This ensures that the classifier receives sufficient information about the allowed regions, which is critical for accurately delineating the decision boundary. This balanced approach helps maintain robustness in the model's learning process. The training set is then divided into a labeled pool, an unlabeled pool for training, and a test dataset. During this splitting process, hyperparameters are tuned to maintain an equivalent ratio of allowed to disallowed points across subsets. The complete Active Learning (AL) algorithm is illustrated in Fig.~\ref{fig:AL}. 
The iterative process of selecting high-entropy points, annotating them, and updating the labeled pool continues until the model’s performance on the test data reaches a plateau, indicating that further iterations do not significantly enhance accuracy. We choose our satisficing metric as a $100 \%$ recall over the test data, ensuring that our model correctly identifies all allowed points. Once this criterion is met, we use accuracy as the optimization metric, selecting the model that performs best on the test dataset. This dual-focus on recall and accuracy ensures that the model is both sensitive to true positives and reliable in its overall predictions.

Our decision to prioritize recall stems from the need to ensure that all theoretically allowed points are identified, given the extreme imbalance in our dataset. In scenarios where positive (allowed) points are exceedingly rare, maximizing recall minimizes the risk of missing these critical points, even if it means accepting a higher rate of false positives initially. Out of $20$ million randomly generated points, only a handful are permitted, indicating that the allowed region is confined to a specific, narrow area within the parameter space. In such a scenario, it is crucial for our classifier to identify all the allowed points even if they are interspersed within a vast number of disallowed points. This ensures that no potentially valid points are missed. However, to mitigate the impact of false positives, our algorithm incorporates a secondary verification step, outlined in Figure ~\ref{fig:parameterspace}, which details the entire process. Initially, randomly sampled points are evaluated by our machine learning classifier. Points classified as positive are then subjected to verification using traditional numerical techniques. This verification process acts as a filter, significantly reducing the number of false positives while ensuring that no allowed points are missed. The efficiency gains from our algorithm are substantial. The reduction in analysis time from $4$ hours to just $10$ minutes is achieved without compromising the accuracy of identifying allowed points. This reduction in computational time highlights the effectiveness of combining machine learning with traditional methods to handle large-scale data analysis in complex parameter spaces.

The dataset, which satisfies the stability, unitarity, and perturbativity constraints, is subsequently subjected to a comprehensive analysis using the \texttt{HiggsTools} package. This package includes \texttt{HiggsBounds-6} for direct search constraint testing and \texttt{HiggsSignals-3} for goodness-of-fit testing. Points that pass both the \texttt{HiggsBounds-6} and \texttt{HiggsSignals-3} tests are further scrutinized using additional constraints. One of these is the $b \rightarrow s \gamma $ flavor constraint, which examines the implications of the model on rare flavor-changing neutral current processes.Finally, the electroweak parameter constraints are calculated using the \texttt{SPheno} package. This involves evaluating parameters such as the oblique parameters $S, T,$ and $U$, which provide a stringent test of the model's compatibility with precision electroweak measurements.

\begin{figure}[h]
  \centering
  \begin{minipage}[b]{0.4\textwidth}
    \begin{tikzpicture}[node distance=1.5cm,scale=0.73,transform shape]

    \node (annotate) [startstop] {Annotate the queried samples};
    \node (labeled) [process, below left=1.0cm and 1.5cm of annotate] {Labeled Dataset};
    \node (model) [process, right=2.0cm of labeled] {ML model};
    \node (test_unlabeled) [process, right=2.0cm of model] {Test on Unlabeled Dataset};
    \node (test_dev) [startstop, below=1.5cm of model] {Test on Development Set};

    \draw [arrow] (annotate) -- (labeled);
    \draw [arrow] (labeled) -- (model);
    \draw [arrow] (model) -- (test_unlabeled);
    \draw [arrow] (test_unlabeled) -- (annotate);
    \draw [arrow] (model) -- (test_dev);

    \end{tikzpicture}
    \caption{Flow of Active Learning}
    \label{fig:AL}
  \end{minipage}
  \hfill
  \begin{minipage}[b]{0.4\textwidth}
    \begin{tikzpicture}[node distance=2cm,scale=0.8,transform shape]

    \node (sampling) [startstop] {Random Sampling};
    \node (model2) [process, below=2cm of sampling] {ML model};
    \node (numerical) [startstop, below=2cm of model2] {Traditional Numerical Methods};

    \draw [arrow] (sampling) -- (model2);
    \draw [arrow] (model2) -- (numerical);

    \end{tikzpicture}
    \caption{Flow of Parameter Space Extraction}
    \label{fig:parameterspace}
  \end{minipage}
\end{figure}

\section{Results}
\label{sec:results}
We are now in a position to present and discuss the main results of the paper. In what follows, we have displayed the constraints on the parameters of the model in by looking at correlations between the various masses and angles. The color coding of the plots is as follows: regions shaded in blue is allowed by the stability, perturbativity and unitarity constraints while the red region is allowed by the direct search constraints and the goodness of fit test implemented through with the help of the package \texttt{HiggsTools}, and also the $b \rightarrow s \gamma$ flavor constraint. The electroweak precision constraints that are calculated with the help of the \texttt{SPheno} package are indicated by the green shaded region. We note here that the the constraints are placed sequentially, i.e., the red regions are laid on top of the blue, and the green on top of the red. Thus, it is the final regions marked in green that simultaneously satisfy the theoretical, experimental, and the precision electroweak constraints and thus provide us data about interesting regions of the parameter space that can be probed by further phenomenological studies. 

\subsection{Regular Hierarchy}
\label{sec:regular}

\begin{figure}[h!]
        \includegraphics[scale=0.7]{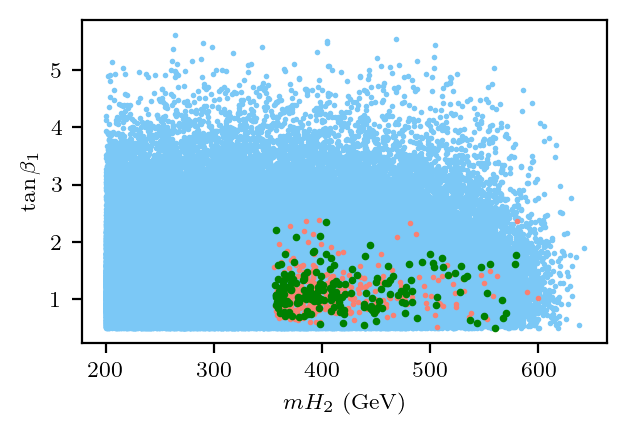}
        \includegraphics[scale=0.7]{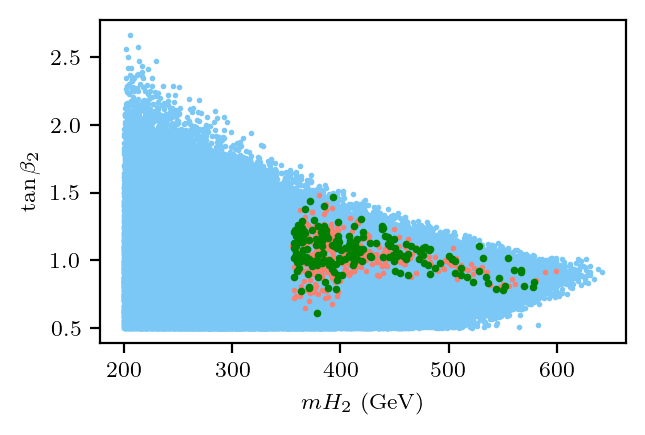}
        \includegraphics[scale=0.7]{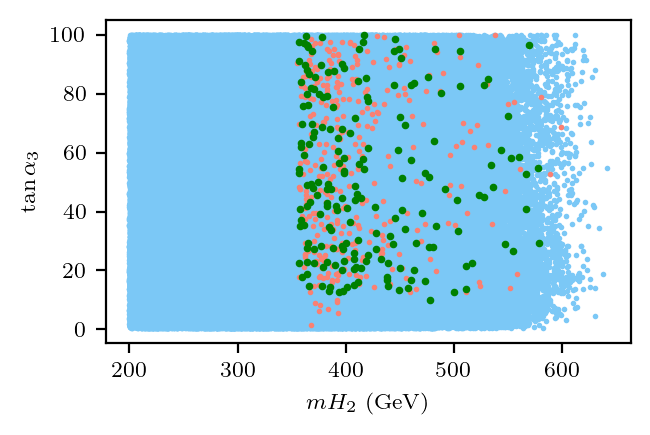}
        \includegraphics[scale=0.7]{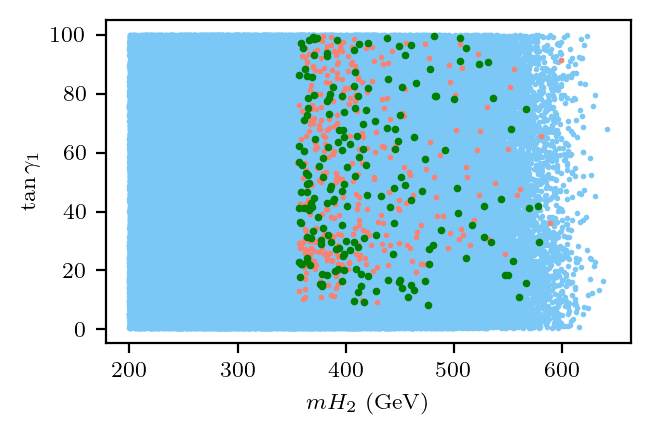}
        \includegraphics[scale=0.7]{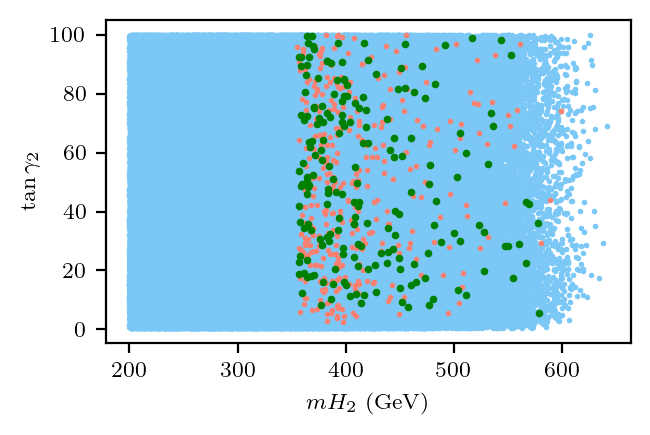}
        
       \caption{The allowed regions for the Alignment 1 case with the lightest of the three CP-even Higgses identified as the 125 GeV SM-like Higgs boson.  The plots show how the allowed region in $m_{H_2}$ correlates with the various angles. The color coding of the plots is as follows: regions shaded in blue is allowed by the stability, perturbativity and unitarity constraints while the red region is allowed by the direct search constraints and the goodness of fit test, and also the $b \rightarrow s \gamma$ flavor constraint. The electroweak precision constraints are indicated by the green shaded region.}

       \label{fig:mH2-al1}
\end{figure}

We begin with the results for the regular hierarchy. The results in this section represent the allowed parameter space for the alignment-1 case, wherein we identify the lightest Higgs (labeled $H_1$) with the SM-like 125 GeV Higgs boson. Note that we are operating under the assumption that $k=\cos(\alpha_1-\beta_1)=1$, and thus for Eqn.~\ref{eq:al1} to be satisfied, $\alpha_2$ is fixed to be $ \beta_2 + 2n\pi$ as explained in Sec.~\ref{sec:regularorder}. Thus, the free parameters here are the masses $m_{H_2}, m_{H_3}, m_{H_2^{\pm}}, m_{H_3^{\pm}}, m_{A_2},$ and $m_{A_3}$, and the angles $\alpha_3,\beta_1,\beta_2,\gamma_1,$ and $\gamma_2$.  We begin by analyzing the extent of regions possible in the mass of $H_2$ as it correlates with various angles in the theory - these are shown collectively in Fig.~\ref{fig:mH2-al1}. 
A few important observations can be made immediately upon inspection of Fig.~\ref{fig:mH2-al1}: the mass of $H_2$ is tightly constrained to lie in the window $350\, \textrm{GeV}<m_{H_2}<580\, \textrm{GeV}$. While the angles $\gamma_1$, $\gamma_2$, and $\alpha_3$ are largely unconstrained, the same is certainly not true of $\beta_1$ and $\beta_2$. Specifically, we have $\tan\beta_1<2.4$ and $\tan\beta_2<1.5$. Understandably, very low or high $\tan\beta$ is disallowed - it can be seen from Eqn.~\ref{eqn:lambda-masses} that the $\lambda$'s are a rather complicated function of $\tan\beta_1$ and $\tan\beta_2$, and perturbativity constraints force strict upper and lower bounds on $\tan\beta_{1,2}$.

\begin{figure}[h!]
        \includegraphics[scale=0.7]{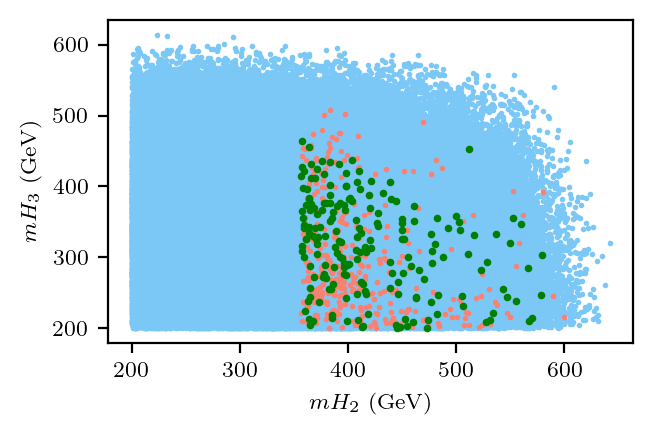}
        \includegraphics[scale=0.7]{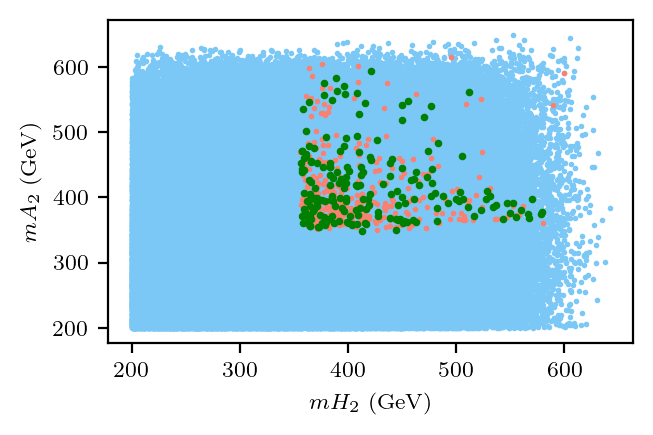}
        \includegraphics[scale=0.7]{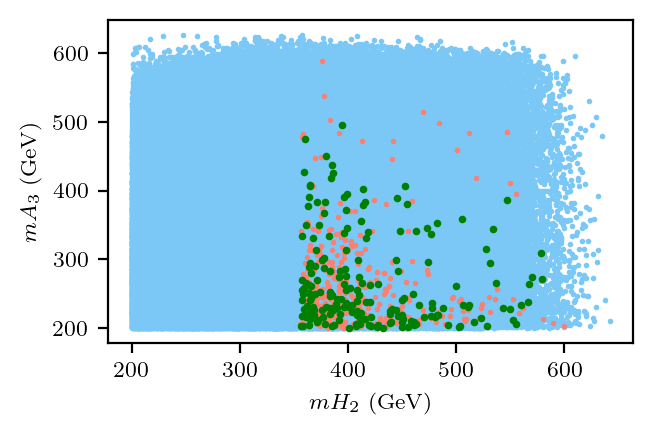}
        \includegraphics[scale=0.7]{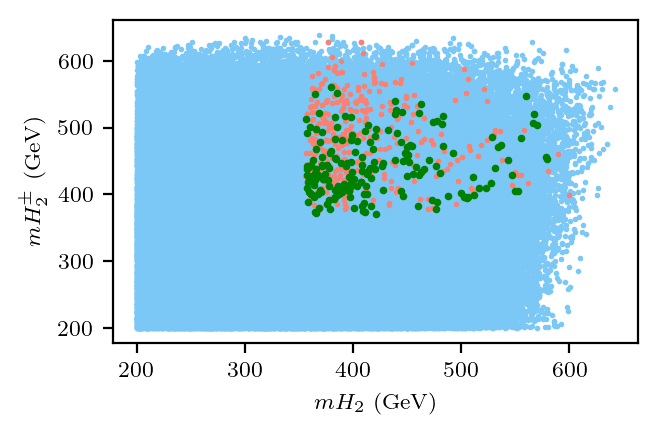}
        \includegraphics[scale=0.7]{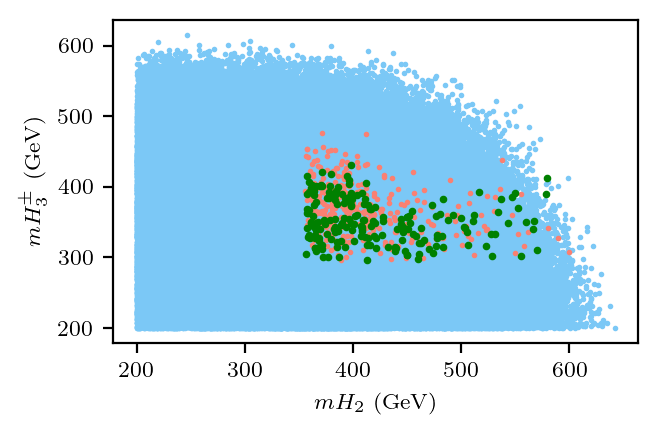}
         \caption{The allowed regions for the Alignment 1 case with the lightest of the three CP-even Higgses identified as the 125 GeV SM-like Higgs boson.  The plots show how the allowed region in $m_{H_2}$ correlates with the masses of the other Higgses in the theory.}
         \label{fig:mH2-al1-masses}
\end{figure}

\begin{figure}[h!]
        \centering
        \includegraphics[scale=0.8]{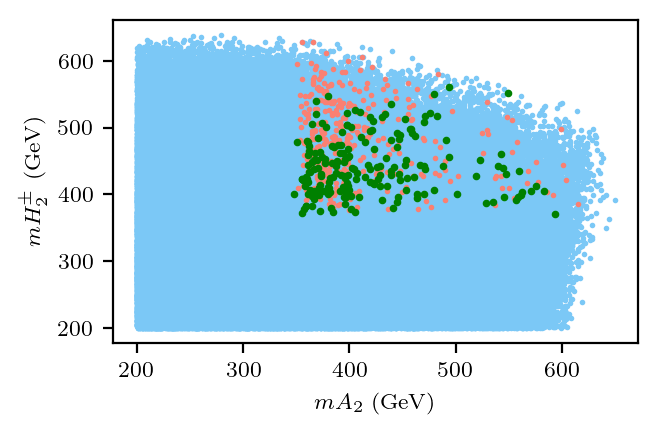}
        \includegraphics[scale=0.8]{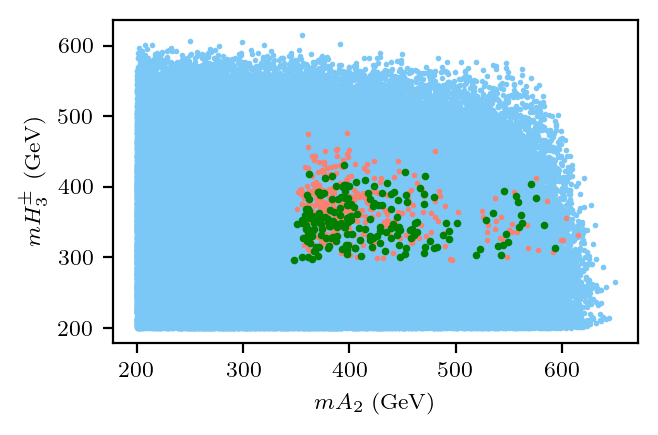}
        \includegraphics[scale=0.8]{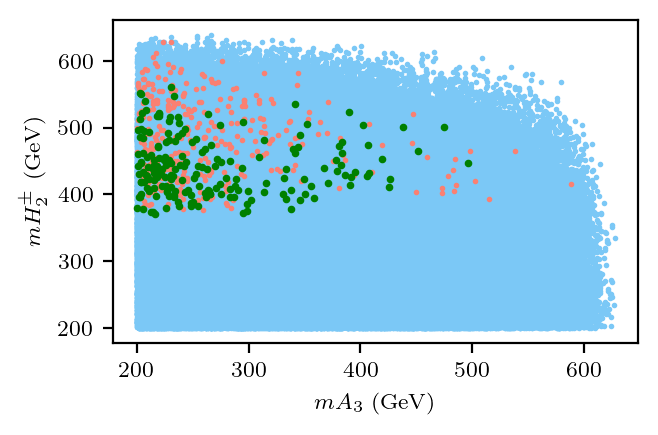}
        \includegraphics[scale=0.8]{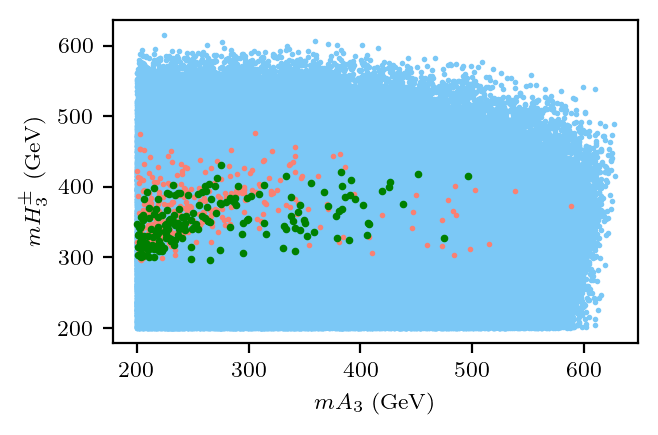}
        \caption{The allowed regions for the Alignment 1 case with the lightest of the three CP-even Higgses identified as the 125 GeV SM-like Higgs boson.  The plots show the correlation of the charged and pseudoscalar Higgs bosons.}
        \label{fig:charged-pseudo-correlation}
\end{figure}

In Fig.~\ref{fig:mH2-al1-masses}, we display the ranges allowed for the masses of all the other Higgses in the theory. We note that all masses are bounded from above. Since the perturbativity constraint demands that all the $\lambda$'s lie below $4\pi$, and the $\lambda$'s were traded for the masses via Eqn.~\ref{eqn:lambda-masses}, demanding that the model be perturbative now imposes strict upper limits on all the Higgs masses. While fixing $H_1$ to be the 125 GeV Higgs in the normal hierarchy, we did not impose any mass ordering between the other two CP-even Higgses. Thus, we see from the first plot of Fig.~\ref{fig:mH2-al1-masses}, that regions with $m_{H_3}>m_{H_2}$ and $m_{H_3} <m_{H_2}$ are both populated. Specifically, we find that while the $H_2$ mass is constrained to be $350\,\textrm{GeV}<m_{H_2}\lessapprox 600$ GeV, somewhat similar to the constraints on $m_{A_2}$ and $m_{H_2^\pm}$, the majority of the allowed points prefer a slightly lighter $A_3$, around $200\,\textrm{GeV}<m_{A_3}< 400$ GeV (though higher values are not completely ruled out). The other charged Higgs boson mass lies in the interval $300\,\textrm{GeV}<m_{H_3^\pm}< 420$ GeV.
In the 2HDM, the constraints from the $T$ parameter (or equivalently, the $\rho$ parameter) forces the masses of the charged and pseudoscalar Higgses to be degenerate. In the case of the 3HDM, because of the presence of multiple $A$'s and $H^\pm$'s, there are additional contributions to the $\rho$ parameter arising from different combinations. For example, the $W$ propagator can be corrected because of loops involving $H_2^+A_2$, $H_2^+A_3$ etc. This means that the allowed points do not make all pairs of charged and pseudoscalar Higgs bosons degenerate, but only approximately so. We can observe this in Fig.~\ref{fig:charged-pseudo-correlation} where we exhibit the correlation between the various pairs of the $H^\pm$'s and the pseudoscalar $A$'s in this set-up.

\subsection{Medial Hierarchy}

In this case, we align our intermediate Higgs with the known SM Higgs Boson, which means that we have one Higgs lighter than 125 GeV and the other heavier. In what follows, we designate the $H_2$ to be the SM-like Higgs - we reiterate that this is again a specific choice and other possibilities do exist. The color coding for all the plots in this section (and the next) are the same as described in the beginning of Sec.~\ref{sec:results}.

\begin{figure}[h!]
        \includegraphics[scale=0.8]{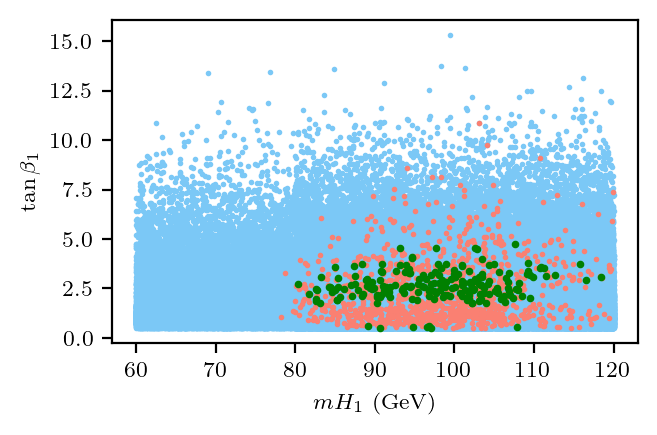}
        \includegraphics[scale=0.8]{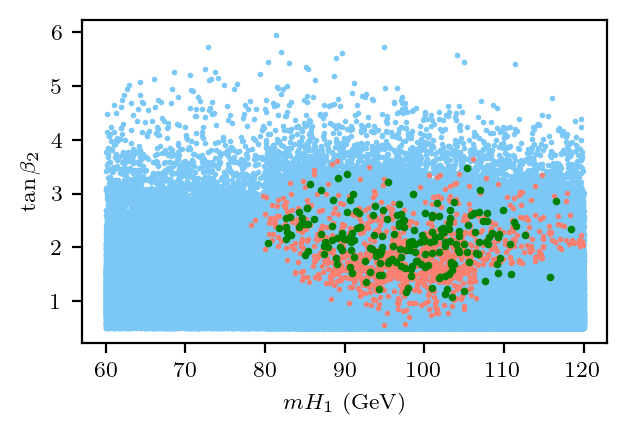}
        \includegraphics[scale=0.8]{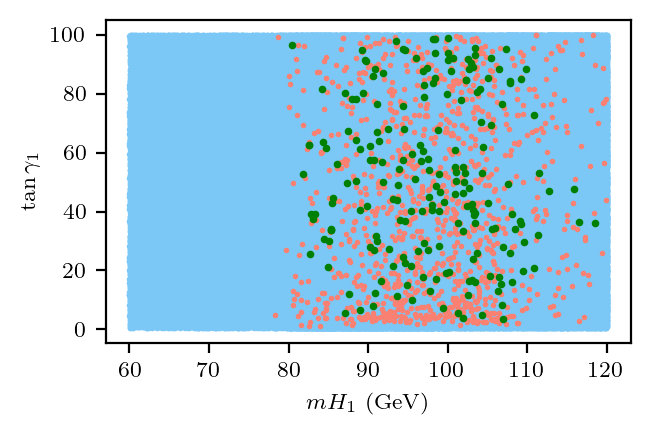}
        \includegraphics[scale=0.8]{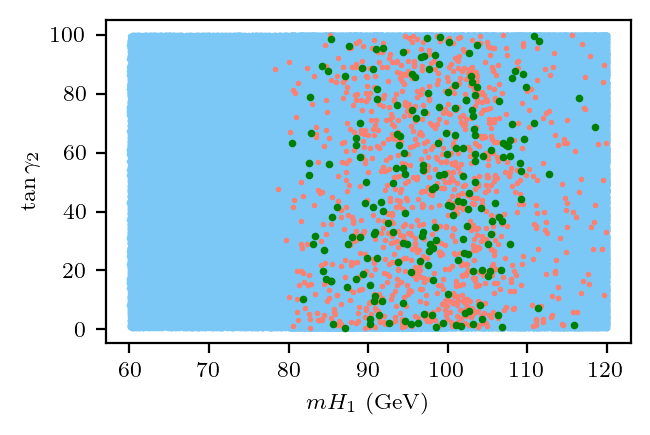} 
       \caption{The allowed regions for the Alignment 2 case with the intermediate mass CP-even Higgs identified as the 125 GeV SM-like Higgs boson.  The plots show how the allowed region in $m_{H_1}$ correlates with the various angles.}
       \label{fig:mH1-angles}
\end{figure}

We begin, in Fig.~\ref{fig:mH1-angles}, by displaying the allowed regions in the parameter space of $m_{H_1}$ and the various angles. Strikingly, the combined theoretical and experimental constraints still allow for a CP-even Higges \emph{lighter} than 125 GeV in this model. Specifically, for the democratic 3HDM discussed in this paper, a second, light Higgs in the mass range $82\, \textrm{GeV}\leq m_{H_1} \leq 120\, \textrm{GeV}$ is permissible. As with the regular hierarchy discussed above, both $\tan\beta_1$ and $\tan\beta_2$ are constrained to lie in the $1-4$ range, while the mixing angles $\gamma_{1,2}$ are relatively unconstrained.

\begin{figure}[h!]
        \includegraphics[scale=0.7]{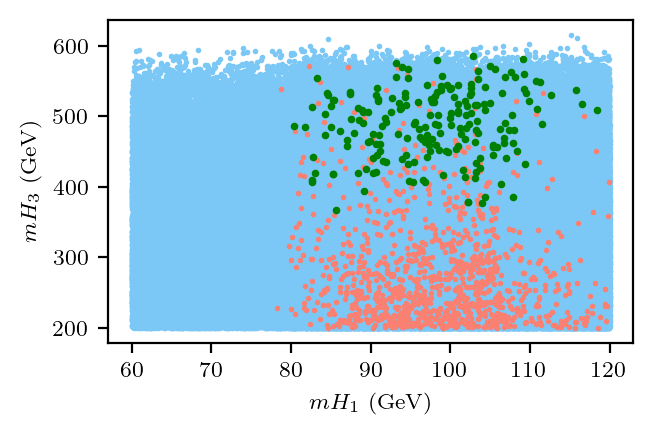}
        \includegraphics[scale=0.7]{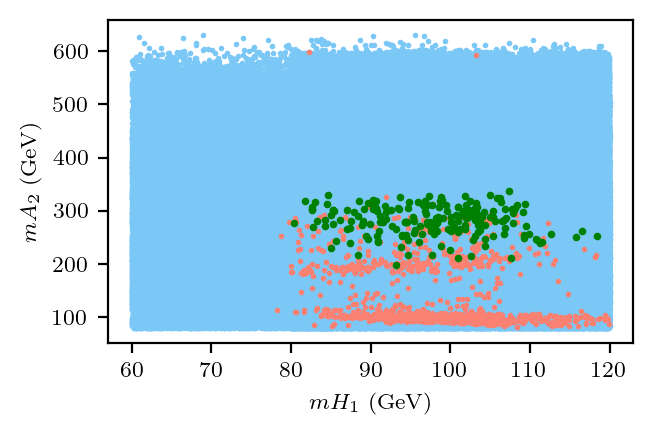}
        \includegraphics[scale=0.7]{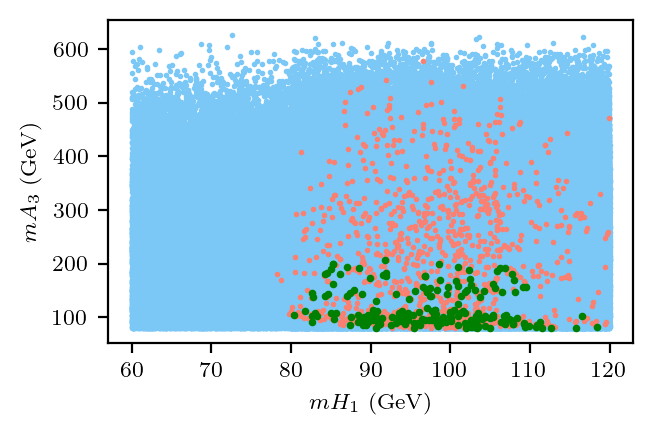}
        \includegraphics[scale=0.7]{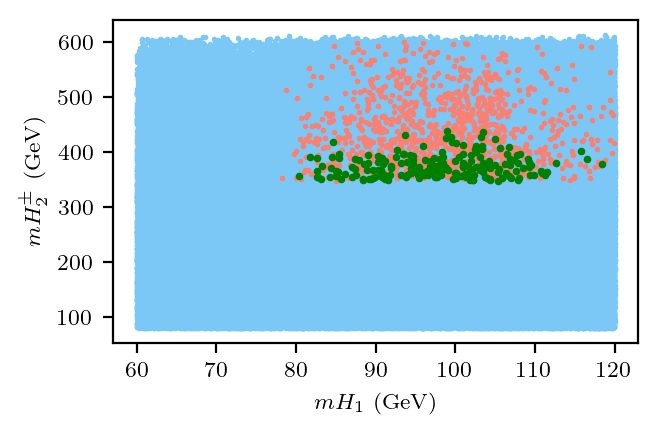}
        \includegraphics[scale=0.7]{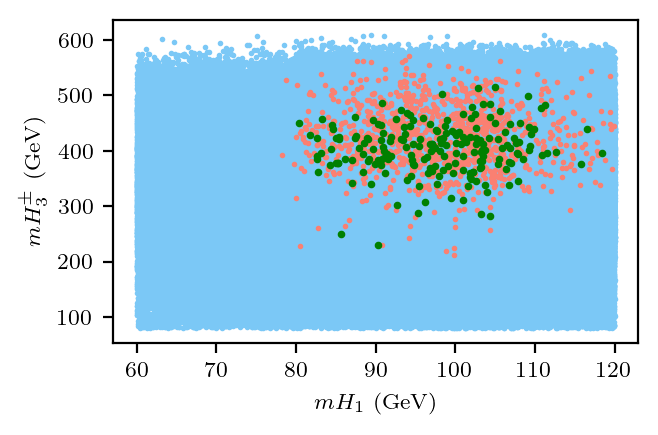}
         \caption{The allowed regions for the Alignment 2 case with the second lightest CP-even Higgses identified as the 125 GeV SM-like Higgs boson.  The plots show how the allowed region in $m_{H_1}$ correlates with the masses of the other Higgses in the theory.}
         \label{fig:mH1-al1-masses}
\end{figure}

Next in Fig.~\ref{fig:mH1-al1-masses}, we display the allowed regions in the masses of the various Higgses. To begin with, we find that the allowed range of the heaviest CP-even Higgs ($H_3$ in this case) is around $400-600$ GeV, which is roughly similar to the $H_2$ range in the previous case. However, we see that the allowed range of masses for the pseudoscalar bosons is much more constrained in this case. Specifically, we have $200\, \textrm{GeV}\lesssim m_{A_2} \lesssim 320\, \textrm{GeV}$ and $90\, \textrm{GeV}\lesssim m_{A_3} \lesssim 200\, \textrm{GeV}$. Both the charged Higgs masses are constrained to lie above 350 GeV, and thus the degree of ``non-degeneracy", particularly between the charged Higgses and the $A_3$ is a little more striking in this scenario.

\subsection{Inverted Hierarchy}

\begin{figure}[h!]
        \includegraphics[scale=0.7]{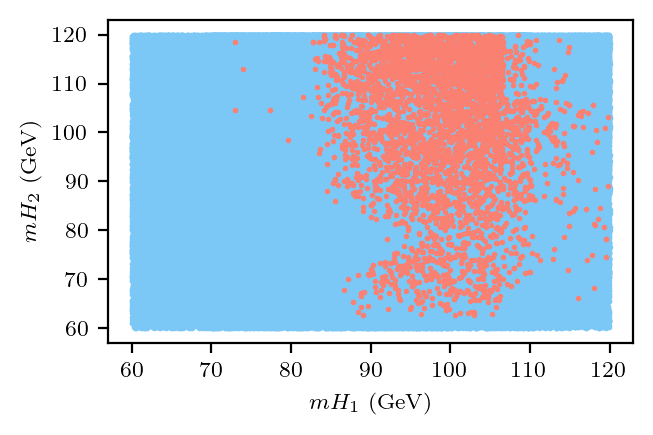}
        \includegraphics[scale=0.7]{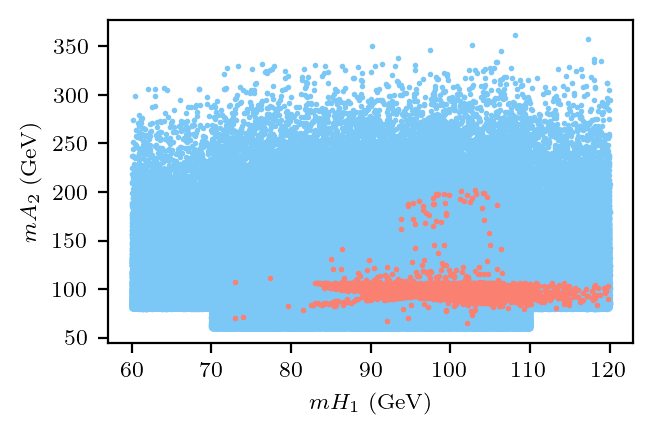}
        \includegraphics[scale=0.7]{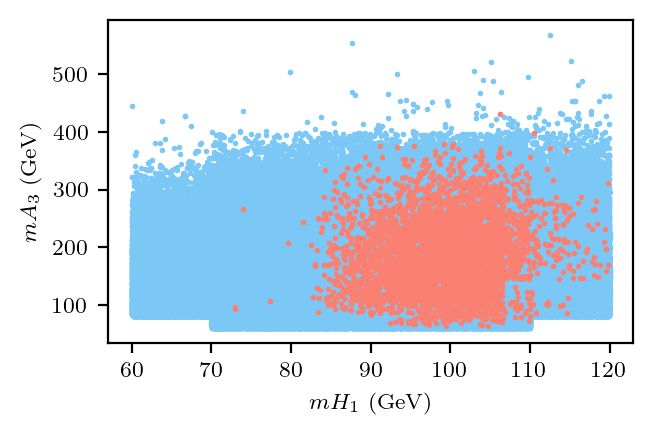}
        \includegraphics[scale=0.7]{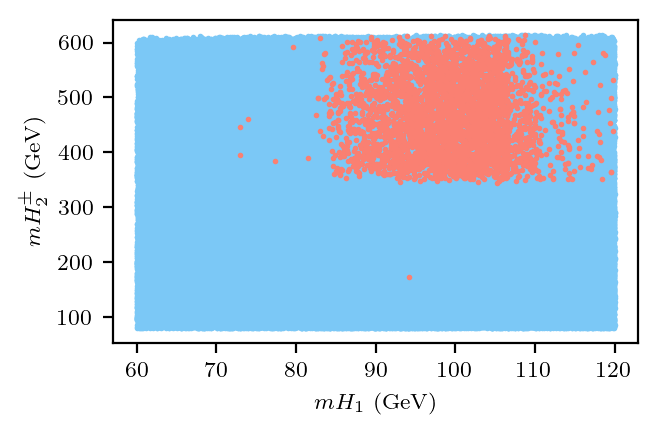}
        \includegraphics[scale=0.7]{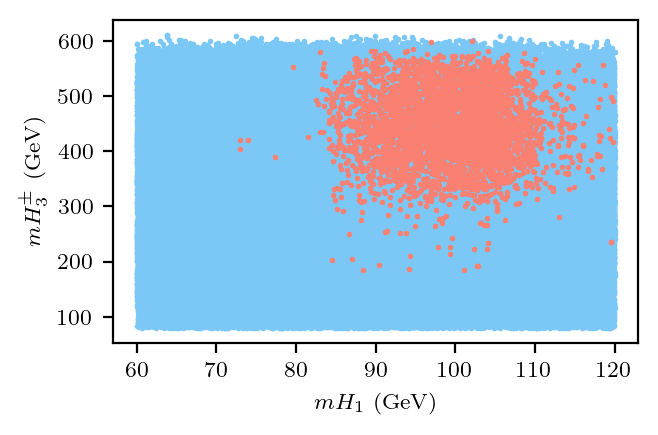}
         \caption{The allowed regions for the Alignment 3 case with the heaviest of the three CP-even Higgses identified as the 125 GeV SM-like Higgs boson.  The plots show how the allowed region in $m_{H_1}$ correlates with the masses of the other Higgses in the theory. As we can see, this scenario is not allowed as it does not pass the electroweak precision test constraints.}
         \label{fig:mH1-al1-masses-al3}
\end{figure}

Here, we align the heaviest Higgs Boson with the known SM Higgs boson, which means this case explores the possibility of having 2 CP-Even Higgs lighter than the 125 GeV Higgs Boson. While this would be very interesting to probe at the LHC, we find that that the data do not support this case, at least in the democratic 3HDM set-up we are working with. We display in Fig.~\ref{fig:mH1-al1-masses-al3} the allowed ranges of masses in this case - we see clearly that while the model does pass all the theoretical and experimental constraints, there is no subset of these points that also have the experimentally observed values of the $S$, $T$, and $U$ parameters.
\section{Discussion and Conclusions}
\label{sec:conclusions}
The 3HDM is a theoretically interesting, and well motivated extension of the SM. It is certainly interesting to explore this scenario wherein the number of scalar doublets exactly mirrors the quark and lepton sector. The presence of the third doublet makes the particle spectrum of this model much more rich than, say, the 2HDM. And accordingly, based on how the Yukawa sector is chosen, one could have many variants of this model, each with presumably their own constraints and associated phenomenology. Nevertheless, in this paper, we studied the so-called ``democratic" set-up of the 3HDM  wherein each doublet contributes to the mass of one species of fermion, i.e., one each for the up-type quarks, down-type quarks, and the charged leptons. We also imposed a discrete $Z_3$ symmetry which simplifies the scalar Lagrangian. Since there are three CP-even Higgses after EWSB in this model, the identification of one of them as the 125-GeV SM-like Higgs boson would still leave two (heavier or lighter) neutral Higgs bosons in the spectrum (in addition to two sets charged and pseudoscalar Higgs bosons). 

To understand the regions of parameter space that are allowed by all experimental and theoretical constraints, we studied three cases: Regular Hierarchy wherein the lightest of the three Higgses is the SM-like Higgs, Medial Hierarchy with the second lightest being the 125 GeV Higgs, and Inverted Hierarchy in which the heaviest of the three is identified as the SM-like Higgs. Imposing all theoretical and experimental and theoretical constraints even in this simplified set-up is still a daunting task given the multidimensional nature of the 3HDM parameter space. To mitigate this, we adopt a machine learning strategy by designing an active learning algorithm to effectively delineate the decision boundary between allowed and disallowed regions. This reduces the computational time significantly, while also maximizing accuracy.

We find, under this set-up, that this model - though highly constrained, still does have a lot of parameter space that can potentially be probed at the LHC and other future colliders. Both the Regular and Medial Hierarchy cases have heavy scalar and pseudoscalar Higgs bosons in the range of $100-600$ GeV, potentially in the right ball park to warrant a dedicated LHC study. Interestingly, the case of having at least one CP-even Higgs \emph{lighter} than the SM-like 125 GeV Higgs is not ruled out, as our results for the Medial Hierarchy case demonstrate. However, the more radical scenario of having two Higgses lighter than the 125 GeV SM-like Higgs is not favored by electroweak precision data. Nevertheless, the 3HDM remains a very interesting possibility in its own right, both from the point of view of the theoretical formulation (that can take many avatars depending on the discrete symmetries and the choice of Yukawas) and the associated collider phenomenology.

\begin{acknowledgments}
A.K. acknowledges the support from the Director's Fellowship at IIT Gandhinagar. A.K. gratefully acknowledges Kousik Loho, Rameswar Sahu, and Lalit Pathak for their insightful discussions and valuable inputs during the course of this work. The authors also thank Dipankar Das for pointing out a typographical error in the manuscript. A.S. thanks Anusandhan National Research Foundation (ANRF) for providing necessary financial support through the SERB-NPDF grant (Ref No: PDF/2023/002572). S.K.R. acknowledges the support from the Department of Atomic Energy (DAE), India, for the Regional Centre for Accelerator-based Particle Physics (RECAPP), Harish Chandra Research Institute. 
\end{acknowledgments}

\appendix

\section{Charged Higgs Mass Matrix Calculation}
\label{sec:appcharged}

The mass term for the Charged Higgs can be extracted from the scalar potential,	
\begin{equation*}
    V_C^{mass} \supset \begin{pmatrix}
        \phi_1^- & \phi_2^- & \phi_3^-
    \end{pmatrix} \mathcal{M}^2_{\phi^{\pm}} \begin{pmatrix}
        \phi_1^+ \\ \phi_2^+ \\ \phi_3^+ 
    \end{pmatrix} 
\end{equation*}
where, $\mathcal{M}^2_{\phi^{\pm}}$ is the mass matrix which is to be diagonalized
\begin{equation*}
    \mathcal{M}^2_{\phi^{\pm}} = \begin{pmatrix}
        (\mathcal{M}^2_{\phi^{\pm}})_{11} & (\mathcal{M}^2_{\phi^{\pm}})_{12} & (\mathcal{M}^2_{\phi^{\pm}})_{13} \\ (\mathcal{M}^2_{\phi^{\pm}})_{12}^* & (\mathcal{M}^2_{\phi^{\pm}})_{22} & (\mathcal{M}^2_{\phi^{\pm}})_{23} \\ (\mathcal{M}^2_{\phi^{\pm}})_{13}^* & (\mathcal{M}^2_{\phi^{\pm}})_{23}^* & (\mathcal{M}^2_{\phi^{\pm}})_{33}
    \end{pmatrix}
\end{equation*}
The elements of the matrix are given as,	
\begin{equation*}
    \begin{split}
        (\mathcal{M}^2_{\phi^{\pm}})_{11} & = -\frac{v_2^2}{2}\lambda_{7} - \frac{v_3^2}{2}\lambda_{8} - Re(\lambda_{10})v_2v_3 - \frac{v_2v_3}{2v_1}(Re(\lambda_{11})v_2+Re(\lambda_{12})v_3) \\  (\mathcal{M}^2_{\phi^{\pm}})_{12} & = \frac{v_1v_2}{2}\lambda_{7} + \frac{v_1v_3}{2}\lambda_{10} + \frac{v_2v_3}{2}\lambda_{11}\\
        (\mathcal{M}^2_{\phi^{\pm}})_{13} & = \frac{v_1v_2}{2}\lambda_{10} + \frac{v_1v_3}{2}\lambda_{8} + \frac{v_2v_3}{2}\lambda_{12}\\
        (\mathcal{M}^2_{\phi^{\pm}})_{22} & = -\frac{v_1^2}{2}\lambda_{7} - \frac{v_3^2}{2}\lambda_{9} - Re(\lambda_{11})v_1v_3 - \frac{v_1v_3}{2v_2}(Re(\lambda_{10})v_1+Re(\lambda_{12})v_3) \\
        (\mathcal{M}^2_{\phi^{\pm}})_{23} & = \frac{v_1v_2}{2}\lambda_{11}^* + \frac{v_1v_3}{2}\lambda_{12} + \frac{v_2v_3}{2}\lambda_{9} \\
        (\mathcal{M}^2_{\phi^{\pm}})_{33} & = -\frac{v_1^2}{2}\lambda_{8} - \frac{v_2^2}{2}\lambda_{9} - Re(\lambda_{12})v_1v_2 - \frac{v_1v_2}{2v_3}(Re(\lambda_{10})v_1+Re(\lambda_{11})v_2)
    \end{split}
\end{equation*}
Under a transformation by matrix $O_\beta$
\begin{equation*}
    (B_C)^2 = O_\beta.\mathcal{M}^2_{\phi^{\pm}}.O_\beta^T
\end{equation*}
where the matrix $O_\beta$ is given by,
\begin{equation}
    O_\beta = \begin{pmatrix}
        c_{\beta 2} & 0 & s_{\beta 2} \\ 0 & 1 & 0 \\ -s_{\beta 2} & 0 & c_{\beta 2}
    \end{pmatrix} \begin{pmatrix} c_{\beta 1} & s_{\beta 1} & 0 \\ -s_{\beta 1} & c_{\beta 1} & 0 \\ 0 & 0 & 1
    \end{pmatrix}
\end{equation}

\begin{equation*}
    O_\beta = \begin{pmatrix}
        c_{\beta 2}c_{\beta 1} & c_{\beta 2}s_{\beta 1} & s_{\beta 2} \\ -s_{\beta 1} & c_{\beta 1} & 0 \\ -c_{\beta 1}s_{\beta 2} & -s_{\beta 1}s_{\beta 2} & c_{\beta 2}
    \end{pmatrix}
\end{equation*}
$(B_C)^2$ is then written as 
\begin{equation*}
    (B_C)^2 =  \begin{pmatrix}
        0 & 0 & 0 \\ 0 & \mathcal{M}^2_{22} & \mathcal{M}^2_{23} \\ 0 & \mathcal{M}^{2*}_{23} & \mathcal{M}^2_{33}\\
    \end{pmatrix}
\end{equation*} 
where,
\begin{equation*}	
    \begin{split}
        \mathcal{M}^2_{22} & = - \frac{v_3}{2v_1v_2(v_1^2+v_2^2)}\left(Re(\lambda_{11})v_2(2v_1^4+2v_1^2v_2^2+v_2^4)+Re(\lambda_{10})(v_1^5+2v_1^3v_2^2+2v_1v_2^4)\right) \\ & - \frac{v_1^2+v_2^2}{2}\lambda_{7} - \frac{v_3}{2v_1v_2(v_1^2+v_2^2)}\left(v_3(\lambda_{9}v_1^3v_2+\lambda_{8}v_1v_2^3+Re(\lambda_{12})(v_1^4+v_2^4))\right) \\
        \mathcal{M}^2_{33} & = -\frac{v^2}{2v_3(v_1^2+v_2^2)}\left(Re(\lambda_{10})v_1^2v_2+Re(\lambda_{11})v_1v_2^2+2Re(\lambda_{12})v_1v_2v_3 + \lambda_{8}v_1^2v_3+\lambda_{9}v_2^2v_3\right) \\
        \mathcal{M}^2_{23} & = \frac{v}{2(v_1^2+v_2^2)}\left(-Re(\lambda_{10})v_1v_2^2+Re(\lambda_{11})v_1^2v_2+Re(\lambda_{12})v_3(v_1^2-v_2^2)-\lambda_{8}v_1v_2v_3-\lambda_{9}v_1v_2v_3\right)
    \end{split}
\end{equation*}

The terms $\mathcal{M}^2_{12}$ and $\mathcal{M}^2_{13}$ are given by 
	
\begin{equation*}
\begin{split}
    \mathcal{M}^2_{12} & = \frac{iv_3}{2v_{12}v}\left(Im(\lambda_{10})v_1(v_1^2+2v_2^2) + Im(\lambda_{11})v_2(v_2^2+2v_1^2)+Im(\lambda_{12})v_3(v_2^2-v_1^2)\right) \\
    \mathcal{M}^2_{13} & = \frac{iv_1v_2}{2v_{12}}\left(Im(\lambda_{10})v_1-Im(\lambda_{11})v_2+2Im(\lambda_{12})v_3\right)   
\end{split}
\end{equation*}
These can be canceled off by a phase rotation of the imaginary parts of $\lambda_{10}$ and $\lambda_{11}$
\begin{equation}	
\label{eq:lambdacons}
\begin{split}
    Im(\lambda_{10}) & = - \frac{v_3}{v_1}Im(\lambda_{12}) \\
    Im(\lambda_{11}) & =   \frac{v_3}{v_2}Im(\lambda_{12}) \\
\end{split}
\end{equation} 
Now, under a transformation by matrix $O_{\gamma 2}$,
\begin{equation*}
    O_{\gamma 2}.(B_C)^2.O_{\gamma 2}^\dagger = \begin{pmatrix}
        0 & 0 & 0 \\ 0 & m_{H^{\pm}_2}^2 & 0 \\ 0 & 0 & m_{H^{\pm}_3}^2
    \end{pmatrix}
\end{equation*}
where the matrix $O_{\gamma 2}$ is defined as, 
\begin{equation}
    \label{gam2trans}
    O_{\gamma 2} =  \begin{pmatrix} 1 & 0 & 0 \\ 0 & c_{\gamma 2} & -s_{\gamma 2} \\ 0 & s_{\gamma 2} & c_{\gamma 2}
    \end{pmatrix}
\end{equation}
Hence,
\begin{equation*}
    (B_C)^2 = O_{\gamma 2}^\dagger . \begin{pmatrix}
        0 & 0 & 0 \\ 0 & m_{H^{\pm}_2}^2 & 0 \\ 0 & 0 & m_{H^{\pm}_3}^2
    \end{pmatrix} . O_{\gamma 2}
\end{equation*}
Equating the two terms, we get
\begin{equation*}
    \begin{split}
        \mathcal{M}^2_{22} & = m_{H^{\pm}_2}^2 c_{\gamma 2}^2 + m_{H^{\pm}_3}^2 s_{\gamma 2}^2 \\
        \mathcal{M}^2_{33} & = m_{H^{\pm}_2}^2 s_{\gamma 2}^2 + m_{H^{\pm}_3}^2 c_{\gamma 2}^2 \\
        \mathcal{M}^2_{23} & = c_{\gamma 2}s_{\gamma 2} (m_{H^{\pm}_3}^2-m_{H^{\pm}_2}^2) 
    \end{split} 
\end{equation*}
We can now invert these relations to solve for $\lambda_{7}, \lambda_{8}$ and $\lambda_{9}$  
\begin{equation*}
    \begin{split}
        \lambda_{7} & = m_{H^{\pm}_2}^2\left(\frac{2(-c_{\gamma 2}s_{\beta 1}+c_{\beta 1}s_{\beta 2}s_{\gamma 2})(c_{\beta 1}c_{\gamma 2}+s_{\beta 1}s_{\beta 2}s_{\gamma 2})}{c_{\beta 1}c_{\beta 2}^2s_{\beta 1}v^2}\right) +
        m_{H^{\pm}_3}^2\left(\frac{2(-c_{\beta 1}s_{\gamma 2}+s_{\beta 1}s_{\beta 2}c_{\gamma 2})(s_{\beta 1}s_{\gamma 2}+c_{\beta 1}s_{\beta 2}c_{\gamma 2})}{c_{\beta 1}c_{\beta 2}^2s_{\beta 1}v^2}\right)  \\ & - \frac{s_{\beta 2}}{s_{\beta 1}c_{\beta 2}}Re(\lambda_{10}) - \frac{s_{\beta 2}}{c_{\beta 1}c_{\beta 2}}Re(\lambda_{11})
        \\ 
        \lambda_{8} & = m_{H^{\pm}_2}^2\left(\frac{2s_{\gamma 2}(c_{\gamma 2}s_{\beta 1}-c_{\beta 1}s_{\beta 2}s_{\gamma 2})}{c_{\beta 1}s_{\beta 2}v^2}\right) -
        m_{H^{\pm}_3}^2\left(\frac{2c_{\gamma 2}(s_{\beta 1}s_{\gamma 2}+c_{\beta 1}s_{\beta 2}c_{\gamma 2})}{c_{\beta 1}s_{\beta 2}v^2}\right) - \frac{s_{\beta 1}c_{\beta 2}}{s_{\beta 2}}Re(\lambda_{10}) - \frac{s_{\beta 1}}{c_{\beta 1}}Re(\lambda_{12})  \\  \lambda_{9} & = - m_{H^{\pm}_2}^2\left(\frac{2s_{\gamma 2}(c_{\beta 1}c_{\gamma 2}+s_{\beta 1}s_{\beta 2}s_{\gamma 2})}{s_{\beta 1}s_{\beta 2}v^2}\right) + m_{H^{\pm}_3}^2\left(\frac{2c_{\gamma 2}(c_{\beta 1}s_{\gamma 2}-s_{\beta 1}s_{\beta 2}c_{\gamma 2})}{s_{\beta 1}s_{\beta 2}v^2}\right) - \frac{c_{\beta 1}c_{\beta 2}}{s_{\beta 2}}Re(\lambda_{11}) - \frac{c_{\beta 1}}{s_{\beta 1}}Re(\lambda_{12})
    \end{split}
\end{equation*}

\subsection{Transformation}

The transformation rules from gauge eigenstates to mass eigenstates are given as following:
\begin{equation*}
    V_C^{mass} =\begin{pmatrix}
        \phi_1^- & \phi_2^- & \phi_3^-
    \end{pmatrix} \mathcal{M}^2_{\phi^{\pm}} \begin{pmatrix}
        \phi_1^+ \\ \phi_2^+ \\ \phi_3^+ 
    \end{pmatrix} 
\end{equation*}

\begin{equation*}
    V = \begin{pmatrix}
        \phi_1^- & \phi_2^- & \phi_3^-
    \end{pmatrix} (O_{\gamma 2}.O_\beta)^\dagger (O_{\gamma 2}.O_\beta) \mathcal{M}^2_{\phi^{\pm}} (O_{\gamma 2}.O_\beta)^\dagger (O_{\gamma 2}.O_\beta) \begin{pmatrix}
        \phi_1^+ \\ \phi_2^+ \\ \phi_3^+
    \end{pmatrix}
\end{equation*}
Hence, the transformation relation between the gauge and mass eigenstates is written as 
\begin{equation*}
    \begin{pmatrix}
        G^\pm \\ H_2^\pm \\ H_3^\pm
    \end{pmatrix} = O_{\gamma 2}.O_\beta \begin{pmatrix}
        \phi_1^\pm \\ \phi_2^\pm \\ \phi_3^\pm
    \end{pmatrix}
\end{equation*}

\begin{equation*}
    \begin{pmatrix}
        \phi_1^\pm \\ \phi_2^\pm \\ \phi_3^\pm
    \end{pmatrix} = (O_{\gamma 2}.O_\beta)^\dagger \begin{pmatrix}
        G^\pm \\ H_2^\pm \\ H_3^\pm
    \end{pmatrix}
\end{equation*}
where, 
\begin{equation}
    \label{chargetransmat}
    (O_{\gamma 2}.O_\beta)^\dagger = \begin{pmatrix}
        c_{\beta 1}c_{\beta 2} & -c_{\gamma 2 }s_{\beta 1} + c_{\beta 1}s_{\beta 2}s_{\gamma 2} & -c_{\beta 1}c_{\gamma 2}s_{\beta 2}-s_{\beta 1}s_{\gamma 2} \\
        c_{\beta 2}s_{\beta 1} & c_{\beta 1}c_{\gamma 2} + s_{\beta 1}s_{\beta 2}s_{\gamma 2} & -c_{\gamma 2}s_{\beta 1}s_{\beta 2}+c_{\beta 1}s_{\gamma 2} \\
        s_{\beta 2} & -c_{\beta 2}s_{\gamma 2} & c_{\beta 2}c_{\gamma 2} 
    \end{pmatrix}
\end{equation}
The gauge eigenstates can hence be represented in terms of mass eigenstates as:
\begin{equation}
    \label{chargetransrelat}
    \begin{split}
        \phi_1^\pm & = c_{\beta 1}c_{\beta 2} G^\pm + (-c_{\gamma 2 }s_{\beta 1} + c_{\beta 1}s_{\beta 2}s_{\gamma 2})H_2^\pm + (-c_{\beta 1}c_{\gamma 2}s_{\beta 2}-s_{\beta 1}s_{\gamma 2})H_3^\pm \\
        \phi_2^\pm & = c_{\beta 2}s_{\beta 1} G^\pm + (c_{\beta 1}c_{\gamma 2} + s_{\beta 1}s_{\beta 2}s_{\gamma 2})H_2^\pm +(-c_{\gamma 2}s_{\beta 1}s_{\beta 2}+c_{\beta 1}s_{\gamma 2})H_3^\pm \\
        \phi_3^\pm & = s_{\beta 2} G^\pm + (-c_{\beta 2}s_{\gamma 2}) H_2^\pm + (c_{\beta 2}c_{\gamma 2}) H_3^\pm
    \end{split}
\end{equation}

\section{CP-Odd Sector Mass Matrix Calculation}
\label{sec:appendpseudo}

The mass term for the CP-odd Higgs can be extracted from the scalar potential,
\begin{equation*}
    V_n^{mass} \supset \begin{pmatrix}
        n_1 & n_2 & n_3
    \end{pmatrix} \frac{\mathcal{M}^2_n}{2} \begin{pmatrix}
        n_1 \\ n_2 \\ n_3 
    \end{pmatrix} 
\end{equation*}
where, $\mathcal{M}^2_n$ is the mass matrix which is to be diagonalized
\begin{equation*}
    \mathcal{M}^2_n = \begin{pmatrix}
        (\mathcal{M}^2_n)_{11} & (\mathcal{M}^2_n)_{12} & (\mathcal{M}^2_n)_{13} \\ (\mathcal{M}^2_n)_{12} & (\mathcal{M}^2_n)_{22} & (\mathcal{M}^2_n)_{23} \\ (\mathcal{M}^2_n)_{13} & (\mathcal{M}^2_n)_{23} & (\mathcal{M}^2_n)_{33}
    \end{pmatrix}
\end{equation*}
The elements of the matrix are given as,
\begin{equation*}
    \begin{split}
        (\mathcal{M}^2_n)_{11} & = -2Re(\lambda_{10})v_2v_3 - \frac{v_2v_3}{2v_1}(Re(\lambda_{11})v_2+Re(\lambda_{12})v_3)\\ (\mathcal{M}^2_n)_{12} & = Re(\lambda_{10})v_1v_3 + Re(\lambda_{11})v_2v_3 - \frac{v_3^2}{2}Re(\lambda_{12})\\
        (\mathcal{M}^2_n)_{13} & = Re(\lambda_{10})v_1v_2 + Re(\lambda_{12})v_2v_3 - \frac{v_2^2}{2}Re(\lambda_{11})\\
        (\mathcal{M}^2_n)_{22} & = -2Re(\lambda_{11})v_1v_3 - \frac{v_1v_3}{2v_2}(Re(\lambda_{10})v_1+Re(\lambda_{12})v_3)\\
        (\mathcal{M}^2_n)_{23} & = Re(\lambda_{11})v_1v_2 + Re(\lambda_{12})v_1v_3 - \frac{v_1^2}{2}Re(\lambda_{10}) \\
        (\mathcal{M}^2_n)_{33} & = -2Re(\lambda_{12})v_1v_2 - \frac{v_1v_2}{2v_3}(Re(\lambda_{10})v_1+Re(\lambda_{11})v_2)
    \end{split}
\end{equation*}
Under a transformation by matrix $O_\beta$, which is given in equation ~\ref*{eq:betatrans}
\begin{equation*}
    (B_P)^2 = O_\beta.\mathcal{M}^2_{n}.O_\beta^T
\end{equation*}

\begin{equation*}
    (B_P)^2 =  \begin{pmatrix}
        0 & 0 & 0 \\ 0 & (B_P^2)_{22} & (B_P^2)_{23} \\ 0 & (B_P^2)_{23} & (B_P^2)_{33}\\
    \end{pmatrix}
\end{equation*} 
where, 
\begin{equation*}
    \begin{split}
        (B_P^2)_{22} & = -\frac{v_3}{2v_1v_2(v_1^2+v_2^2)}\left(Re(\lambda_{10})v_1(v_1^2+2v_2^2)^2+Re(\lambda_{11})v_2(2v_1^2+v_2^2)^2+Re(\lambda_{12})v_3(v_1^2-v_2^2)^2\right)  \\ (B_P^2)_{23} & = \frac{v}{2(v_1^2+v_2^2)}\left(-Re(\lambda_{10})v_1(v_1^2+2v_2^2)+Re(\lambda_{11})v_2(2v_1^2+v_2^2)+2Re(\lambda_{12})v_3(v_1^2-v_2^2)\right) \\ (B_P^2)_{33} & = -\frac{v^2v_1v_2}{2v_3(v_1^2+v_2^2)}\left(Re(\lambda_{10})v_1+Re(\lambda_{11})v_2+4Re(\lambda_{12})v_3\right)
    \end{split}
\end{equation*}
a subsequent transformation by matrix $O_{\gamma 1}$, which is defined as, 
\begin{equation}
    \label{eq:gamma1trans}
    O_{\gamma 1} = \begin{pmatrix} 1 & 0 & 0 \\ 0 & c_{\gamma 1} & -s_{\gamma 1} \\ 0 & s_{\gamma 1} & c_{\gamma 1}
    \end{pmatrix}
\end{equation}
gives
\begin{equation*}
    O_{\gamma 1}.(B_P)^2.O_{\gamma 1}^\dagger = \begin{pmatrix}
        0 & 0 & 0 \\ 0 & m_{A_1}^2 & 0 \\ 0 & 0 & m_{A_2}^2
    \end{pmatrix}
\end{equation*}
Hence,
\begin{equation*}
    (B_P)^2 = O_{\gamma 1}^\dagger . \begin{pmatrix}
        0 & 0 & 0 \\ 0 & m_{A_1}^2 & 0 \\ 0 & 0 & m_{A_2}^2
    \end{pmatrix} . O_{\gamma 1}
\end{equation*}
Equating the two terms, we get
\begin{equation*}
    \begin{split}
        (B_P^2)_{22} & = m_{A_1}^2 c_{\gamma 1}^2 + m_{A_2}^2 s_{\gamma 1}^2 \\
        (B_P^2)_{33} & = m_{A_1}^2 s_{\gamma 1}^2 + m_{A_2}^2 c_{\gamma 1}^2 \\
        (B_P^2)_{23} & = c_{\gamma 1}s_{\gamma 1} (m_{A_2}^2-m_{A_1}^2) 
    \end{split} 
\end{equation*}
We can now invert these relations to solve for $Re(\lambda_{10}), Re(\lambda_{11})$ and $Re(\lambda_{12})$ 
\begin{equation*}
    \begin{split}
        Re(\lambda_{10}) & = -\frac{2m_{A_1}^2}{9c_{\beta 1}^2c_{\beta 2}s_{\beta 1}s_{\beta 2}v^2}\left((-2c_{\beta 1}s_{\beta 1}c_{\gamma 1}+(c_{\beta 1}^2-s_{\beta 1}^2)s_{\beta 2}s_{\gamma 1})(-s_{\beta 1}c_{\beta 1}c_{\gamma 1}+(2c_{\beta 1}^2+s_{\beta 1}^2)s_{\beta 2}s_{\gamma 1})\right) \\ & -\frac{2m_{A_2}^2}{9c_{\beta 1}^2c_{\beta 2}s_{\beta 1}s_{\beta 2}v^2}\left((s_{\beta 1}c_{\beta 1}s_{\gamma 1}+(2c_{\beta 1}^2+s_{\beta 1}^2)s_{\beta 2}c_{\gamma 1})(2s_{\beta 1}c_{\beta 1}s_{\gamma 1}+(c_{\beta 1}^2-s_{\beta 1}^2)s_{\beta 2}c_{\gamma 1})\right)   \\ 
        Re(\lambda_{11}) & = \frac{2m_{A_1}^2}{9c_{\beta 1}c_{\beta 2}s_{\beta 1}^2s_{\beta 2}v^2}\left((s_{\beta 1}c_{\beta 1}c_{\gamma 1}+(1+s_{\beta 1}^2)s_{\beta 2}s_{\gamma 1})(-2s_{\beta 1}c_{\beta 1}c_{\gamma 1}+(c_{\beta 1}^2-s_{\beta 1}^2)s_{\beta 2}s_{\gamma 1})\right) \\ & +\frac{2m_{A_2}^2}{9c_{\beta 1}c_{\beta 2}s_{\beta 1}^2s_{\beta 2}v^2}\left(((1+s_{\beta 1}^2)s_{\beta 2}c_{\gamma 1}-s_{\beta 1}c_{\beta 1}s_{\gamma 1})(2s_{\beta 1}c_{\beta 1}s_{\gamma 1}+(c_{\beta 1}^2-s_{\beta 1}^2)s_{\beta 2}c_{\gamma 1})\right)   \\ Re(\lambda_{12}) & = \frac{2m_{A_1}^2}{9c_{\beta 1}s_{\beta 1}s_{\beta 2}^2v^2}\left((s_{\beta 1}c_{\beta 1}c_{\gamma 1}-(1+c_{\beta 1}^2)s_{\beta 2}s_{\gamma 1})(s_{\beta 1}c_{\beta 1}c_{\gamma 1}+(1+s_{\beta 1}^2)s_{\beta 2}s_{\gamma 1})\right) \\ & + \frac{2m_{A_2}^2}{9c_{\beta 1}s_{\beta 1}s_{\beta 2}^2v^2}\left((s_{\beta 1}c_{\beta 1}s_{\gamma 1}-(1+s_{\beta 1}^2)s_{\beta 2}c_{\gamma 1})(s_{\beta 1}c_{\beta 1}s_{\gamma 1}+(1+c_{\beta 1}^2)s_{\beta 2}c_{\gamma 1})\right)   
    \end{split}
\end{equation*}

\subsection{Transformation}
    
The transformation rules from gauge eigenstates to mass eigenstates are given as following:
\begin{equation*}
    V_n^{mass} = \begin{pmatrix}
        n_1 & n_2 & n_3
    \end{pmatrix} \frac{\mathcal{M}^2_n}{2} \begin{pmatrix}
        n_1 \\ n_2 \\ n_3 
    \end{pmatrix} 
\end{equation*}

\begin{equation*}
    V = \begin{pmatrix}
        n_1 & n_2 & n_3
    \end{pmatrix} (O_{\gamma 1}.O_\beta)^T (O_{\gamma 1}.O_\beta) \mathcal{M}^2_{n} (O_{\gamma 1}.O_\beta)^T (O_{\gamma 1}.O_\beta) \begin{pmatrix}
        n_1 \\ n_2 \\ n_3 
    \end{pmatrix}
\end{equation*}
Hence, the transformation relation between the gauge and mass eigenstates is written as 
\begin{equation*}
    \begin{pmatrix}
        G_0 \\ A_1 \\ A_2
    \end{pmatrix} = O_{\gamma 1}.O_\beta \begin{pmatrix}
        n_1 \\ n_2 \\ n_3
    \end{pmatrix}
\end{equation*}

\begin{equation*}
    \begin{pmatrix}
        n_1 \\ n_2 \\ n_3
    \end{pmatrix} = (O_{\gamma 1}.O_\beta)^T \begin{pmatrix}
        G_0 \\ A_1 \\ A_2
    \end{pmatrix}
\end{equation*}
where, 
\begin{equation}
    (O_{\gamma 1}.O_\beta)^T = \begin{pmatrix}
        c_{\beta 1}c_{\beta 2} & -c_{\gamma 1 }s_{\beta 1} + c_{\beta 1}s_{\beta 2}s_{\gamma 1} & -c_{\beta 1}c_{\gamma 1}s_{\beta 2}-s_{\beta 1}s_{\gamma 1} \\
        c_{\beta 2}s_{\beta 1} & c_{\beta 1}c_{\gamma 1} + s_{\beta 1}s_{\beta 2}s_{\gamma 1} & -c_{\gamma 1}s_{\beta 1}s_{\beta 2}+c_{\beta 1}s_{\gamma 1} \\
        s_{\beta 2} & -c_{\beta 2}s_{\gamma 1} & c_{\beta 2}c_{\gamma 1} 
    \end{pmatrix}
\end{equation}
The gauge eigenstates can hence be represented in terms of mass eigenstates as:
\begin{equation}
    \begin{split}
        n_1 & = c_{\beta 1}c_{\beta 2} G_0 + (-c_{\gamma 1 }s_{\beta 1} + c_{\beta 1}s_{\beta 2}s_{\gamma 1})A_1 + (-c_{\beta 1}c_{\gamma 1}s_{\beta 2}-s_{\beta 1}s_{\gamma 1})A_2 \\
        n_2 & = c_{\beta 2}s_{\beta 1} G_0 + (c_{\beta 1}c_{\gamma 1} + s_{\beta 1}s_{\beta 2}s_{\gamma 1})A_1+(-c_{\gamma 1}s_{\beta 1}s_{\beta 2}+c_{\beta 1}s_{\gamma 1})A_2 \\
        n_3 & = s_{\beta 2} G_0 + (-c_{\beta 2}s_{\gamma 1}) A_1 + (c_{\beta 2}c_{\gamma 1}) A_2
    \end{split}
\end{equation}

\section{Mixing Terms}
\label{sec:appmixing}

The presence of complex phases in the potential is responsible for CP-violation in the form of mixing between the 2 CP-Odd and the 3 CP-even neutral Higgs states.
\begin{equation*}
    V_{np} \supset \begin{pmatrix}
        n_1 & n_2 & n_3
    \end{pmatrix} \frac{\mathcal{M}^2_{np}}{2} \begin{pmatrix}
        p_1 \\ p_2 \\ p_3 
    \end{pmatrix} 
\end{equation*}
where, $\mathcal{M}^2_{np}$ is given as
\begin{equation*}
    \mathcal{M}^2_{np} = \begin{pmatrix}
        (\mathcal{M}^2_{np})_{11} & (\mathcal{M}^2_{np})_{12} & (\mathcal{M}^2_{np})_{13} \\ (\mathcal{M}^2_{np})_{21} & (\mathcal{M}^2_{np})_{22} & (\mathcal{M}^2_{np})_{23} \\ (\mathcal{M}^2_{np})_{31} & (\mathcal{M}^2_{np})_{32} & (\mathcal{M}^2_{np})_{33}
    \end{pmatrix}
\end{equation*}
The elements of the matrix are given as,
\begin{equation*}
    \begin{split}
        (\mathcal{M}^2_{np})_{11} & = 2v_2v_3Im(\lambda_{10})\\ (\mathcal{M}^2_{np})_{12} & = v_1v_3Im(\lambda_{10}) + v_2v_3Im(\lambda_{11}) + \frac{v_3^2}{2}Im(\lambda_{12})\\
        (\mathcal{M}^2_{np})_{13} & = v_1v_2Im(\lambda_{10}) + v_2v_3Im(\lambda_{12}) + \frac{v_2^2}{2}Im(\lambda_{11})\\
        (\mathcal{M}^2_{np})_{21} & = -v_1v_3Im(\lambda_{10}) - v_2v_3Im(\lambda_{11}) + \frac{v_3^2}{2}Im(\lambda_{12})\\
        (\mathcal{M}^2_{np})_{22} & = -2v_1v_3Im(\lambda_{11})\\
        (\mathcal{M}^2_{np})_{23} & = -v_1v_2Im(\lambda_{11}) + v_1v_3Im(\lambda_{12}) - \frac{v_1^2}{2}Im(\lambda_{10}) \\
        (\mathcal{M}^2_{np})_{31} & = -v_1v_2Im(\lambda_{10}) - v_2v_3Im(\lambda_{12}) + \frac{v_2^2}{2}Im(\lambda_{11}) \\
        (\mathcal{M}^2_{np})_{32} & = v_1v_2Im(\lambda_{11}) - v_1v_3Im(\lambda_{12}) - \frac{v_1^2}{2}Im(\lambda_{10}) \\
        (\mathcal{M}^2_{np})_{33} & = -2v_1v_2Im(\lambda_{12})
    \end{split}
\end{equation*}
We can substitute for $Im(\lambda_{10})$ and $Im(\lambda_{11})$ using equations ~\ref{eq:lambdacons}. We see that the only solution possible to turn these CP-violating terms off and hence exactly diagonalize the mass matrices is, 
\begin{equation}
    Im(\lambda_{12}) = 0
\end{equation} 

\section{Couplings}
\label{sec:couplings-app}

\begin{table}[h]
    \centering
    \begin{tabular}{l l}
    \toprule[1pt]
        Coupling & \ \ \ \ \ \ \ Coefficient \\
        \midrule[1pt]
        $A_2H_1Z$ & \ \ \ \ \ \ \ $\frac{e(p_1^\mu-p_2^\mu)(c_{\alpha 2}c_{\gamma 1}k_2-(s_{\beta 2}c_{\alpha 2}k_1-c_{\beta 2}s_{\alpha 2})s_{\gamma 1})}{2c_ws_w}$ \\
        $A_2H_2Z$ & \ \ \ \ \ \ \ $\frac{e(p_1^\mu-p_2^\mu)(c_{\alpha_3}(-k_1c_{\gamma_1}-k_2s_{\beta_2}s_{\gamma_1}) + s_{\alpha_3}(-c_{\gamma_1}s_{\alpha_2}k_2 + (c_{\alpha_2}c_{\beta_2}+k_1s_{\alpha_2}s_{\beta_2})s_{\gamma_1}))}{2c_ws_w}$ \\
        $A_2H_3Z$ & \ \ \ \ \ \ \ $\frac{e(p_1^\mu-p_2^\mu)(s_{\alpha_3}k_2s_{\beta_2}s_{\gamma_1} + c_{\alpha_3}(-c_{\gamma_1}s_{\alpha_2}k_2 + c_{\alpha_2}c_{\beta_2}s_{\gamma_1}) + k_1(c_{\gamma_1}s_{\alpha_3} + c_{\alpha_3}s_{\alpha_2}s_{\beta_2}s_{\gamma_1})}{2c_ws_w}$ \\
        $A_3H_1Z$ & \ \ \ \ \ \ \ $\frac{-e(p_1^\mu-p_2^\mu)(c_{\beta_2}c_{\gamma_1}s_{\alpha_2} + c_{\alpha_2}(-k_1c_{\gamma_1}s_{\beta_2} - k_2s_{\gamma_1}))}{2c_ws_w}$ \\
        $A_3H_2Z$ & \ \ \ \ \ \ \ $\frac{-e(p_1^\mu-p_2^\mu)(c_{\alpha_2}c_{\beta_2}c_{\gamma_1}s_{\alpha_3} + k_1(c_{\gamma_1}s_{\alpha_2}s_{\alpha_3}s_{\beta_2} + c_{\alpha_3}s_{\gamma_1}) - k_2(c_{\alpha_3}c_{\gamma_1}s_{\beta_2} - s_{\alpha_2}s_{\alpha_3}s_{\gamma_1}))}{2c_ws_w}$ \\
        $A_3H_3Z$ & \ \ \ \ \ \ \ $\frac{e(p_1^\mu-p_2^\mu)(-c_{\alpha_2}c_{\alpha_3}c_{\beta_2}c_{\gamma_1} + s_{\alpha_3}(-c_{\gamma_1}k_2s_{\beta_2} + k_1s_{\gamma_1}) - c_{\alpha_3}s_{\alpha_2}(k_1c_{\gamma_1}s_{\beta_2} + k_2s_{\gamma_1}))}{2c_ws_w}$ \\
        $A_2H_2^-W^+$ & \ \ \ \ \ \ \ $\frac{e(p_1^\mu-p_2^\mu)(\cos(\gamma_1-\gamma_2))}{2s_w}$ \\
        $A_2H_3^-W^+$ & \ \ \ \ \ \ \ $\frac{-e(p_1^\mu-p_2^\mu)(\sin(\gamma_1-\gamma_2))}{2s_w}$ \\
        $A_3H_2^-W^+$ & \ \ \ \ \ \ \ $\frac{e(p_1^\mu-p_2^\mu)(\sin(\gamma_1-\gamma_2))}{2s_w}$ \\
        $A_3H_3^-W^+$ & \ \ \ \ \ \ \ $\frac{e(p_1^\mu-p_2^\mu)(\cos(\gamma_1-\gamma_2))}{2s_w}$ \\
        $H_1H_2^-W^+$ & \ \ \ \ \ \ \ $\frac{-e(p_1^\mu-p_2^\mu)(-c_{\beta_2}s_{\alpha_2}s_{\gamma_2} + c_{\alpha_2}(-c_{\gamma_2}k_2 + k_1s_{\beta_2}s_{\gamma_2}))}{2s_w}$ \\
        $H_1H_3^-W^+$ & \ \ \ \ \ \ \ $\frac{-e(p_1^\mu-p_2^\mu)(c_{\beta_2}c_{\gamma_2}s_{\alpha_2} + c_{\alpha_2}(-k_1c_{\gamma_2}s_{\beta_2} - k_2s_{\gamma_2}))}{2s_w}$ \\
        $H_2H_2^-W^+$ & \ \ \ \ \ \ \ $\frac{-e(p_1^\mu-p_2^\mu)(c_{\alpha_3}(k_1c_{\gamma_2} + k_2s_{\beta_2}s_{\gamma_2}) - s_{\alpha_3}(-c_{\gamma_2}s_{\alpha_2}k_2 + (c_{\alpha_2}c_{\beta_2} + k_1s_{\alpha_2}s_{\beta_2})s_{\gamma_2}))}{2s_w}$ \\
        $H_2H_3^-W^+$ & \ \ \ \ \ \ \ $\frac{-e(p_1^\mu-p_2^\mu)(c_{\alpha_2}c_{\beta_2}c_{\gamma_2}s_{\alpha_3} + k_1(c_{\gamma_2}s_{\alpha_2}s_{\alpha_3}s_{\beta_2} + c_{\alpha_3}s_{\gamma_2}) - k_2(c_{\alpha_3}c_{\gamma_2}s_{\beta_2} - s_{\alpha_2}s_{\alpha_3}s_{\gamma_2}))}{2s_w}$ \\
        $H_3H_2^-W^+$ & \ \ \ \ \ \ \ $\frac{e(p_1^\mu-p_2^\mu)(s_{\alpha_3}k_2s_{\beta_2}s_{\gamma_2} + c_{\alpha_3}(-c_{\gamma_2}s_{\alpha_2}k_2 + c_{\alpha_2}c_{\beta_2}s_{\gamma_2}) + k_1(c_{\gamma_2}s_{\alpha_3} + c_{\alpha_3}s_{\alpha_2}s_{\beta_2}s_{\gamma_2}))}{2s_w}$ \\
        $H_3H_3^-W^+$ & \ \ \ \ \ \ \ $\frac{-e(p_1^\mu-p_2^\mu)(c_{\alpha_2}c_{\alpha_3}c_{\beta_2}c_{\gamma_2} - s_{\alpha_3}(-c_{\gamma_2}k_2s_{\beta_2} + k_1s_{\gamma_1}) + c_{\alpha_3}s_{\alpha_2}(k_1c_{\gamma_2}s_{\beta_2} + k_2s_{\gamma_2}))}{2s_w}$ \\
        $H_2^+H_2^-Z$ & \ \ \ \ \ \ \ $-e(p_1^\mu-p_2^\mu)\cot(2\theta_w)$ \\
        $H_3^+H_3^-Z$ & \ \ \ \ \ \ \ $-e(p_1^\mu-p_2^\mu)\cot(2\theta_w)$ \\
    \bottomrule[1pt]
    \bottomrule[1pt]
    \end{tabular}
    \caption{List of all three point couplings involving the scalar and vector bosons (of the form $SSV$) in the model.}
    \label{tab:couplingsadd}
\end{table}

\bibliography{bibliography}

\begin{thebibliography}{60}%
\makeatletter
\providecommand \@ifxundefined [1]{%
 \@ifx{#1\undefined}
}%
\providecommand \@ifnum [1]{%
 \ifnum #1\expandafter \@firstoftwo
 \else \expandafter \@secondoftwo
 \fi
}%
\providecommand \@ifx [1]{%
 \ifx #1\expandafter \@firstoftwo
 \else \expandafter \@secondoftwo
 \fi
}%
\providecommand \natexlab [1]{#1}%
\providecommand \enquote  [1]{``#1''}%
\providecommand \bibnamefont  [1]{#1}%
\providecommand \bibfnamefont [1]{#1}%
\providecommand \citenamefont [1]{#1}%
\providecommand \href@noop [0]{\@secondoftwo}%
\providecommand \href [0]{\begingroup \@sanitize@url \@href}%
\providecommand \@href[1]{\@@startlink{#1}\@@href}%
\providecommand \@@href[1]{\endgroup#1\@@endlink}%
\providecommand \@sanitize@url [0]{\catcode `\\12\catcode `\$12\catcode
  `\&12\catcode `\#12\catcode `\^12\catcode `\_12\catcode `\%12\relax}%
\providecommand \@@startlink[1]{}%
\providecommand \@@endlink[0]{}%
\providecommand \url  [0]{\begingroup\@sanitize@url \@url }%
\providecommand \@url [1]{\endgroup\@href {#1}{\urlprefix }}%
\providecommand \urlprefix  [0]{URL }%
\providecommand \Eprint [0]{\href }%
\providecommand \doibase [0]{https://doi.org/}%
\providecommand \selectlanguage [0]{\@gobble}%
\providecommand \bibinfo  [0]{\@secondoftwo}%
\providecommand \bibfield  [0]{\@secondoftwo}%
\providecommand \translation [1]{[#1]}%
\providecommand \BibitemOpen [0]{}%
\providecommand \bibitemStop [0]{}%
\providecommand \bibitemNoStop [0]{.\EOS\space}%
\providecommand \EOS [0]{\spacefactor3000\relax}%
\providecommand \BibitemShut  [1]{\csname bibitem#1\endcsname}%
\let\auto@bib@innerbib\@empty
\bibitem [{\citenamefont {Salam}(1968)}]{Salam:1968rm}%
  \BibitemOpen
  \bibfield  {author} {\bibinfo {author} {\bibfnamefont {A.}~\bibnamefont
  {Salam}},\ }\bibfield  {title} {\bibinfo {title} {{Weak and Electromagnetic
  Interactions}},\ }\href {https://doi.org/10.1142/9789812795915_0034}
  {\bibfield  {journal} {\bibinfo  {journal} {Conf. Proc. C}\ }\textbf
  {\bibinfo {volume} {680519}},\ \bibinfo {pages} {367} (\bibinfo {year}
  {1968})}\BibitemShut {NoStop}%
\bibitem [{\citenamefont {Weinberg}(1967)}]{Weinberg:1967tq}%
  \BibitemOpen
  \bibfield  {author} {\bibinfo {author} {\bibfnamefont {S.}~\bibnamefont
  {Weinberg}},\ }\bibfield  {title} {\bibinfo {title} {{A Model of Leptons}},\
  }\href {https://doi.org/10.1103/PhysRevLett.19.1264} {\bibfield  {journal}
  {\bibinfo  {journal} {Phys. Rev. Lett.}\ }\textbf {\bibinfo {volume} {19}},\
  \bibinfo {pages} {1264} (\bibinfo {year} {1967})}\BibitemShut {NoStop}%
\bibitem [{\citenamefont {Glashow}(1961)}]{Glashow:1961tr}%
  \BibitemOpen
  \bibfield  {author} {\bibinfo {author} {\bibfnamefont {S.~L.}\ \bibnamefont
  {Glashow}},\ }\bibfield  {title} {\bibinfo {title} {{Partial Symmetries of
  Weak Interactions}},\ }\href {https://doi.org/10.1016/0029-5582(61)90469-2}
  {\bibfield  {journal} {\bibinfo  {journal} {Nucl. Phys.}\ }\textbf {\bibinfo
  {volume} {22}},\ \bibinfo {pages} {579} (\bibinfo {year} {1961})}\BibitemShut
  {NoStop}%
\bibitem [{\citenamefont {Aad}\ \emph {et~al.}(2012)\citenamefont {Aad} \emph
  {et~al.}}]{ATLAS:2012yve}%
  \BibitemOpen
  \bibfield  {author} {\bibinfo {author} {\bibfnamefont {G.}~\bibnamefont
  {Aad}} \emph {et~al.} (\bibinfo {collaboration} {ATLAS}),\ }\bibfield
  {title} {\bibinfo {title} {{Observation of a new particle in the search for
  the Standard Model Higgs boson with the ATLAS detector at the LHC}},\ }\href
  {https://doi.org/10.1016/j.physletb.2012.08.020} {\bibfield  {journal}
  {\bibinfo  {journal} {Phys. Lett. B}\ }\textbf {\bibinfo {volume} {716}},\
  \bibinfo {pages} {1} (\bibinfo {year} {2012})},\ \Eprint
  {https://arxiv.org/abs/1207.7214} {arXiv:1207.7214 [hep-ex]} \BibitemShut
  {NoStop}%
\bibitem [{\citenamefont {Chatrchyan}\ \emph {et~al.}(2012)\citenamefont
  {Chatrchyan} \emph {et~al.}}]{CMS:2012qbp}%
  \BibitemOpen
  \bibfield  {author} {\bibinfo {author} {\bibfnamefont {S.}~\bibnamefont
  {Chatrchyan}} \emph {et~al.} (\bibinfo {collaboration} {CMS}),\ }\bibfield
  {title} {\bibinfo {title} {{Observation of a New Boson at a Mass of 125 GeV
  with the CMS Experiment at the LHC}},\ }\href
  {https://doi.org/10.1016/j.physletb.2012.08.021} {\bibfield  {journal}
  {\bibinfo  {journal} {Phys. Lett. B}\ }\textbf {\bibinfo {volume} {716}},\
  \bibinfo {pages} {30} (\bibinfo {year} {2012})},\ \Eprint
  {https://arxiv.org/abs/1207.7235} {arXiv:1207.7235 [hep-ex]} \BibitemShut
  {NoStop}%
\bibitem [{\citenamefont {Branco}\ \emph {et~al.}(2012)\citenamefont {Branco},
  \citenamefont {Ferreira}, \citenamefont {Lavoura}, \citenamefont {Rebelo},
  \citenamefont {Sher},\ and\ \citenamefont {Silva}}]{Branco:2011iw}%
  \BibitemOpen
  \bibfield  {author} {\bibinfo {author} {\bibfnamefont {G.~C.}\ \bibnamefont
  {Branco}}, \bibinfo {author} {\bibfnamefont {P.~M.}\ \bibnamefont
  {Ferreira}}, \bibinfo {author} {\bibfnamefont {L.}~\bibnamefont {Lavoura}},
  \bibinfo {author} {\bibfnamefont {M.~N.}\ \bibnamefont {Rebelo}}, \bibinfo
  {author} {\bibfnamefont {M.}~\bibnamefont {Sher}},\ and\ \bibinfo {author}
  {\bibfnamefont {J.~P.}\ \bibnamefont {Silva}},\ }\bibfield  {title} {\bibinfo
  {title} {{Theory and phenomenology of two-Higgs-doublet models}},\ }\href
  {https://doi.org/10.1016/j.physrep.2012.02.002} {\bibfield  {journal}
  {\bibinfo  {journal} {Phys. Rept.}\ }\textbf {\bibinfo {volume} {516}},\
  \bibinfo {pages} {1} (\bibinfo {year} {2012})},\ \Eprint
  {https://arxiv.org/abs/1106.0034} {arXiv:1106.0034 [hep-ph]} \BibitemShut
  {NoStop}%
\bibitem [{\citenamefont {Coleppa}\ \emph {et~al.}(2014)\citenamefont
  {Coleppa}, \citenamefont {Kling},\ and\ \citenamefont
  {Su}}]{Coleppa:2013dya}%
  \BibitemOpen
  \bibfield  {author} {\bibinfo {author} {\bibfnamefont {B.}~\bibnamefont
  {Coleppa}}, \bibinfo {author} {\bibfnamefont {F.}~\bibnamefont {Kling}},\
  and\ \bibinfo {author} {\bibfnamefont {S.}~\bibnamefont {Su}},\ }\bibfield
  {title} {\bibinfo {title} {{Constraining Type II 2HDM in Light of LHC Higgs
  Searches}},\ }\href {https://doi.org/10.1007/JHEP01(2014)161} {\bibfield
  {journal} {\bibinfo  {journal} {JHEP}\ }\textbf {\bibinfo {volume} {01}},\
  \bibinfo {pages} {161}},\ \Eprint {https://arxiv.org/abs/1305.0002}
  {arXiv:1305.0002 [hep-ph]} \BibitemShut {NoStop}%
\bibitem [{\citenamefont {Chang}\ \emph {et~al.}(2013)\citenamefont {Chang},
  \citenamefont {Kang}, \citenamefont {Lee}, \citenamefont {Lee}, \citenamefont
  {Park},\ and\ \citenamefont {Song}}]{Chang:2012ve}%
  \BibitemOpen
  \bibfield  {author} {\bibinfo {author} {\bibfnamefont {S.}~\bibnamefont
  {Chang}}, \bibinfo {author} {\bibfnamefont {S.~K.}\ \bibnamefont {Kang}},
  \bibinfo {author} {\bibfnamefont {J.-P.}\ \bibnamefont {Lee}}, \bibinfo
  {author} {\bibfnamefont {K.~Y.}\ \bibnamefont {Lee}}, \bibinfo {author}
  {\bibfnamefont {S.~C.}\ \bibnamefont {Park}},\ and\ \bibinfo {author}
  {\bibfnamefont {J.}~\bibnamefont {Song}},\ }\bibfield  {title} {\bibinfo
  {title} {{Comprehensive study of two Higgs doublet model in light of the new
  boson with mass around 125 GeV}},\ }\href
  {https://doi.org/10.1007/JHEP05(2013)075} {\bibfield  {journal} {\bibinfo
  {journal} {JHEP}\ }\textbf {\bibinfo {volume} {05}},\ \bibinfo {pages}
  {075}},\ \Eprint {https://arxiv.org/abs/1210.3439} {arXiv:1210.3439 [hep-ph]}
  \BibitemShut {NoStop}%
\bibitem [{\citenamefont {Grinstein}\ and\ \citenamefont
  {Uttayarat}(2013)}]{Grinstein:2013npa}%
  \BibitemOpen
  \bibfield  {author} {\bibinfo {author} {\bibfnamefont {B.}~\bibnamefont
  {Grinstein}}\ and\ \bibinfo {author} {\bibfnamefont {P.}~\bibnamefont
  {Uttayarat}},\ }\bibfield  {title} {\bibinfo {title} {{Carving Out Parameter
  Space in Type-II Two Higgs Doublets Model}},\ }\href
  {https://doi.org/10.1007/JHEP06(2013)094} {\bibfield  {journal} {\bibinfo
  {journal} {JHEP}\ }\textbf {\bibinfo {volume} {06}},\ \bibinfo {pages}
  {094}},\ \bibinfo {note} {[Erratum: JHEP 09, 110 (2013)]},\ \Eprint
  {https://arxiv.org/abs/1304.0028} {arXiv:1304.0028 [hep-ph]} \BibitemShut
  {NoStop}%
\bibitem [{\citenamefont {Georgi}\ and\ \citenamefont
  {Machacek}(1985)}]{Georgi:1985nv}%
  \BibitemOpen
  \bibfield  {author} {\bibinfo {author} {\bibfnamefont {H.}~\bibnamefont
  {Georgi}}\ and\ \bibinfo {author} {\bibfnamefont {M.}~\bibnamefont
  {Machacek}},\ }\bibfield  {title} {\bibinfo {title} {{DOUBLY CHARGED HIGGS
  BOSONS}},\ }\href {https://doi.org/10.1016/0550-3213(85)90325-6} {\bibfield
  {journal} {\bibinfo  {journal} {Nucl. Phys. B}\ }\textbf {\bibinfo {volume}
  {262}},\ \bibinfo {pages} {463} (\bibinfo {year} {1985})}\BibitemShut
  {NoStop}%
\bibitem [{\citenamefont {Chanowitz}\ and\ \citenamefont
  {Golden}(1985)}]{Chanowitz:1985ug}%
  \BibitemOpen
  \bibfield  {author} {\bibinfo {author} {\bibfnamefont {M.~S.}\ \bibnamefont
  {Chanowitz}}\ and\ \bibinfo {author} {\bibfnamefont {M.}~\bibnamefont
  {Golden}},\ }\bibfield  {title} {\bibinfo {title} {{Higgs Boson Triplets With
  M ($W$) = M ($Z$) $\cos \theta \omega$}},\ }\href
  {https://doi.org/10.1016/0370-2693(85)90700-2} {\bibfield  {journal}
  {\bibinfo  {journal} {Phys. Lett. B}\ }\textbf {\bibinfo {volume} {165}},\
  \bibinfo {pages} {105} (\bibinfo {year} {1985})}\BibitemShut {NoStop}%
\bibitem [{\citenamefont {de~Lima}\ and\ \citenamefont
  {Logan}(2022)}]{deLima:2022yvn}%
  \BibitemOpen
  \bibfield  {author} {\bibinfo {author} {\bibfnamefont {C.~H.}\ \bibnamefont
  {de~Lima}}\ and\ \bibinfo {author} {\bibfnamefont {H.~E.}\ \bibnamefont
  {Logan}},\ }\bibfield  {title} {\bibinfo {title} {{Unavoidable Higgs coupling
  deviations in the Z2-symmetric Georgi-Machacek model}},\ }\href
  {https://doi.org/10.1103/PhysRevD.106.115020} {\bibfield  {journal} {\bibinfo
   {journal} {Phys. Rev. D}\ }\textbf {\bibinfo {volume} {106}},\ \bibinfo
  {pages} {115020} (\bibinfo {year} {2022})},\ \Eprint
  {https://arxiv.org/abs/2209.08393} {arXiv:2209.08393 [hep-ph]} \BibitemShut
  {NoStop}%
\bibitem [{\citenamefont {Drozd}\ \emph {et~al.}(2014)\citenamefont {Drozd},
  \citenamefont {Grzadkowski}, \citenamefont {Gunion},\ and\ \citenamefont
  {Jiang}}]{Drozd:2014yla}%
  \BibitemOpen
  \bibfield  {author} {\bibinfo {author} {\bibfnamefont {A.}~\bibnamefont
  {Drozd}}, \bibinfo {author} {\bibfnamefont {B.}~\bibnamefont {Grzadkowski}},
  \bibinfo {author} {\bibfnamefont {J.~F.}\ \bibnamefont {Gunion}},\ and\
  \bibinfo {author} {\bibfnamefont {Y.}~\bibnamefont {Jiang}},\ }\bibfield
  {title} {\bibinfo {title} {{Extending two-Higgs-doublet models by a singlet
  scalar field - the Case for Dark Matter}},\ }\href
  {https://doi.org/10.1007/JHEP11(2014)105} {\bibfield  {journal} {\bibinfo
  {journal} {JHEP}\ }\textbf {\bibinfo {volume} {11}},\ \bibinfo {pages}
  {105}},\ \Eprint {https://arxiv.org/abs/1408.2106} {arXiv:1408.2106 [hep-ph]}
  \BibitemShut {NoStop}%
\bibitem [{\citenamefont {Bhattacharya}\ \emph {et~al.}(2024)\citenamefont
  {Bhattacharya}, \citenamefont {Dey}, \citenamefont {Lahiri},\ and\
  \citenamefont {Mukhopadhyaya}}]{Bhattacharya:2023qfs}%
  \BibitemOpen
  \bibfield  {author} {\bibinfo {author} {\bibfnamefont {S.}~\bibnamefont
  {Bhattacharya}}, \bibinfo {author} {\bibfnamefont {A.}~\bibnamefont {Dey}},
  \bibinfo {author} {\bibfnamefont {J.}~\bibnamefont {Lahiri}},\ and\ \bibinfo
  {author} {\bibfnamefont {B.}~\bibnamefont {Mukhopadhyaya}},\ }\bibfield
  {title} {\bibinfo {title} {{High scale validity of two-Higgs-doublet
  scenarios with a real scalar singlet dark matter}},\ }\href
  {https://doi.org/10.1103/PhysRevD.110.055034} {\bibfield  {journal} {\bibinfo
   {journal} {Phys. Rev. D}\ }\textbf {\bibinfo {volume} {110}},\ \bibinfo
  {pages} {055034} (\bibinfo {year} {2024})},\ \Eprint
  {https://arxiv.org/abs/2308.12473} {arXiv:2308.12473 [hep-ph]} \BibitemShut
  {NoStop}%
\bibitem [{\citenamefont {Muhlleitner}\ \emph {et~al.}(2017)\citenamefont
  {Muhlleitner}, \citenamefont {Sampaio}, \citenamefont {Santos},\ and\
  \citenamefont {Wittbrodt}}]{Muhlleitner:2016mzt}%
  \BibitemOpen
  \bibfield  {author} {\bibinfo {author} {\bibfnamefont {M.}~\bibnamefont
  {Muhlleitner}}, \bibinfo {author} {\bibfnamefont {M.~O.~P.}\ \bibnamefont
  {Sampaio}}, \bibinfo {author} {\bibfnamefont {R.}~\bibnamefont {Santos}},\
  and\ \bibinfo {author} {\bibfnamefont {J.}~\bibnamefont {Wittbrodt}},\
  }\bibfield  {title} {\bibinfo {title} {{The N2HDM under Theoretical and
  Experimental Scrutiny}},\ }\href {https://doi.org/10.1007/JHEP03(2017)094}
  {\bibfield  {journal} {\bibinfo  {journal} {JHEP}\ }\textbf {\bibinfo
  {volume} {03}},\ \bibinfo {pages} {094}},\ \Eprint
  {https://arxiv.org/abs/1612.01309} {arXiv:1612.01309 [hep-ph]} \BibitemShut
  {NoStop}%
\bibitem [{\citenamefont {Keus}\ \emph {et~al.}(2018)\citenamefont {Keus},
  \citenamefont {Koivunen},\ and\ \citenamefont {Tuominen}}]{Keus:2017ioh}%
  \BibitemOpen
  \bibfield  {author} {\bibinfo {author} {\bibfnamefont {V.}~\bibnamefont
  {Keus}}, \bibinfo {author} {\bibfnamefont {N.}~\bibnamefont {Koivunen}},\
  and\ \bibinfo {author} {\bibfnamefont {K.}~\bibnamefont {Tuominen}},\
  }\bibfield  {title} {\bibinfo {title} {{Singlet scalar and 2HDM extensions of
  the Standard Model: CP-violation and constraints from $(g-2)_\mu$ and
  $e$EDM}},\ }\href {https://doi.org/10.1007/JHEP09(2018)059} {\bibfield
  {journal} {\bibinfo  {journal} {JHEP}\ }\textbf {\bibinfo {volume} {09}},\
  \bibinfo {pages} {059}},\ \Eprint {https://arxiv.org/abs/1712.09613}
  {arXiv:1712.09613 [hep-ph]} \BibitemShut {NoStop}%
\bibitem [{\citenamefont {Logan}\ \emph {et~al.}(2021)\citenamefont {Logan},
  \citenamefont {Moretti}, \citenamefont {Rojas-Ciofalo},\ and\ \citenamefont
  {Song}}]{Logan:2020mdz}%
  \BibitemOpen
  \bibfield  {author} {\bibinfo {author} {\bibfnamefont {H.~E.}\ \bibnamefont
  {Logan}}, \bibinfo {author} {\bibfnamefont {S.}~\bibnamefont {Moretti}},
  \bibinfo {author} {\bibfnamefont {D.}~\bibnamefont {Rojas-Ciofalo}},\ and\
  \bibinfo {author} {\bibfnamefont {M.}~\bibnamefont {Song}},\ }\bibfield
  {title} {\bibinfo {title} {{CP violation from charged Higgs bosons in the
  three Higgs doublet model}},\ }\href
  {https://doi.org/10.1007/JHEP07(2021)158} {\bibfield  {journal} {\bibinfo
  {journal} {JHEP}\ }\textbf {\bibinfo {volume} {07}},\ \bibinfo {pages}
  {158}},\ \Eprint {https://arxiv.org/abs/2012.08846} {arXiv:2012.08846
  [hep-ph]} \BibitemShut {NoStop}%
\bibitem [{\citenamefont {Akeroyd}\ \emph
  {et~al.}(2021{\natexlab{a}})\citenamefont {Akeroyd}, \citenamefont {Logan},
  \citenamefont {Moretti}, \citenamefont {Rojas-Ciofalo}, \citenamefont
  {Shindou},\ and\ \citenamefont {Song}}]{Akeroyd:2021fpf}%
  \BibitemOpen
  \bibfield  {author} {\bibinfo {author} {\bibfnamefont {A.~G.}\ \bibnamefont
  {Akeroyd}}, \bibinfo {author} {\bibfnamefont {H.~E.}\ \bibnamefont {Logan}},
  \bibinfo {author} {\bibfnamefont {S.}~\bibnamefont {Moretti}}, \bibinfo
  {author} {\bibfnamefont {D.}~\bibnamefont {Rojas-Ciofalo}}, \bibinfo {author}
  {\bibfnamefont {T.}~\bibnamefont {Shindou}},\ and\ \bibinfo {author}
  {\bibfnamefont {M.}~\bibnamefont {Song}},\ }\bibfield  {title} {\bibinfo
  {title} {{CP-Violation in the 3-Higgs Doublet Model: CP-Asymmetries from
  Charged Higgs Bosons and Electric Dipole Moments}},\ }\href@noop {} {\
  (\bibinfo {year} {2021}{\natexlab{a}})},\ \Eprint
  {https://arxiv.org/abs/2111.11931} {arXiv:2111.11931 [hep-ph]} \BibitemShut
  {NoStop}%
\bibitem [{\citenamefont {Dey}\ \emph {et~al.}(2024{\natexlab{a}})\citenamefont
  {Dey}, \citenamefont {Keus}, \citenamefont {Moretti},\ and\ \citenamefont
  {Shepherd-Themistocleous}}]{Dey:2023exa}%
  \BibitemOpen
  \bibfield  {author} {\bibinfo {author} {\bibfnamefont {A.}~\bibnamefont
  {Dey}}, \bibinfo {author} {\bibfnamefont {V.}~\bibnamefont {Keus}}, \bibinfo
  {author} {\bibfnamefont {S.}~\bibnamefont {Moretti}},\ and\ \bibinfo {author}
  {\bibfnamefont {C.}~\bibnamefont {Shepherd-Themistocleous}},\ }\bibfield
  {title} {\bibinfo {title} {{A smoking gun signature of the 3HDM}},\ }\href
  {https://doi.org/10.1007/JHEP07(2024)038} {\bibfield  {journal} {\bibinfo
  {journal} {JHEP}\ }\textbf {\bibinfo {volume} {07}},\ \bibinfo {pages}
  {038}},\ \Eprint {https://arxiv.org/abs/2310.06593} {arXiv:2310.06593
  [hep-ph]} \BibitemShut {NoStop}%
\bibitem [{\citenamefont {C\'arcamo~Hern\'andez}\ \emph
  {et~al.}(2024)\citenamefont {C\'arcamo~Hern\'andez}, \citenamefont
  {Espinoza}, \citenamefont {G\'omez-Izquierdo}, \citenamefont
  {Marchant~Gonz\'alez},\ and\ \citenamefont
  {Mondrag\'on}}]{CarcamoHernandez:2022vjk}%
  \BibitemOpen
  \bibfield  {author} {\bibinfo {author} {\bibfnamefont {A.~E.}\ \bibnamefont
  {C\'arcamo~Hern\'andez}}, \bibinfo {author} {\bibfnamefont {C.}~\bibnamefont
  {Espinoza}}, \bibinfo {author} {\bibfnamefont {J.~C.}\ \bibnamefont
  {G\'omez-Izquierdo}}, \bibinfo {author} {\bibfnamefont {J.}~\bibnamefont
  {Marchant~Gonz\'alez}},\ and\ \bibinfo {author} {\bibfnamefont
  {M.}~\bibnamefont {Mondrag\'on}},\ }\bibfield  {title} {\bibinfo {title}
  {{Phenomenology of extended multiHiggs doublet models with $S_4$ family
  symmetry}},\ }\href {https://doi.org/10.1140/epjc/s10052-024-13633-5}
  {\bibfield  {journal} {\bibinfo  {journal} {Eur. Phys. J. C}\ }\textbf
  {\bibinfo {volume} {84}},\ \bibinfo {pages} {1239} (\bibinfo {year}
  {2024})},\ \Eprint {https://arxiv.org/abs/2212.12000} {arXiv:2212.12000
  [hep-ph]} \BibitemShut {NoStop}%
\bibitem [{\citenamefont {Boto}\ \emph {et~al.}(2024)\citenamefont {Boto},
  \citenamefont {Figueiredo}, \citenamefont {Rom\~ao},\ and\ \citenamefont
  {Silva}}]{Boto:2024tzp}%
  \BibitemOpen
  \bibfield  {author} {\bibinfo {author} {\bibfnamefont {R.}~\bibnamefont
  {Boto}}, \bibinfo {author} {\bibfnamefont {P.~N.}\ \bibnamefont
  {Figueiredo}}, \bibinfo {author} {\bibfnamefont {J.~C.}\ \bibnamefont
  {Rom\~ao}},\ and\ \bibinfo {author} {\bibfnamefont {J.~a.~P.}\ \bibnamefont
  {Silva}},\ }\bibfield  {title} {\bibinfo {title} {{Novel two component dark
  matter features in the Z$_{2}$ \texttimes{} Z$_{2}$ 3HDM}},\ }\href
  {https://doi.org/10.1007/JHEP11(2024)108} {\bibfield  {journal} {\bibinfo
  {journal} {JHEP}\ }\textbf {\bibinfo {volume} {11}},\ \bibinfo {pages}
  {108}},\ \Eprint {https://arxiv.org/abs/2407.15933} {arXiv:2407.15933
  [hep-ph]} \BibitemShut {NoStop}%
\bibitem [{\citenamefont {Kun\v{c}inas}\ \emph {et~al.}(2024)\citenamefont
  {Kun\v{c}inas}, \citenamefont {Osland},\ and\ \citenamefont
  {Rebelo}}]{Kuncinas:2024zjq}%
  \BibitemOpen
  \bibfield  {author} {\bibinfo {author} {\bibfnamefont {A.}~\bibnamefont
  {Kun\v{c}inas}}, \bibinfo {author} {\bibfnamefont {P.}~\bibnamefont
  {Osland}},\ and\ \bibinfo {author} {\bibfnamefont {M.~N.}\ \bibnamefont
  {Rebelo}},\ }\bibfield  {title} {\bibinfo {title} {{U(1)-charged Dark Matter
  in three-Higgs-doublet models}},\ }\href
  {https://doi.org/10.1007/JHEP11(2024)086} {\bibfield  {journal} {\bibinfo
  {journal} {JHEP}\ }\textbf {\bibinfo {volume} {11}},\ \bibinfo {pages}
  {086}},\ \Eprint {https://arxiv.org/abs/2408.02728} {arXiv:2408.02728
  [hep-ph]} \BibitemShut {NoStop}%
\bibitem [{\citenamefont {Deng}\ \emph {et~al.}(2025)\citenamefont {Deng},
  \citenamefont {Boto}, \citenamefont {Ivanov},\ and\ \citenamefont
  {Silva}}]{Deng:2025dcq}%
  \BibitemOpen
  \bibfield  {author} {\bibinfo {author} {\bibfnamefont {H.}~\bibnamefont
  {Deng}}, \bibinfo {author} {\bibfnamefont {R.}~\bibnamefont {Boto}}, \bibinfo
  {author} {\bibfnamefont {I.~P.}\ \bibnamefont {Ivanov}},\ and\ \bibinfo
  {author} {\bibfnamefont {J.~a.~P.}\ \bibnamefont {Silva}},\ }\bibfield
  {title} {\bibinfo {title} {{Dark matter stabilized by a non-Abelian group:
  Lessons from the \ensuremath{\Sigma}(36)\,3HDM}},\ }\href
  {https://doi.org/10.1103/PhysRevD.111.055006} {\bibfield  {journal} {\bibinfo
   {journal} {Phys. Rev. D}\ }\textbf {\bibinfo {volume} {111}},\ \bibinfo
  {pages} {055006} (\bibinfo {year} {2025})},\ \Eprint
  {https://arxiv.org/abs/2501.05929} {arXiv:2501.05929 [hep-ph]} \BibitemShut
  {NoStop}%
\bibitem [{\citenamefont {Cordero}\ \emph {et~al.}(2018)\citenamefont
  {Cordero}, \citenamefont {Hernandez-Sanchez}, \citenamefont {Keus},
  \citenamefont {King}, \citenamefont {Moretti}, \citenamefont {Rojas},\ and\
  \citenamefont {Sokolowska}}]{Cordero:2017owj}%
  \BibitemOpen
  \bibfield  {author} {\bibinfo {author} {\bibfnamefont {A.}~\bibnamefont
  {Cordero}}, \bibinfo {author} {\bibfnamefont {J.}~\bibnamefont
  {Hernandez-Sanchez}}, \bibinfo {author} {\bibfnamefont {V.}~\bibnamefont
  {Keus}}, \bibinfo {author} {\bibfnamefont {S.~F.}\ \bibnamefont {King}},
  \bibinfo {author} {\bibfnamefont {S.}~\bibnamefont {Moretti}}, \bibinfo
  {author} {\bibfnamefont {D.}~\bibnamefont {Rojas}},\ and\ \bibinfo {author}
  {\bibfnamefont {D.}~\bibnamefont {Sokolowska}},\ }\bibfield  {title}
  {\bibinfo {title} {{Dark Matter Signals at the LHC from a 3HDM}},\ }\href
  {https://doi.org/10.1007/JHEP05(2018)030} {\bibfield  {journal} {\bibinfo
  {journal} {JHEP}\ }\textbf {\bibinfo {volume} {05}},\ \bibinfo {pages}
  {030}},\ \Eprint {https://arxiv.org/abs/1712.09598} {arXiv:1712.09598
  [hep-ph]} \BibitemShut {NoStop}%
\bibitem [{\citenamefont {Cordero-Cid}\ \emph {et~al.}(2020)\citenamefont
  {Cordero-Cid}, \citenamefont {Hern\'andez-S\'anchez}, \citenamefont {Keus},
  \citenamefont {Moretti}, \citenamefont {Rojas},\ and\ \citenamefont
  {Soko\l{}owska}}]{Cordero-Cid:2018man}%
  \BibitemOpen
  \bibfield  {author} {\bibinfo {author} {\bibfnamefont {A.}~\bibnamefont
  {Cordero-Cid}}, \bibinfo {author} {\bibfnamefont {J.}~\bibnamefont
  {Hern\'andez-S\'anchez}}, \bibinfo {author} {\bibfnamefont {V.}~\bibnamefont
  {Keus}}, \bibinfo {author} {\bibfnamefont {S.}~\bibnamefont {Moretti}},
  \bibinfo {author} {\bibfnamefont {D.}~\bibnamefont {Rojas}},\ and\ \bibinfo
  {author} {\bibfnamefont {D.}~\bibnamefont {Soko\l{}owska}},\ }\bibfield
  {title} {\bibinfo {title} {{Lepton collider indirect signatures of dark
  CP-violation}},\ }\href {https://doi.org/10.1140/epjc/s10052-020-7689-0}
  {\bibfield  {journal} {\bibinfo  {journal} {Eur. Phys. J. C}\ }\textbf
  {\bibinfo {volume} {80}},\ \bibinfo {pages} {135} (\bibinfo {year} {2020})},\
  \Eprint {https://arxiv.org/abs/1812.00820} {arXiv:1812.00820 [hep-ph]}
  \BibitemShut {NoStop}%
\bibitem [{\citenamefont {Aranda}\ \emph {et~al.}(2021)\citenamefont {Aranda},
  \citenamefont {Hern\'andez-Otero}, \citenamefont {Hern\'andez-Sanchez},
  \citenamefont {Keus}, \citenamefont {Moretti}, \citenamefont
  {Rojas-Ciofalo},\ and\ \citenamefont {Shindou}}]{Aranda:2019vda}%
  \BibitemOpen
  \bibfield  {author} {\bibinfo {author} {\bibfnamefont {A.}~\bibnamefont
  {Aranda}}, \bibinfo {author} {\bibfnamefont {D.}~\bibnamefont
  {Hern\'andez-Otero}}, \bibinfo {author} {\bibfnamefont {J.}~\bibnamefont
  {Hern\'andez-Sanchez}}, \bibinfo {author} {\bibfnamefont {V.}~\bibnamefont
  {Keus}}, \bibinfo {author} {\bibfnamefont {S.}~\bibnamefont {Moretti}},
  \bibinfo {author} {\bibfnamefont {D.}~\bibnamefont {Rojas-Ciofalo}},\ and\
  \bibinfo {author} {\bibfnamefont {T.}~\bibnamefont {Shindou}},\ }\bibfield
  {title} {\bibinfo {title} {{Z$_3$ symmetric inert ( 2+1 )-Higgs-doublet
  model}},\ }\href {https://doi.org/10.1103/PhysRevD.103.015023} {\bibfield
  {journal} {\bibinfo  {journal} {Phys. Rev. D}\ }\textbf {\bibinfo {volume}
  {103}},\ \bibinfo {pages} {015023} (\bibinfo {year} {2021})},\ \Eprint
  {https://arxiv.org/abs/1907.12470} {arXiv:1907.12470 [hep-ph]} \BibitemShut
  {NoStop}%
\bibitem [{\citenamefont {Hernandez-Sanchez}\ \emph {et~al.}(2020)\citenamefont
  {Hernandez-Sanchez}, \citenamefont {Keus}, \citenamefont {Moretti},
  \citenamefont {Rojas-Ciofalo},\ and\ \citenamefont
  {Sokolowska}}]{Hernandez-Sanchez:2020aop}%
  \BibitemOpen
  \bibfield  {author} {\bibinfo {author} {\bibfnamefont {J.}~\bibnamefont
  {Hernandez-Sanchez}}, \bibinfo {author} {\bibfnamefont {V.}~\bibnamefont
  {Keus}}, \bibinfo {author} {\bibfnamefont {S.}~\bibnamefont {Moretti}},
  \bibinfo {author} {\bibfnamefont {D.}~\bibnamefont {Rojas-Ciofalo}},\ and\
  \bibinfo {author} {\bibfnamefont {D.}~\bibnamefont {Sokolowska}},\ }\bibfield
   {title} {\bibinfo {title} {{Complementary Probes of Two-component Dark
  Matter}},\ }\href@noop {} {\  (\bibinfo {year} {2020})},\ \Eprint
  {https://arxiv.org/abs/2012.11621} {arXiv:2012.11621 [hep-ph]} \BibitemShut
  {NoStop}%
\bibitem [{\citenamefont {Keus}\ \emph
  {et~al.}(2014{\natexlab{a}})\citenamefont {Keus}, \citenamefont {King},
  \citenamefont {Moretti},\ and\ \citenamefont {Sokolowska}}]{Keus:2014jha}%
  \BibitemOpen
  \bibfield  {author} {\bibinfo {author} {\bibfnamefont {V.}~\bibnamefont
  {Keus}}, \bibinfo {author} {\bibfnamefont {S.~F.}\ \bibnamefont {King}},
  \bibinfo {author} {\bibfnamefont {S.}~\bibnamefont {Moretti}},\ and\ \bibinfo
  {author} {\bibfnamefont {D.}~\bibnamefont {Sokolowska}},\ }\bibfield  {title}
  {\bibinfo {title} {{Dark Matter with Two Inert Doublets plus One Higgs
  Doublet}},\ }\href {https://doi.org/10.1007/JHEP11(2014)016} {\bibfield
  {journal} {\bibinfo  {journal} {JHEP}\ }\textbf {\bibinfo {volume} {11}},\
  \bibinfo {pages} {016}},\ \Eprint {https://arxiv.org/abs/1407.7859}
  {arXiv:1407.7859 [hep-ph]} \BibitemShut {NoStop}%
\bibitem [{\citenamefont {Keus}\ \emph
  {et~al.}(2014{\natexlab{b}})\citenamefont {Keus}, \citenamefont {King},\ and\
  \citenamefont {Moretti}}]{Keus:2014isa}%
  \BibitemOpen
  \bibfield  {author} {\bibinfo {author} {\bibfnamefont {V.}~\bibnamefont
  {Keus}}, \bibinfo {author} {\bibfnamefont {S.~F.}\ \bibnamefont {King}},\
  and\ \bibinfo {author} {\bibfnamefont {S.}~\bibnamefont {Moretti}},\
  }\bibfield  {title} {\bibinfo {title} {{Phenomenology of the inert ( 2+1 )
  and ( 4+2 ) Higgs doublet models}},\ }\href
  {https://doi.org/10.1103/PhysRevD.90.075015} {\bibfield  {journal} {\bibinfo
  {journal} {Phys. Rev. D}\ }\textbf {\bibinfo {volume} {90}},\ \bibinfo
  {pages} {075015} (\bibinfo {year} {2014}{\natexlab{b}})},\ \Eprint
  {https://arxiv.org/abs/1408.0796} {arXiv:1408.0796 [hep-ph]} \BibitemShut
  {NoStop}%
\bibitem [{\citenamefont {Cordero-Cid}\ \emph {et~al.}(2016)\citenamefont
  {Cordero-Cid}, \citenamefont {Hern\'andez-S\'anchez}, \citenamefont {Keus},
  \citenamefont {King}, \citenamefont {Moretti}, \citenamefont {Rojas},\ and\
  \citenamefont {Soko\l{}owska}}]{Cordero-Cid:2016krd}%
  \BibitemOpen
  \bibfield  {author} {\bibinfo {author} {\bibfnamefont {A.}~\bibnamefont
  {Cordero-Cid}}, \bibinfo {author} {\bibfnamefont {J.}~\bibnamefont
  {Hern\'andez-S\'anchez}}, \bibinfo {author} {\bibfnamefont {V.}~\bibnamefont
  {Keus}}, \bibinfo {author} {\bibfnamefont {S.~F.}\ \bibnamefont {King}},
  \bibinfo {author} {\bibfnamefont {S.}~\bibnamefont {Moretti}}, \bibinfo
  {author} {\bibfnamefont {D.}~\bibnamefont {Rojas}},\ and\ \bibinfo {author}
  {\bibfnamefont {D.}~\bibnamefont {Soko\l{}owska}},\ }\bibfield  {title}
  {\bibinfo {title} {{CP violating scalar Dark Matter}},\ }\href
  {https://doi.org/10.1007/JHEP12(2016)014} {\bibfield  {journal} {\bibinfo
  {journal} {JHEP}\ }\textbf {\bibinfo {volume} {12}},\ \bibinfo {pages}
  {014}},\ \Eprint {https://arxiv.org/abs/1608.01673} {arXiv:1608.01673
  [hep-ph]} \BibitemShut {NoStop}%
\bibitem [{\citenamefont {Dey}\ \emph {et~al.}(2024{\natexlab{b}})\citenamefont
  {Dey}, \citenamefont {Hern\'andez-S\'anchez}, \citenamefont {Keus},
  \citenamefont {Moretti},\ and\ \citenamefont {Shindou}}]{Dey:2024epo}%
  \BibitemOpen
  \bibfield  {author} {\bibinfo {author} {\bibfnamefont {A.}~\bibnamefont
  {Dey}}, \bibinfo {author} {\bibfnamefont {J.}~\bibnamefont
  {Hern\'andez-S\'anchez}}, \bibinfo {author} {\bibfnamefont {V.}~\bibnamefont
  {Keus}}, \bibinfo {author} {\bibfnamefont {S.}~\bibnamefont {Moretti}},\ and\
  \bibinfo {author} {\bibfnamefont {T.}~\bibnamefont {Shindou}},\ }\bibfield
  {title} {\bibinfo {title} {{On the CP Properties of Spin-0 Dark Matter}},\
  }\href@noop {} {\  (\bibinfo {year} {2024}{\natexlab{b}})},\ \Eprint
  {https://arxiv.org/abs/2409.16360} {arXiv:2409.16360 [hep-ph]} \BibitemShut
  {NoStop}%
\bibitem [{\citenamefont {Das}\ \emph {et~al.}(2025)\citenamefont {Das},
  \citenamefont {Levy},\ and\ \citenamefont {Prasad}}]{Das:2025mqs}%
  \BibitemOpen
  \bibfield  {author} {\bibinfo {author} {\bibfnamefont {D.}~\bibnamefont
  {Das}}, \bibinfo {author} {\bibfnamefont {M.}~\bibnamefont {Levy}},\ and\
  \bibinfo {author} {\bibfnamefont {A.~M.}\ \bibnamefont {Prasad}},\ }\bibfield
   {title} {\bibinfo {title} {{Flavor puzzle in three Higgs-doublet models:
  Insights from BGL and lessons from flavor data}},\ }\href@noop {} {\
  (\bibinfo {year} {2025})},\ \Eprint {https://arxiv.org/abs/2502.20296}
  {arXiv:2502.20296 [hep-ph]} \BibitemShut {NoStop}%
\bibitem [{\citenamefont {Altmannshofer}\ and\ \citenamefont
  {Toner}(2025)}]{Altmannshofer:2025pjj}%
  \BibitemOpen
  \bibfield  {author} {\bibinfo {author} {\bibfnamefont {W.}~\bibnamefont
  {Altmannshofer}}\ and\ \bibinfo {author} {\bibfnamefont {K.}~\bibnamefont
  {Toner}},\ }\bibfield  {title} {\bibinfo {title} {{Flavor Constraints in a
  Generational Three Higgs Doublet Model}},\ }\href@noop {} {\  (\bibinfo
  {year} {2025})},\ \Eprint {https://arxiv.org/abs/2502.04579}
  {arXiv:2502.04579 [hep-ph]} \BibitemShut {NoStop}%
\bibitem [{\citenamefont {Rom\~ao}\ and\ \citenamefont {Crispim
  Rom\~ao}(2024)}]{Romao:2024gjx}%
  \BibitemOpen
  \bibfield  {author} {\bibinfo {author} {\bibfnamefont {J.~C.}\ \bibnamefont
  {Rom\~ao}}\ and\ \bibinfo {author} {\bibfnamefont {M.}~\bibnamefont {Crispim
  Rom\~ao}},\ }\bibfield  {title} {\bibinfo {title} {{Combining evolutionary
  strategies and novelty detection to go beyond the alignment limit of the Z3
  3HDM}},\ }\href {https://doi.org/10.1103/PhysRevD.109.095040} {\bibfield
  {journal} {\bibinfo  {journal} {Phys. Rev. D}\ }\textbf {\bibinfo {volume}
  {109}},\ \bibinfo {pages} {095040} (\bibinfo {year} {2024})},\ \Eprint
  {https://arxiv.org/abs/2402.07661} {arXiv:2402.07661 [hep-ph]} \BibitemShut
  {NoStop}%
\bibitem [{\citenamefont {Bento}\ \emph {et~al.}(2022)\citenamefont {Bento},
  \citenamefont {Romão},\ and\ \citenamefont {Silva}}]{Bento_2022}%
  \BibitemOpen
  \bibfield  {author} {\bibinfo {author} {\bibfnamefont {M.~P.}\ \bibnamefont
  {Bento}}, \bibinfo {author} {\bibfnamefont {J.~C.}\ \bibnamefont {Romão}},\
  and\ \bibinfo {author} {\bibfnamefont {J.~P.}\ \bibnamefont {Silva}},\
  }\bibfield  {title} {\bibinfo {title} {Unitarity bounds for all
  symmetry-constrained 3hdms},\ }\bibfield  {journal} {\bibinfo  {journal}
  {Journal of High Energy Physics}\ }\textbf {\bibinfo {volume} {2022}},\ \href
  {https://doi.org/10.1007/jhep08(2022)273} {10.1007/jhep08(2022)273} (\bibinfo
  {year} {2022})\BibitemShut {NoStop}%
\bibitem [{\citenamefont {Boto}\ \emph {et~al.}(2021)\citenamefont {Boto},
  \citenamefont {Romão},\ and\ \citenamefont {Silva}}]{Boto_2021}%
  \BibitemOpen
  \bibfield  {author} {\bibinfo {author} {\bibfnamefont {R.}~\bibnamefont
  {Boto}}, \bibinfo {author} {\bibfnamefont {J.~C.}\ \bibnamefont {Romão}},\
  and\ \bibinfo {author} {\bibfnamefont {J.~P.}\ \bibnamefont {Silva}},\
  }\bibfield  {title} {\bibinfo {title} {Current bounds on the type-z $z_3$
  three-higgs-doublet model},\ }\bibfield  {journal} {\bibinfo  {journal}
  {Physical Review D}\ }\textbf {\bibinfo {volume} {104}},\ \href
  {https://doi.org/10.1103/physrevd.104.095006} {10.1103/physrevd.104.095006}
  (\bibinfo {year} {2021})\BibitemShut {NoStop}%
\bibitem [{\citenamefont {Das}\ \emph {et~al.}(2023)\citenamefont {Das},
  \citenamefont {Levy}, \citenamefont {Pal}, \citenamefont {Prasad},
  \citenamefont {Saha},\ and\ \citenamefont {Srivastava}}]{Das:2022gbm}%
  \BibitemOpen
  \bibfield  {author} {\bibinfo {author} {\bibfnamefont {D.}~\bibnamefont
  {Das}}, \bibinfo {author} {\bibfnamefont {M.}~\bibnamefont {Levy}}, \bibinfo
  {author} {\bibfnamefont {P.~B.}\ \bibnamefont {Pal}}, \bibinfo {author}
  {\bibfnamefont {A.~M.}\ \bibnamefont {Prasad}}, \bibinfo {author}
  {\bibfnamefont {I.}~\bibnamefont {Saha}},\ and\ \bibinfo {author}
  {\bibfnamefont {A.}~\bibnamefont {Srivastava}},\ }\bibfield  {title}
  {\bibinfo {title} {{Democratic three-Higgs-doublet models: The custodial
  limit and wrong-sign Yukawa coupling}},\ }\href
  {https://doi.org/10.1103/PhysRevD.107.055035} {\bibfield  {journal} {\bibinfo
   {journal} {Phys. Rev. D}\ }\textbf {\bibinfo {volume} {107}},\ \bibinfo
  {pages} {055035} (\bibinfo {year} {2023})},\ \Eprint
  {https://arxiv.org/abs/2301.00231} {arXiv:2301.00231 [hep-ph]} \BibitemShut
  {NoStop}%
\bibitem [{\citenamefont {Chakraborti}\ \emph {et~al.}(2021)\citenamefont
  {Chakraborti}, \citenamefont {Das}, \citenamefont {Levy}, \citenamefont
  {Mukherjee},\ and\ \citenamefont {Saha}}]{Chakraborti:2021bpy}%
  \BibitemOpen
  \bibfield  {author} {\bibinfo {author} {\bibfnamefont {M.}~\bibnamefont
  {Chakraborti}}, \bibinfo {author} {\bibfnamefont {D.}~\bibnamefont {Das}},
  \bibinfo {author} {\bibfnamefont {M.}~\bibnamefont {Levy}}, \bibinfo {author}
  {\bibfnamefont {S.}~\bibnamefont {Mukherjee}},\ and\ \bibinfo {author}
  {\bibfnamefont {I.}~\bibnamefont {Saha}},\ }\bibfield  {title} {\bibinfo
  {title} {{Prospects for light charged scalars in a three-Higgs-doublet model
  with Z3 symmetry}},\ }\href {https://doi.org/10.1103/PhysRevD.104.075033}
  {\bibfield  {journal} {\bibinfo  {journal} {Phys. Rev. D}\ }\textbf {\bibinfo
  {volume} {104}},\ \bibinfo {pages} {075033} (\bibinfo {year} {2021})},\
  \Eprint {https://arxiv.org/abs/2104.08146} {arXiv:2104.08146 [hep-ph]}
  \BibitemShut {NoStop}%
\bibitem [{\citenamefont {Akeroyd}\ \emph {et~al.}(2017)\citenamefont
  {Akeroyd}, \citenamefont {Moretti}, \citenamefont {Yagyu},\ and\
  \citenamefont {Yildirim}}]{Akeroyd:2016ssd}%
  \BibitemOpen
  \bibfield  {author} {\bibinfo {author} {\bibfnamefont {A.~G.}\ \bibnamefont
  {Akeroyd}}, \bibinfo {author} {\bibfnamefont {S.}~\bibnamefont {Moretti}},
  \bibinfo {author} {\bibfnamefont {K.}~\bibnamefont {Yagyu}},\ and\ \bibinfo
  {author} {\bibfnamefont {E.}~\bibnamefont {Yildirim}},\ }\bibfield  {title}
  {\bibinfo {title} {{Light charged Higgs boson scenario in 3-Higgs doublet
  models}},\ }\href {https://doi.org/10.1142/S0217751X17501457} {\bibfield
  {journal} {\bibinfo  {journal} {Int. J. Mod. Phys. A}\ }\textbf {\bibinfo
  {volume} {32}},\ \bibinfo {pages} {1750145} (\bibinfo {year} {2017})},\
  \Eprint {https://arxiv.org/abs/1605.05881} {arXiv:1605.05881 [hep-ph]}
  \BibitemShut {NoStop}%
\bibitem [{\citenamefont {Akeroyd}\ \emph {et~al.}(2018)\citenamefont
  {Akeroyd}, \citenamefont {Moretti},\ and\ \citenamefont
  {Song}}]{Akeroyd:2018axd}%
  \BibitemOpen
  \bibfield  {author} {\bibinfo {author} {\bibfnamefont {A.~G.}\ \bibnamefont
  {Akeroyd}}, \bibinfo {author} {\bibfnamefont {S.}~\bibnamefont {Moretti}},\
  and\ \bibinfo {author} {\bibfnamefont {M.}~\bibnamefont {Song}},\ }\bibfield
  {title} {\bibinfo {title} {{Light charged Higgs boson with dominant decay to
  quarks and its search at the LHC and future colliders}},\ }\href
  {https://doi.org/10.1103/PhysRevD.98.115024} {\bibfield  {journal} {\bibinfo
  {journal} {Phys. Rev. D}\ }\textbf {\bibinfo {volume} {98}},\ \bibinfo
  {pages} {115024} (\bibinfo {year} {2018})},\ \Eprint
  {https://arxiv.org/abs/1810.05403} {arXiv:1810.05403 [hep-ph]} \BibitemShut
  {NoStop}%
\bibitem [{\citenamefont {Akeroyd}\ \emph {et~al.}(2020)\citenamefont
  {Akeroyd}, \citenamefont {Moretti},\ and\ \citenamefont
  {Song}}]{Akeroyd:2019mvt}%
  \BibitemOpen
  \bibfield  {author} {\bibinfo {author} {\bibfnamefont {A.~G.}\ \bibnamefont
  {Akeroyd}}, \bibinfo {author} {\bibfnamefont {S.}~\bibnamefont {Moretti}},\
  and\ \bibinfo {author} {\bibfnamefont {M.}~\bibnamefont {Song}},\ }\bibfield
  {title} {\bibinfo {title} {{Light charged Higgs boson with dominant decay to
  a charm quark and a bottom quark and its search at LEP2 and future $e^+e^-$
  colliders}},\ }\href {https://doi.org/10.1103/PhysRevD.101.035021} {\bibfield
   {journal} {\bibinfo  {journal} {Phys. Rev. D}\ }\textbf {\bibinfo {volume}
  {101}},\ \bibinfo {pages} {035021} (\bibinfo {year} {2020})},\ \Eprint
  {https://arxiv.org/abs/1908.00826} {arXiv:1908.00826 [hep-ph]} \BibitemShut
  {NoStop}%
\bibitem [{\citenamefont {Akeroyd}\ \emph {et~al.}(2022)\citenamefont
  {Akeroyd}, \citenamefont {Moretti},\ and\ \citenamefont
  {Song}}]{Akeroyd:2022ouy}%
  \BibitemOpen
  \bibfield  {author} {\bibinfo {author} {\bibfnamefont {A.~G.}\ \bibnamefont
  {Akeroyd}}, \bibinfo {author} {\bibfnamefont {S.}~\bibnamefont {Moretti}},\
  and\ \bibinfo {author} {\bibfnamefont {M.}~\bibnamefont {Song}},\ }\bibfield
  {title} {\bibinfo {title} {{Slight excess at 130 GeV in search for a charged
  Higgs boson decaying to a charm quark and a bottom quark at the Large Hadron
  Collider}},\ }\href {https://doi.org/10.1088/1361-6471/ac77a6} {\bibfield
  {journal} {\bibinfo  {journal} {J. Phys. G}\ }\textbf {\bibinfo {volume}
  {49}},\ \bibinfo {pages} {085004} (\bibinfo {year} {2022})},\ \Eprint
  {https://arxiv.org/abs/2202.03522} {arXiv:2202.03522 [hep-ph]} \BibitemShut
  {NoStop}%
\bibitem [{\citenamefont {Das}\ and\ \citenamefont {Saha}(2019)}]{Das:2019yad}%
  \BibitemOpen
  \bibfield  {author} {\bibinfo {author} {\bibfnamefont {D.}~\bibnamefont
  {Das}}\ and\ \bibinfo {author} {\bibfnamefont {I.}~\bibnamefont {Saha}},\
  }\bibfield  {title} {\bibinfo {title} {{Alignment limit in three
  Higgs-doublet models}},\ }\href {https://doi.org/10.1103/PhysRevD.100.035021}
  {\bibfield  {journal} {\bibinfo  {journal} {Phys. Rev. D}\ }\textbf {\bibinfo
  {volume} {100}},\ \bibinfo {pages} {035021} (\bibinfo {year} {2019})},\
  \Eprint {https://arxiv.org/abs/1904.03970} {arXiv:1904.03970 [hep-ph]}
  \BibitemShut {NoStop}%
\bibitem [{\citenamefont {Cree}\ and\ \citenamefont
  {Logan}(2011)}]{Cree:2011uy}%
  \BibitemOpen
  \bibfield  {author} {\bibinfo {author} {\bibfnamefont {G.}~\bibnamefont
  {Cree}}\ and\ \bibinfo {author} {\bibfnamefont {H.~E.}\ \bibnamefont
  {Logan}},\ }\bibfield  {title} {\bibinfo {title} {{Yukawa alignment from
  natural flavor conservation}},\ }\href
  {https://doi.org/10.1103/PhysRevD.84.055021} {\bibfield  {journal} {\bibinfo
  {journal} {Phys. Rev. D}\ }\textbf {\bibinfo {volume} {84}},\ \bibinfo
  {pages} {055021} (\bibinfo {year} {2011})},\ \Eprint
  {https://arxiv.org/abs/1106.4039} {arXiv:1106.4039 [hep-ph]} \BibitemShut
  {NoStop}%
\bibitem [{\citenamefont {Keus}\ \emph
  {et~al.}(2014{\natexlab{c}})\citenamefont {Keus}, \citenamefont {King},\ and\
  \citenamefont {Moretti}}]{Keus:2013hya}%
  \BibitemOpen
  \bibfield  {author} {\bibinfo {author} {\bibfnamefont {V.}~\bibnamefont
  {Keus}}, \bibinfo {author} {\bibfnamefont {S.~F.}\ \bibnamefont {King}},\
  and\ \bibinfo {author} {\bibfnamefont {S.}~\bibnamefont {Moretti}},\
  }\bibfield  {title} {\bibinfo {title} {{Three-Higgs-doublet models:
  symmetries, potentials and Higgs boson masses}},\ }\href
  {https://doi.org/10.1007/JHEP01(2014)052} {\bibfield  {journal} {\bibinfo
  {journal} {JHEP}\ }\textbf {\bibinfo {volume} {01}},\ \bibinfo {pages}
  {052}},\ \Eprint {https://arxiv.org/abs/1310.8253} {arXiv:1310.8253 [hep-ph]}
  \BibitemShut {NoStop}%
\bibitem [{\citenamefont {Glashow}\ and\ \citenamefont
  {Weinberg}(1977)}]{Glashow:1976nt}%
  \BibitemOpen
  \bibfield  {author} {\bibinfo {author} {\bibfnamefont {S.~L.}\ \bibnamefont
  {Glashow}}\ and\ \bibinfo {author} {\bibfnamefont {S.}~\bibnamefont
  {Weinberg}},\ }\bibfield  {title} {\bibinfo {title} {{Natural Conservation
  Laws for Neutral Currents}},\ }\href
  {https://doi.org/10.1103/PhysRevD.15.1958} {\bibfield  {journal} {\bibinfo
  {journal} {Phys. Rev. D}\ }\textbf {\bibinfo {volume} {15}},\ \bibinfo
  {pages} {1958} (\bibinfo {year} {1977})}\BibitemShut {NoStop}%
\bibitem [{\citenamefont {Akeroyd}\ \emph {et~al.}(2000)\citenamefont
  {Akeroyd}, \citenamefont {Arhrib},\ and\ \citenamefont
  {Naimi}}]{Akeroyd:2000wc}%
  \BibitemOpen
  \bibfield  {author} {\bibinfo {author} {\bibfnamefont {A.~G.}\ \bibnamefont
  {Akeroyd}}, \bibinfo {author} {\bibfnamefont {A.}~\bibnamefont {Arhrib}},\
  and\ \bibinfo {author} {\bibfnamefont {E.-M.}\ \bibnamefont {Naimi}},\
  }\bibfield  {title} {\bibinfo {title} {{Note on tree level unitarity in the
  general two Higgs doublet model}},\ }\href
  {https://doi.org/10.1016/S0370-2693(00)00962-X} {\bibfield  {journal}
  {\bibinfo  {journal} {Phys. Lett. B}\ }\textbf {\bibinfo {volume} {490}},\
  \bibinfo {pages} {119} (\bibinfo {year} {2000})},\ \Eprint
  {https://arxiv.org/abs/hep-ph/0006035} {arXiv:hep-ph/0006035} \BibitemShut
  {NoStop}%
\bibitem [{\citenamefont {Bahl}\ \emph {et~al.}(2023)\citenamefont {Bahl},
  \citenamefont {Biekötter}, \citenamefont {Heinemeyer}, \citenamefont {Li},
  \citenamefont {Paasch}, \citenamefont {Weiglein},\ and\ \citenamefont
  {Wittbrodt}}]{Bahl_2023}%
  \BibitemOpen
  \bibfield  {author} {\bibinfo {author} {\bibfnamefont {H.}~\bibnamefont
  {Bahl}}, \bibinfo {author} {\bibfnamefont {T.}~\bibnamefont {Biekötter}},
  \bibinfo {author} {\bibfnamefont {S.}~\bibnamefont {Heinemeyer}}, \bibinfo
  {author} {\bibfnamefont {C.}~\bibnamefont {Li}}, \bibinfo {author}
  {\bibfnamefont {S.}~\bibnamefont {Paasch}}, \bibinfo {author} {\bibfnamefont
  {G.}~\bibnamefont {Weiglein}},\ and\ \bibinfo {author} {\bibfnamefont
  {J.}~\bibnamefont {Wittbrodt}},\ }\bibfield  {title} {\bibinfo {title}
  {Higgstools: Bsm scalar phenomenology with new versions of higgsbounds and
  higgssignals},\ }\href {https://doi.org/10.1016/j.cpc.2023.108803} {\bibfield
   {journal} {\bibinfo  {journal} {Computer Physics Communications}\ }\textbf
  {\bibinfo {volume} {291}},\ \bibinfo {pages} {108803} (\bibinfo {year}
  {2023})}\BibitemShut {NoStop}%
\bibitem [{\citenamefont {Benbrik}\ \emph {et~al.}(2024)\citenamefont
  {Benbrik}, \citenamefont {Boukidi},\ and\ \citenamefont
  {Moretti}}]{Benbrik:2024ptw}%
  \BibitemOpen
  \bibfield  {author} {\bibinfo {author} {\bibfnamefont {R.}~\bibnamefont
  {Benbrik}}, \bibinfo {author} {\bibfnamefont {M.}~\bibnamefont {Boukidi}},\
  and\ \bibinfo {author} {\bibfnamefont {S.}~\bibnamefont {Moretti}},\
  }\bibfield  {title} {\bibinfo {title} {{Superposition of CP-even and CP-odd
  Higgs resonances: Explaining the 95~GeV excesses within a two-Higgs-doublet
  model}},\ }\href {https://doi.org/10.1103/PhysRevD.110.115030} {\bibfield
  {journal} {\bibinfo  {journal} {Phys. Rev. D}\ }\textbf {\bibinfo {volume}
  {110}},\ \bibinfo {pages} {115030} (\bibinfo {year} {2024})},\ \Eprint
  {https://arxiv.org/abs/2405.02899} {arXiv:2405.02899 [hep-ph]} \BibitemShut
  {NoStop}%
\bibitem [{\citenamefont {Belyaev}\ \emph {et~al.}(2024)\citenamefont
  {Belyaev}, \citenamefont {Benbrik}, \citenamefont {Boukidi}, \citenamefont
  {Chakraborti}, \citenamefont {Moretti},\ and\ \citenamefont
  {Semlali}}]{Belyaev:2023xnv}%
  \BibitemOpen
  \bibfield  {author} {\bibinfo {author} {\bibfnamefont {A.}~\bibnamefont
  {Belyaev}}, \bibinfo {author} {\bibfnamefont {R.}~\bibnamefont {Benbrik}},
  \bibinfo {author} {\bibfnamefont {M.}~\bibnamefont {Boukidi}}, \bibinfo
  {author} {\bibfnamefont {M.}~\bibnamefont {Chakraborti}}, \bibinfo {author}
  {\bibfnamefont {S.}~\bibnamefont {Moretti}},\ and\ \bibinfo {author}
  {\bibfnamefont {S.}~\bibnamefont {Semlali}},\ }\bibfield  {title} {\bibinfo
  {title} {{Explanation of the hints for a 95 GeV Higgs boson within a 2-Higgs
  Doublet Model}},\ }\href {https://doi.org/10.1007/JHEP05(2024)209} {\bibfield
   {journal} {\bibinfo  {journal} {JHEP}\ }\textbf {\bibinfo {volume} {05}},\
  \bibinfo {pages} {209}},\ \Eprint {https://arxiv.org/abs/2306.09029}
  {arXiv:2306.09029 [hep-ph]} \BibitemShut {NoStop}%
\bibitem [{\citenamefont {Borzumati}\ and\ \citenamefont
  {Greub}(1998)}]{Borzumati_1998}%
  \BibitemOpen
  \bibfield  {author} {\bibinfo {author} {\bibfnamefont {F.~M.}\ \bibnamefont
  {Borzumati}}\ and\ \bibinfo {author} {\bibfnamefont {C.}~\bibnamefont
  {Greub}},\ }\bibfield  {title} {\bibinfo {title} {Two higgs doublet model
  predictions for $b \rightarrow x_s \gamma$ in nlo qcd},\ }\bibfield
  {journal} {\bibinfo  {journal} {Physical Review D}\ }\textbf {\bibinfo
  {volume} {58}},\ \href {https://doi.org/10.1103/physrevd.58.074004}
  {10.1103/physrevd.58.074004} (\bibinfo {year} {1998})\BibitemShut {NoStop}%
\bibitem [{\citenamefont {Akeroyd}\ \emph
  {et~al.}(2021{\natexlab{b}})\citenamefont {Akeroyd}, \citenamefont {Moretti},
  \citenamefont {Shindou},\ and\ \citenamefont {Song}}]{Akeroyd_2021}%
  \BibitemOpen
  \bibfield  {author} {\bibinfo {author} {\bibfnamefont {A.}~\bibnamefont
  {Akeroyd}}, \bibinfo {author} {\bibfnamefont {S.}~\bibnamefont {Moretti}},
  \bibinfo {author} {\bibfnamefont {T.}~\bibnamefont {Shindou}},\ and\ \bibinfo
  {author} {\bibfnamefont {M.}~\bibnamefont {Song}},\ }\bibfield  {title}
  {\bibinfo {title} {Cp asymmetries of $\bar{B} \rightarrow x_s/x_d \gamma$ in
  models with three higgs doublets},\ }\bibfield  {journal} {\bibinfo
  {journal} {Physical Review D}\ }\textbf {\bibinfo {volume} {103}},\ \href
  {https://doi.org/10.1103/physrevd.103.015035} {10.1103/physrevd.103.015035}
  (\bibinfo {year} {2021}{\natexlab{b}})\BibitemShut {NoStop}%
\bibitem [{\citenamefont {Workman}\ and\ \citenamefont
  {Others}(2022)}]{Workman:2022ynf}%
  \BibitemOpen
  \bibfield  {author} {\bibinfo {author} {\bibfnamefont {R.~L.}\ \bibnamefont
  {Workman}}\ and\ \bibinfo {author} {\bibnamefont {Others}} (\bibinfo
  {collaboration} {Particle Data Group}),\ }\bibfield  {title} {\bibinfo
  {title} {{Review of Particle Physics}},\ }\href
  {https://doi.org/10.1093/ptep/ptac097} {\bibfield  {journal} {\bibinfo
  {journal} {PTEP}\ }\textbf {\bibinfo {volume} {2022}},\ \bibinfo {pages}
  {083C01} (\bibinfo {year} {2022})}\BibitemShut {NoStop}%
\bibitem [{\citenamefont {Porod}(2003)}]{Porod_2003}%
  \BibitemOpen
  \bibfield  {author} {\bibinfo {author} {\bibfnamefont {W.}~\bibnamefont
  {Porod}},\ }\bibfield  {title} {\bibinfo {title} {Spheno, a program for
  calculating supersymmetric spectra, susy particle decays and susy particle
  production at e+e- colliders},\ }\href
  {https://doi.org/10.1016/s0010-4655(03)00222-4} {\bibfield  {journal}
  {\bibinfo  {journal} {Computer Physics Communications}\ }\textbf {\bibinfo
  {volume} {153}},\ \bibinfo {pages} {275–315} (\bibinfo {year}
  {2003})}\BibitemShut {NoStop}%
\bibitem [{\citenamefont {Porod}\ and\ \citenamefont
  {Staub}(2012)}]{Porod_2012}%
  \BibitemOpen
  \bibfield  {author} {\bibinfo {author} {\bibfnamefont {W.}~\bibnamefont
  {Porod}}\ and\ \bibinfo {author} {\bibfnamefont {F.}~\bibnamefont {Staub}},\
  }\bibfield  {title} {\bibinfo {title} {Spheno 3.1: extensions including
  flavour, cp-phases and models beyond the mssm},\ }\href
  {https://doi.org/10.1016/j.cpc.2012.05.021} {\bibfield  {journal} {\bibinfo
  {journal} {Computer Physics Communications}\ }\textbf {\bibinfo {volume}
  {183}},\ \bibinfo {pages} {2458–2469} (\bibinfo {year} {2012})}\BibitemShut
  {NoStop}%
\bibitem [{\citenamefont {Staub}(2014)}]{Staub_2014}%
  \BibitemOpen
  \bibfield  {author} {\bibinfo {author} {\bibfnamefont {F.}~\bibnamefont
  {Staub}},\ }\bibfield  {title} {\bibinfo {title} {Sarah 4: A tool for (not
  only susy) model builders},\ }\href
  {https://doi.org/10.1016/j.cpc.2014.02.018} {\bibfield  {journal} {\bibinfo
  {journal} {Computer Physics Communications}\ }\textbf {\bibinfo {volume}
  {185}},\ \bibinfo {pages} {1773–1790} (\bibinfo {year} {2014})}\BibitemShut
  {NoStop}%
\bibitem [{\citenamefont {Grimus}\ \emph {et~al.}(2008)\citenamefont {Grimus},
  \citenamefont {Lavoura}, \citenamefont {Ogreid},\ and\ \citenamefont
  {Osland}}]{Grimus_2008}%
  \BibitemOpen
  \bibfield  {author} {\bibinfo {author} {\bibfnamefont {W.}~\bibnamefont
  {Grimus}}, \bibinfo {author} {\bibfnamefont {L.}~\bibnamefont {Lavoura}},
  \bibinfo {author} {\bibfnamefont {O.~M.}\ \bibnamefont {Ogreid}},\ and\
  \bibinfo {author} {\bibfnamefont {P.}~\bibnamefont {Osland}},\ }\bibfield
  {title} {\bibinfo {title} {A precision constraint on multi-higgs-doublet
  models},\ }\href {https://doi.org/10.1088/0954-3899/35/7/075001} {\bibfield
  {journal} {\bibinfo  {journal} {Journal of Physics G: Nuclear and Particle
  Physics}\ }\textbf {\bibinfo {volume} {35}},\ \bibinfo {pages} {075001}
  (\bibinfo {year} {2008})}\BibitemShut {NoStop}%
\bibitem [{\citenamefont {Feickert}\ and\ \citenamefont
  {Nachman}(2021)}]{Feickert:2021ajf}%
  \BibitemOpen
  \bibfield  {author} {\bibinfo {author} {\bibfnamefont {M.}~\bibnamefont
  {Feickert}}\ and\ \bibinfo {author} {\bibfnamefont {B.}~\bibnamefont
  {Nachman}},\ }\bibfield  {title} {\bibinfo {title} {{A Living Review of
  Machine Learning for Particle Physics}},\ }\href@noop {} {\  (\bibinfo {year}
  {2021})},\ \Eprint {https://arxiv.org/abs/2102.02770} {arXiv:2102.02770
  [hep-ph]} \BibitemShut {NoStop}%
\bibitem [{\citenamefont {Plehn}\ \emph {et~al.}(2022)\citenamefont {Plehn},
  \citenamefont {Butter}, \citenamefont {Dillon}, \citenamefont {Heimel},
  \citenamefont {Krause},\ and\ \citenamefont {Winterhalder}}]{Plehn:2022ftl}%
  \BibitemOpen
  \bibfield  {author} {\bibinfo {author} {\bibfnamefont {T.}~\bibnamefont
  {Plehn}}, \bibinfo {author} {\bibfnamefont {A.}~\bibnamefont {Butter}},
  \bibinfo {author} {\bibfnamefont {B.}~\bibnamefont {Dillon}}, \bibinfo
  {author} {\bibfnamefont {T.}~\bibnamefont {Heimel}}, \bibinfo {author}
  {\bibfnamefont {C.}~\bibnamefont {Krause}},\ and\ \bibinfo {author}
  {\bibfnamefont {R.}~\bibnamefont {Winterhalder}},\ }\bibfield  {title}
  {\bibinfo {title} {{Modern Machine Learning for LHC Physicists}},\
  }\href@noop {} {\  (\bibinfo {year} {2022})},\ \Eprint
  {https://arxiv.org/abs/2211.01421} {arXiv:2211.01421 [hep-ph]} \BibitemShut
  {NoStop}%
\bibitem [{\citenamefont {Settles}(2009)}]{Settles2009ActiveLL}%
  \BibitemOpen
  \bibfield  {author} {\bibinfo {author} {\bibfnamefont {B.}~\bibnamefont
  {Settles}},\ }\href {http://digital.library.wisc.edu/1793/60660} {\emph
  {\bibinfo {title} {Active Learning Literature Survey}}},\ \bibinfo {type}
  {Tech. Rep.}\ \bibinfo {number} {TR1648}\ (\bibinfo  {institution}
  {University of Wisconsin-Madison Department of Computer Sciences},\ \bibinfo
  {year} {2009})\BibitemShut {NoStop}%
\end{thebibliography}%

\end{document}